\documentclass[twocolumn]{aa}  
\usepackage{graphicx,natbib}
\usepackage{txfonts}
\bibpunct{(}{)}{;}{a}{}{,}
\begin{document}
   \title{Analysis of 26 Barium Stars 
   \thanks{Based on spectroscopic observations collected at the 
           European Southern Observatory (ESO), within the 
           Observat\'orio Nacional ON/ESO and ON/IAG agreements, 
           under FAPESP project n$^{\circ}$ 1998/10138-8. Photometric 
           observations collected at the Observat\'orio do Pico dos Dias (LNA/MCT).}
           \fnmsep\thanks{Tables \ref{normais} and \ref{resproc} are only available in elecronic 
           at the CDS via anonymous ftp to cdsarc.u-strasbg.fr/Abstract.html}}

   \subtitle{II. Contributions of s-, r- and p-processes in the production of
                 heavy elements}

   \author{D.M. Allen
          \inst{}\thanks{Present address: Observat\'orio do Valongo/UFRJ,
              Ladeira do Pedro Antonio 43, 20080-090 Rio de Janeiro, RJ, Brazil}
          \and
          B. Barbuy
	  \inst{}
          }

   \offprints{D.M. Allen}

   \institute{Instituto de Astronomia, Geof\'\i sica e Ci\^encias
    Atmosf\'ericas, Universidade de S\~ao Paulo,
              Rua do Mat\~ao 1226, 05508-900 S\~ao Paulo, Brazil,
              \email{dinah@astro.iag.usp.br, barbuy@astro.iag.usp.br}
              }

   \date{Received, 2006; accepted, 2006}

 
  \abstract
   {Barium stars show enhanced abundances of the slow neutron capture (s-process)
   heavy elements, and for this reason they are suitable objects for
   the study of s-process elements.}
   {The aim of this work is to quantify the contributions
    of the s-, r- and p-processes for the total abundance of heavy elements from  
    abundances derived for a sample of 26 barium stars. The abundance ratios between these 
    processes and neutron exposures were studied.}
   {The abundances of the sample stars were compared to those of normal stars 
   thus identifying the fraction relative to the s-process main component.}
   {The fittings of the $\sigma$N curves (neutron capture cross section times 
    abundance, plotted against atomic mass number) for the sample stars suggest 
    that the material from the companion asymptotic giant branch star had 
    approximately the solar isotopic composition as concerns 
    fractions of abundances relative to the s-process main component. The abundance 
    ratios of heavy elements, hs, ls and s and the computed neutron exposure
    are similar to those of post-AGB stars. For some
    sample stars, an exponential neutron exposure fits well the observed data, 
    whereas for others, a single neutron exposure provides a better fit.}
   {The comparison between barium and AGB stars supports the hypothesis of binarity 
    for the barium star formation. Abundances of r-elements that are part of the 
    s-process path in barium stars are usually higher than those in normal stars,
    and for this reason, barium stars seemed to be also 
    enriched in r-elements, although in a lower degree than s-elements. 
    No dependence on luminosity classes was found in the abundance ratios behaviour 
    among the dwarfs and giants of the sample barium stars.}


   \maketitle
%

\section{Introduction}
Nucleosynthesis of elements benefited from the dispute between George Gamow
and Fred Hoyle in explaining how the known chemical elements were formed
\citep{alpher48,hoyle46}. Since then, our knowledge in primordial and stellar 
nucleosynthesis has greatly improved. \citet[][ B$^2$FH]{b2fh} have based
their work on solar abundances, and defined eight nucleosynthetic processes in 
stars, that would be responsible for the formation of the chemical elements.
\citet{wallers97} presented a review of B$^2$FH, including recent and accurate 
experimental results and observations, where they showed that some processes 
defined by B$^2$FH have been confirmed, others redefined and new processes
not known by B$^2$FH have been included in the list.

Three of the eight processes defined by B$^2$FH will be focused
in the present work, the s-, r- and p-processes. We present abundance
ratios between these processes and abundance distribution for a number of heavy 
elements in a sample of barium stars. Abundance determination for the sample stars 
was described in \citet[][paper I hereafter]{papI}.

The abundance ratios of best representatives of s- and r-processes provide
clues on the knowledge of the formation and evolution of the Galaxy, since
each of them is related to a different formation site, including stars of
different characteristics and evolutionary stages. As an example, \citet{mashonk03}
estimated the timescale for the thick disk and halo formation based on 
abundance ratios of [Eu/Ba], [Mg/Fe] and [Eu/Fe], and on the calculations of
chemical evolution by \citet{travaglio99}. \citet{burr00} concluded, from
abundances of metal-poor giant stars, that the contributions of s-processes can be seen at
metallicities as low as [Fe/H] = -2.75, and they are present in stars of
[Fe/H] $>$ -2.3, indicating that the s-process acts at lower metallicities
than predicted by previous work.

Considering the scenario of barium stars formation through enriched material 
transfer from a companion AGB star, it is worth to compare abundances
of AGBs and barium stars. Such study can provide clues for the 
understanding of the enrichment in heavy elements of barium stars, given that 
they should conserve the surface characteristics of AGB stars.

This work intends to address the following questions:

1. What are the s-elements abundances of normal stars?


2. Considering the hypothesis of transfer of material enriched in carbon and 
s-elements from a more evolved companion for barium stars formation, is it
possible to estimate in which proportion these elements are received, from
observed abundances?

3. What is the isotopic composition of the material received? Is it possible
to consider the solar system mix?

4. How is the $\sigma$N curve behaviour for barium stars?

5. What kind of neutron exposures were involved in the nucleosynthesis of the
transferred material?

This paper is organized as follows: Sect. 2 is a brief description of atmospheric
parameters and abundance determination carried out in paper I; Sect. 3 explains 
briefly the s-, r- and p-processes;
Sect. 4 shows the contributions of s-, r- and p-processes in normal stars  
compared with barium stars; Sect. 5 evaluates two kinds of neutron exposures
and shows ratios involving s, hs and ls. In Sect. 6 conclusions are drawn.

%

\section{Atmospheric Parameters and Abundance Calculations}
The observations, atmospheric parameters and abundance calculations were
described in detail in paper I. Here, we only outline the more
important steps.

Photometric data were taken in several runs at the ZEISS 60cm 
telescope at LNA (Laborat\'orio Nacional de Astrof\'\i sica) and from the literature.
Optical spectra were obtained at the 1.52m telescope at ESO, La Silla, 
using the Fiber Fed Extended Range Optical Spectrograph (FEROS) 
\citep{kaufer00}. A set of atmospheric parameters (temperature, surface gravity, 
metallicities and microturbulent velocity) was obtained in an iterative way. 

Photospheric 1D models were extracted from the NMARCS grid \citep{plez92} 
for gravities log g $<$ 3.3, and from \citet{edv93} for less evolved stars, 
with log g $\geq$ 3.3.

The LTE abundance analysis and the spectrum synthesis calculations
were performed using the codes by 
\citet[][ and subsequent improvements in the last thirty years]{Spi67}, 
described in \citet{cay91} and \citet{barb03}. Line lists and respective 
parameters are given in paper I.

%

\section{s-, r- and p-Processes}
Several important references describe in detail the s-, r- and p-processes:
B$\sp 2$FH, \citet{meyer94} and \citet{wallers97} for a more general description,
\citet{kapp89}, \citet{busso99}, \citet{lugaro03}, \citet{lamb77} and 
\citet{raiteri93} for the s-process, \citet{woosley92}, \citet{wanajo03} and 
\citet{argast04} for the r-process, \citet{rayet95} and \citet{goriely05} for the p-process, among others.

Typical s-elements are mainly produced by the s-process, but a smaller fraction of their
abundance is due to the r-process. Analogously, the major production of r-elements
is due to the r-process with a smaller contribution from s-process. Beyond these main
processes, the p-process contributes with a small fraction of heavy elements production,
as shown in Figure 1 of \citet{meyer94}.

In the s-process, neutrons are captured by seed nuclei, which are the iron peak
elements, in a long timescale relative to $\beta$ decay, denominated ``s'' ({\it slow})
by B$\sp 2$FH.
This process has been subdivided into three components according to the site and the 
nucleosynthetic products: main, weak and strong.

The s-process main component is responsible by many isotopes in the range of
atomic mass 63 $\leq$ A $\leq$ 209. Abundance peaks can be observed near
A = 90, 138 and 208.
In the classical analysis \citep{kapp89} the element formation through the s-process
occurs in a chain, starting with the seed nucleus $\sp {56}$Fe.
Following this analysis, it is possible to obtain an expression for the $\sigma$N, 
where $\sigma$ is the neutron capture cross-section and N the abundances due
to the s-process only for each nuclide. This expression involves the mean
distribution of neutron exposure $\tau_o$, the abundance fraction of $\sp {56}$Fe required 
as seed to s-process $G$, and the solar abundance of $\sp {56}$Fe $N_{56}^\odot$,

\begin{equation}
\label{cursin}
\sigma_kN_k = {G N_{56}^\odot\over{\tau_o}}\prod_{i=56}^{k}(1+{1\over {\tau_o\sigma_i}})^{-1}
\end{equation}
where k is the atomic mass number.

For the solar system, the corresponding curve from classical analysis represents very well
the abundances of s-process nuclei for A $>$ 90. For lighter nuclides, this
line appears below the empirical data, suggesting another form of synthesis of s-process
nuclides in stars, being the first one called main component and the second, weak 
component. The good agreement between the classical model and the observed data of this
curve for the solar system represents an interesting characteristic of the s-process, 
taking into account the large number of nuclides between Fe and Bi.

The s-process main component is believed to occur during the 
thermal pulses - asymptotic giant branch (TP-AGB) phase of intermediate or 
low mass stars. In this phase, the star consists of an inert CO core and
the He and H burning shells. The region between the two shells is
the so-called He intershell, where neutrons released by the 
$^{13}$C ($\alpha$,n)$^{16}$O reaction during the interpulse period, and
by the $^{22}$Ne ($\alpha$,n)$^{25}$Mg reaction during the convective 
thermal pulse, are captured by iron peak nuclei.

The s-process weak component is responsible for part of the abundance of nuclides with
atomic mass in the range 23 $\leq$ A $\leq$ 90 \citep{lamb77,raiteri93}. The 
nucleosynthetic site is likely the core helium burning of stars with masses
$\geq$ 10 M$\sb \odot$, where temperature is high enough for
the main neutron source to be the $^{22}$Ne ($\alpha$,n)$^{25}$Mg reaction.
Neutron density is low compared to the main component.

The s-process strong component was postulated in order to provide part of the Pb abundance
\citep[see][]{kapp89}. However, according to \citet{busso99}, it is possible to explain
$^{208}$Pb galactic abundance without it.


The r-process occurs in an environment rich in neutrons, where several of them are 
captured by nuclides in a short timescale compared to $\beta$ decay, and for this 
reason, this process was denominated ``r'' ({\it rapid}) by B$\sp 2$FH. In this case, the
neutron density is larger than that for the s-process. Sites that favour
such high neutron density are final stages of massive stars
as core collapse supernovae (SN II, Ib, Ic)
\citep{wasser00,qian00,qian01} or involving neutron stars 
\citep{woosley92,meyer94,freib99,rossw99,rossw00}.

The fact that r-elements are observed in very metal poor stars suggests that these
elements were produced in supernovae events resulting from the evolution of the
first massive stars in the Galaxy \citep{S96,sneden03,hill02,cowan03,ishim04,honda04}.
Despite these scenarios being promising, some difficulties are found, for instance,
\citet{wanajo01} showed that to reproduce a solar abundance of r-elements,
proto neutron stars must have 2 M$\sb \odot$ and 10 km of radius, characteristics
not observed so far.

The p-process forms nuclei rich in protons. Some s- or r-nuclei, where s- and
r-processes were blocked, capture protons with $\gamma$ emission (p,$\gamma$).
The p-nuclei may also be synthesized by photodesintegration ($\gamma$,n) of
a pre-existent nucleus rich in neutrons (especially s-nuclei), followed by
possible cascades of ($\gamma$,p) and/or ($\gamma$,$\alpha$) reactions.

The p-process site should be rich in hydrogen, with proton density 
$\geq$ 10$\sp 2$ g/cm$\sp 3$, at temperatures of T = 2-3 x 10$^9$ K.
This process site is likely related to SN II, according to \citet{arnould76},
\citet[][ and references therein]{woosley78}, \citet{arnould92} and
\citet{rayet93} or the explosion of a moderately massive white dwarf 
due to the accretion of He-rich matter
\citep[][ and references therein]{goriely05}. 
A quantitative study of the p-process is presented in \citet{rayet95}.

%

\section{Abundance distribution for s-, r- and p-processes}
\subsection{s-, r- and p-processes in normal stars}

\begin{table}[ht!]
\caption{ Results obtained from least-squares fitting of $\log\epsilon_{nor}$(X)
vs. [Fe/H] of normal stars: $\log\epsilon_{nor}$(X) = A[Fe/H] + B; 'cov' is
the covariance between A and B; 'dof' is the number of degrees of freedom.}
{\small
\label{ajustes}
 $$
\setlength\tabcolsep{4pt}
\begin{tabular}{rcccrr}
\hline
\noalign{\smallskip}
X & A & B & $\chi^2_{red}$   & cov (10$^{-4}$) & dof\\
\noalign{\smallskip}
\hline
\noalign{\smallskip}
Sr & 1.013 $\pm$ 0.024 & 2.908 $\pm$ 0.021 & 1.197 &  4.18 &  82 \\
Y  & 1.112 $\pm$ 0.016 & 2.239 $\pm$ 0.013 & 1.002 &  1.58 & 160 \\
Zr & 0.939 $\pm$ 0.019 & 2.625 $\pm$ 0.015 & 1.385 &  2.06 &  66 \\
Mo & 0.970 $\pm$ 0.212 & 1.895 $\pm$ 0.342 & 1.472 & 715.2 &   7 \\
Ba & 0.992 $\pm$ 0.014 & 2.095 $\pm$ 0.011 & 1.978 &  1.07 & 214 \\
La & 1.190 $\pm$ 0.031 & 1.377 $\pm$ 0.039 & 4.891 & 10.91 &  47 \\
Ce & 1.009 $\pm$ 0.047 & 1.500 $\pm$ 0.047 & 0.923 & 20.20 &  34 \\
Pr & 0.867 $\pm$ 0.048 & 0.748 $\pm$ 0.050 & 1.180 & 19.99 &  13 \\
Nd & 0.950 $\pm$ 0.019 & 1.460 $\pm$ 0.017 & 1.635 &  2.46 &  99 \\
Sm & 0.811 $\pm$ 0.045 & 0.951 $\pm$ 0.045 & 1.407 & 18.76 &  34 \\
Eu & 0.783 $\pm$ 0.019 & 0.581 $\pm$ 0.015 & 1.263 &  2.20 & 165 \\
Dy & 0.840 $\pm$ 0.031 & 1.080 $\pm$ 0.044 & 4.312 & 11.70 &  22 \\
\noalign{\smallskip}                         
\hline
\end{tabular}                                
$$                                           
}                                            
\end{table}

\begin{table}[ht!]
\caption{Abundances obtained from least-squares fittings for normal stars with metallicities
corresponding to present sample barium stars. Each five lines correspond to the normal star
with the same metallicity of the barium star indicated in parenthesis. 
$\log\epsilon$(X) correspond to $\log\epsilon_{nor}$(X) and $\epsilon\sb {s,r,p}$(X) correspond to $\epsilon_{s,r,p}$(X)$_{nor}$ from 
equation \ref{nortot}. $\epsilon_o$(X) is from equation \ref{epszero}. Full table is only available in electronic form.}
{\tiny
\label{normais}
 $$
\setlength\tabcolsep{3pt}
\begin{tabular}{rccccccc}
\hline\hline
 &     Sr   &     Y   &   Zr    &   Mo   &   Ba    &   La   &   Ce ...\\
\hline 
 && \multicolumn{4}{c}{[Fe/H] = -0.06 (HD 749)} && \\
\hline
$\log\epsilon$(X)    &  2.85$\pm$0.18 &  2.17$\pm$0.20 &   2.57$\pm$0.17 &  1.84$\pm$0.37 &  2.04$\pm$0.18 &   1.31$\pm$0.22 &  1.44$\pm$0.19 ...  \\
$\epsilon\sb s$(X)   & 594.4$\pm$51.0 & 136.8$\pm$63.2 &  307.4$\pm$20.1 &  34.1$\pm$29.2 &  87.7$\pm$36.1 &   12.5$\pm$6.3  &  20.9$\pm$9.0 ...   \\
$\epsilon\sb r$(X)   &  0.00$\pm$0.00 &  11.9$\pm$5.5  &   58.9$\pm$23.0 &  17.9$\pm$15.4 &  20.6$\pm$8.5  &    7.7$\pm$3.8  &   6.5$\pm$2.8 ...   \\
$\epsilon\sb p$(X)   & 109.0$\pm$46.0 &   0.0$\pm$0.0  &    4.1$\pm$1.6  &  16.7$\pm$14.3 &   0.2$\pm$0.1  &    0.0$\pm$0.0  &   0.1$\pm$0.0 ...   \\
$\epsilon\sb o$(X)   & 109.0$\pm$46.0 &  11.9$\pm$5.5  &   63.0$\pm$24.6 &  34.6$\pm$29.7 &  20.8$\pm$8.6  &    7.7$\pm$3.8  &   6.6$\pm$2.8 ...   \\
.&.&.&.&.&.&.&. ...\\
.&.&.&.&.&.&.&. ...\\
\noalign{\smallskip}                         
\hline
\end{tabular} 
$$                                           
}
\end{table}

\begin{table}[ht!]
\caption{Abundances [$\log\epsilon_{nor}$(X)] of Mo, Dy, Gd and Pb for normal stars. 
Symbol ``*'' indicates that the
column was obtained from a least-squares fitting.}
\label{mogdpb}
 $$
\setlength\tabcolsep{3pt}
\begin{tabular}{lrrrrrr}
\hline
\noalign{\smallskip}
 star & Mo* & Mo & Dy* & Dy & Gd & Pb \\
\noalign{\smallskip}
\hline
\noalign{\smallskip}
HD 749	   & 1.84 & 1.86 & 1.03 & 1.14 &  1.06$\pm$0.19 & 1.89$\pm$0.20 \\
HR 107     & 1.55 & 1.56 & 0.78 & 0.84 &  0.76$\pm$0.07 & 1.59$\pm$0.10 \\
HD 5424    & 1.36 & 1.37 & 0.62 & 0.65 &  0.57$\pm$0.19 & 1.40$\pm$0.20 \\
HD 8270    & 1.49 & 1.50 & 0.73 & 0.78 &  0.70$\pm$0.07 & 1.53$\pm$0.10 \\
HD 12392   & 1.78 & 1.80 & 0.98 & 1.08 &  1.00$\pm$0.19 & 1.83$\pm$0.20 \\
HD 13551   & 1.47 & 1.48 & 0.71 & 0.76 &  0.68$\pm$0.07 & 1.51$\pm$0.10 \\
HD 22589   & 1.63 & 1.65 & 0.85 & 0.93 &  0.85$\pm$0.07 & 1.68$\pm$0.10 \\
HD 27271   & 1.81 & 1.83 & 1.00 & 1.11 &  1.03$\pm$0.19 & 1.86$\pm$0.20 \\
HD 48565   & 1.29 & 1.30 & 0.56 & 0.58 &  0.50$\pm$0.07 & 1.33$\pm$0.10 \\
HD 76225   & 1.59 & 1.61 & 0.82 & 0.89 &  0.81$\pm$0.07 & 1.64$\pm$0.10 \\
HD 87080   & 1.47 & 1.48 & 0.71 & 0.76 &  0.68$\pm$0.07 & 1.51$\pm$0.10 \\
HD 89948   & 1.60 & 1.62 & 0.83 & 0.90 &  0.82$\pm$0.07 & 1.65$\pm$0.10 \\
HD 92545   & 1.78 & 1.80 & 0.98 & 1.08 &  1.00$\pm$0.07 & 1.83$\pm$0.10 \\
HD 106191  & 1.61 & 1.63 & 0.84 & 0.91 &  0.83$\pm$0.07 & 1.66$\pm$0.10 \\
HD 107574  & 1.36 & 1.37 & 0.62 & 0.65 &  0.57$\pm$0.07 & 1.40$\pm$0.10 \\
HD 116869  & 1.58 & 1.60 & 0.81 & 0.88 &  0.80$\pm$0.19 & 1.63$\pm$0.20 \\
HD 123396  & 0.74 & 0.73 & 0.08 & 0.01 & -0.07$\pm$0.19 & 0.76$\pm$0.20 \\
HD 123585  & 1.43 & 1.44 & 0.68 & 0.72 &  0.64$\pm$0.07 & 1.47$\pm$0.10 \\
HD 147609  & 1.46 & 1.47 & 0.70 & 0.75 &  0.67$\pm$0.07 & 1.50$\pm$0.10 \\
HD 150862  & 1.80 & 1.82 & 1.00 & 1.10 &  1.02$\pm$0.07 & 1.85$\pm$0.10 \\
HD 188985  & 1.60 & 1.62 & 0.83 & 0.90 &  0.82$\pm$0.07 & 1.65$\pm$0.10 \\
HD 210709  & 1.86 & 1.88 & 1.05 & 1.16 &  1.08$\pm$0.19 & 1.91$\pm$0.20 \\
HD 210910  & 1.54 & 1.55 & 0.77 & 0.83 &  0.75$\pm$0.19 & 1.58$\pm$0.20 \\
HD 222349  & 1.28 & 1.29 & 0.55 & 0.57 &  0.49$\pm$0.07 & 1.32$\pm$0.10 \\
BD+18 5215 & 1.38 & 1.39 & 0.63 & 0.67 &  0.59$\pm$0.07 & 1.42$\pm$0.10 \\
HD 223938  & 1.56 & 1.57 & 0.79 & 0.85 &  0.77$\pm$0.19 & 1.60$\pm$0.20 \\
\noalign{\smallskip}                         
\hline
\end{tabular}                                
$$                                                                                      
\end{table}

One way to quantify the enrichment of s-elements in barium stars is to compare
their abundances with normal stars.

For the s-, r- and p-elements, the total abundance for the element can be described
by the sum of the abundances corresponding to the three nucleosynthetic processes,
taking into account all contributing isotopes ``i'':

$$\epsilon(X) = \sum_i{\epsilon_s^i} + \sum_i{\epsilon_r^i} + \sum_i{\epsilon_p^i}.$$

\begin{figure*}[ht!]
\centerline{\includegraphics[totalheight=15.0cm]{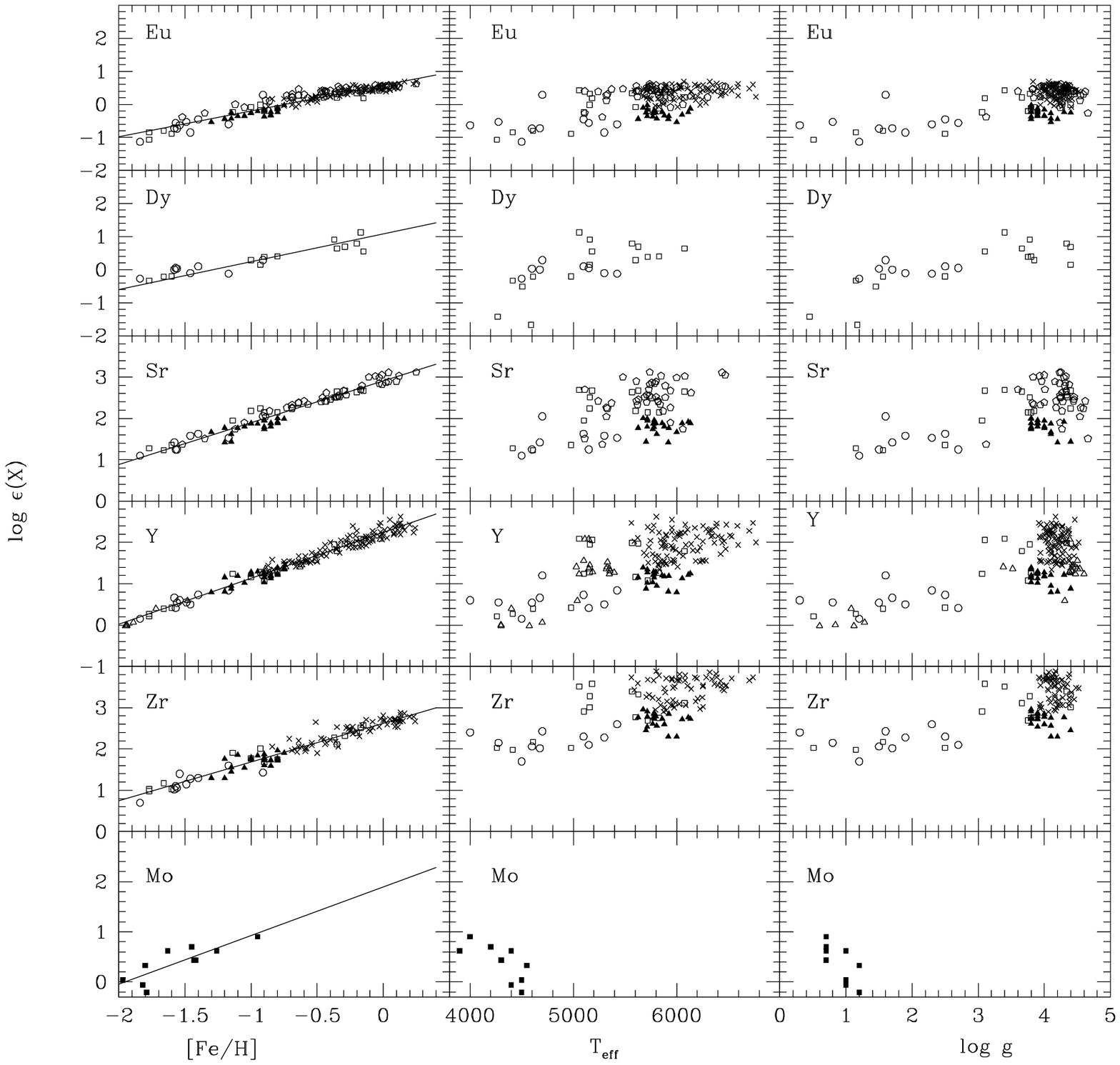}}
\caption{\label{litpatm1} $\log\epsilon$(X) vs. atmospheric parameters 
([Fe/H], T$_{eff}$, $\log$ g) for normal stars. 
Symbols indicate different references:
open squares: \citet{g94}; open circles: \citet{burr00}; open
triangles: \citet{tl99}; filled triangles: \citet{jehin99}; crosses:
\citet{edv93} and \citet{woolf95}; filled squares: \citet{S00}; 
open pentagon \citet{mg01}. Uncertainties were taken from the respective references,
otherwise, the values of $\pm$0.1 for [Fe/H] and $\pm$0.05 for $\log\epsilon$(X) 
were used. These values were also attributed to \citet{jehin99}, given that 
their uncertainties seem underestimated.}
\end{figure*}

\begin{figure*}[ht!]
\centerline{\includegraphics[totalheight=15.0cm]{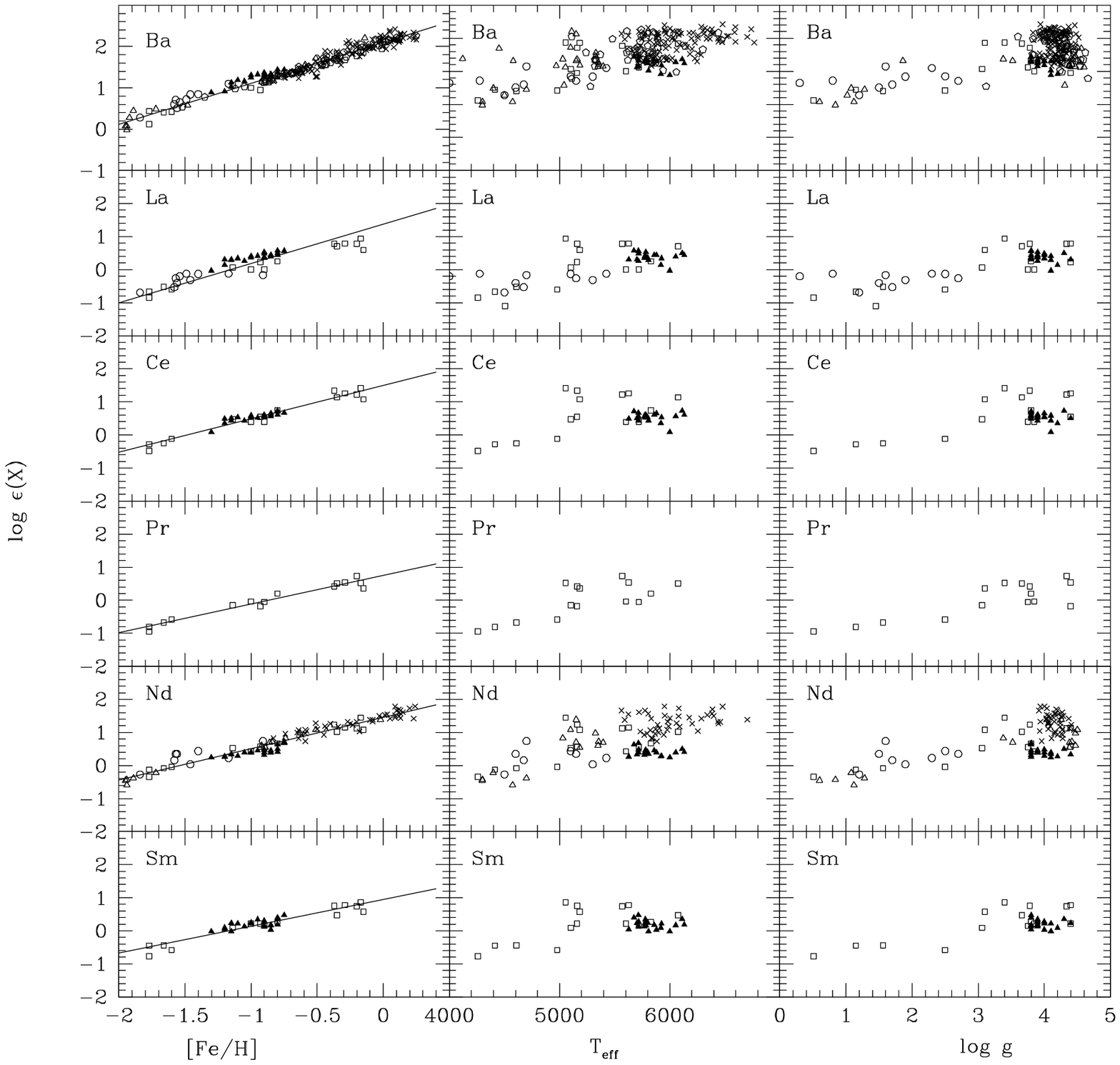}}
\caption{\label{litpatm2} Same as Figure \ref{litpatm1} for other elements.}
\end{figure*}

In order to quantify the contribution of each process to the total abundance of several
heavy elements in normal stars, an extensive set of stars from the 
literature was used. Peculiar stars were withdrawn from these samples in order to obtain
more reliable results. Figures \ref{litpatm1} and \ref{litpatm2} show the 
behaviour of the abundances ($\log\epsilon$(X)) of several heavy elements as a function
of atmospheric parameters (T$_{eff}$, log g, [Fe/H]). It is clear from these figures
that $\log\epsilon$(X) vs. [Fe/H] is a linear correlation ($\log\epsilon_{nor}$(X) = A[Fe/H] + B), 
while $\log\epsilon$(X) vs. T$_{eff}$ and $\log$ g are not. Results of least-squares 
fittings for $\log\epsilon$(X) vs. [Fe/H] are shown in Table \ref{ajustes}.

From $\log\epsilon$(X) vs. [Fe/H] fittings, it is possible 
to determine the total abundance of a certain element in a normal star of a
given metallicity. The first line of Table \ref{normais} shows $\log\epsilon_{nor}$(X)
obtained directly from fittings for normal stars with metallicities corresponding to 
the barium star, indicated in the header. 
Data for molibdenium are rarely available in the literature in the same metallicity range
as the present sample, and for this reason the least-squares fitting was done
using data derived for the globular cluster $\omega$ Centauri by \citet{S00}, with results
shown in column 2 of Table \ref{mogdpb}. In order to verify the reliability
of these results, Mo abundances were calculated through
$\log\epsilon_{nor}$(Mo) = $\log\epsilon_\odot$(Mo) + [Fe/H] + [Mo/Fe], considering
[Mo/Fe] = 0, and results are shown in column 3 of Table \ref{mogdpb}. The agreement
between results from these two columns means that, according to the fitting, 
[Mo/Fe] $\approx$ 0 at metallicities near solar. Data for Gd and Pb are also
rare in the literature, for normal stars in the range of metallicities of the present sample,
and for this reason, $\log\epsilon_{nor}$(Gd) and $\log\epsilon_{nor}$(Pb) were
determined by considering [Gd/Fe] = [Pb/Fe] = 0 $\pm$ 0.05, near the solar metallicity. 
Gd is expected to behave like Dy given that both are produced mainly through the r-process in 
almost the same proportions \citep{arlandini99}. Columns 4 and 5 of Table \ref{mogdpb}
show $\log\epsilon_{nor}$(Dy) values determined by least-squares fitting and by 
summing the star metallicity to the solar value, and columns 6 and 7 show 
$\log\epsilon_{nor}$(Gd) and $\log\epsilon_{nor}$(Pb). 

Considering that the total abundance of an element is the sum of the contributions of 
s-, r- and p-processes, one can write:

\begin{eqnarray}
\label{nortot}
\epsilon_{nor}(X) = & \epsilon_s(X)_{nor} + \epsilon_{sw}(X)_{nor} + \epsilon_{st}(X)_{nor} + \epsilon_r(X)_{nor} + \nonumber \\
& + \epsilon_p(X)_{nor}
\end{eqnarray}
where $\epsilon_s$(X)$_{nor}$, $\epsilon_{sw}$(X)$_{nor}$ and $\epsilon_{st}$(X)$_{nor}$ are, 
respectively, the contributions of main, weak and strong components of s-process and 
$\epsilon_r$(X)$_{nor}$ and $\epsilon_p$(X)$_{nor}$ the r- and p-processes. 
\citet{arlandini99} provide detailed information about s-process main component and
r-process. According to \citet{lugaro03}, around 50\% of $\sp {86}$Sr and $\sp {87}$Sr
in the solar system results from the s-process weak component, hence, the missing abundance
which completes the total abundance in \citet{arlandini99} for these Sr isotopes was attributed
to the weak component. 45\% of $\sp {96}$Zr abundance missing in \citeauthor{arlandini99} 
was also attributed to the s-process weak 
component. A few nuclides are formed mainly through p-process, and they are not in \citet{arlandini99},
such as $\sp {138}$La, $\sp {136,138}$Ce, $\sp {144}$Sm, $\sp {156,158}$Dy \citep{rayet95}.
The total solar abundances adopted by \citet{arlandini99} were those from \citet{andgre89}, 
and it is possible to deduce the abundance of each nuclide p-only. For nuclides partially p,
the difference between the total abundance and the sum of s- and r-fractions from \citet{arlandini99}
was computed, such as for $\sp {94}$Mo, $\sp {142}$Nd, $\sp {152,154}$Gd and $\sp {160}$Dy. The isotopes
$\sp {142,150}$Nd are missing in \citet{rayet95}, however they were considered p-partial or p-only
according to the missing of part or total in \citeauthor{arlandini99}. Once all abundance percentages
were identified, it was possible to derive the abundance fractions of each process relative to 
total abundance of an element, shown in columns 3-7 of Table \ref{sechoque}. The total abundance 
of each element taking into account its $n$ isotopes for normal stars can be written as

\begin{equation}
\label{norfrac}
\epsilon_{nor}(X)= \bigg(\sum_{i}^n g_s^i+\sum_{i}^n g_{sw}^i+\sum_{i}^n g_{st}^i+\sum_{i}^n g_{r}^i+ 
 \sum_{i}^n g_{p}^i \bigg)\epsilon_{nor}(X)
\end{equation}
where $g_s^i$, $g_{sw}^i$, $g_{st}^i$,
$g_{r}^i$ and $g_{p}^i$ are fractions of total abundance relative to the three s-process 
components, r- and p-processes, respectively, for each isotope ``$i$''.

The uncertainty on $\log\epsilon_{nor}$(X) is calculated with 

\begin{eqnarray}
\sigma_{\log\epsilon nor(X)}= & \bigg([Fe/H]^2\sigma_A^2+A^2\sigma^2_{[Fe/H]}+\sigma_B^2+ \nonumber \\
 & + 2[Fe/H]cov(A,B) \bigg)^{0.5}
\end{eqnarray}
where $cov(A,B)$ is the covariance between A and B. The uncertainties on abundances relative to 
s-, r- and p-processes are given by

\begin{equation}
\sigma_{\epsilon j}=\epsilon_j\sigma_{\log\epsilon nor(X)}\ln{10} \nonumber \\
\end{equation}
where ``j'' corresponds to the process involved (s: main, sw: weak, st: strong, 
componentes of s-process, r: r-process, or p: p-process). 

Finally, the sum of abundances from other processes except the s-process main component
was computed, i.e., the sum of fractions 
sw, st, r and p, indicated by subscript ``o'' and its uncertainty are
 
\begin{equation}
\label{epszero}
\epsilon_o(X)= \bigg(\sum_{i}^n g_{sw}^i+\sum_{i}^n g_{st}^i+\sum_{i}^n g_{r}^i+\sum_{i}^n g_{p}^i \bigg)\epsilon_{nor}(X)
\end{equation}
and
\begin{equation}
\sigma_{\epsilon o(X)}=\epsilon_o(X)\sigma_{\log\epsilon nor(X)}\ln{10}. \nonumber \\
\end{equation}

Lines 2, 3 and 4 of Table \ref{normais} show abundances due to s-process main component,
r- and p-process, respectively, and line 5 shows $\epsilon_o$(X).

\begin{table*}[ht!]
\caption{Cross sections in mb$\sp {-1}$ for 30, 23 and 8 keV and abundance fractions
for s-, r- and p-processes relative to total abundance and to total abundance of each process.
Cross sections references: 1 - \citet{bao00}; 2 - \citet{obrien03}; 3 - \citet{arlandini99}} 
\label{sechoque}
{\scriptsize
 $$
\setlength\tabcolsep{2.5pt}
\begin{tabular}{rrcccccccccccccc}
\hline\hline
 A & el & $g_s$ & $g_{sw}$ & $g_{st}$ & $g_{r}$ & $g_{p}$ & $f_s$ &
 $f_{sw}$ & $f_{st}$ & $f_r$ & $f_p$ & $\sigma_c$(30keV) & $\sigma_c$(23keV) & 
 $\sigma_c$(8keV) & ref \\
 \hline 
 84 & Sr &   ...     &   ...     &   ...    &    ...     & 5.650e-03 &   ...     &   ...     &   ...    &    ...     & 1.000e00  &   368$\pm$126 &    ...   &   ...   & 1 \\
 86 & Sr & 4.666e-02 & 5.265e-02 &   ...    &    ...     &   ...     & 5.519e-02 & 3.534e-01 &   ...    &    ...     &   ...     &    64$\pm$3   &    66.80 &   80.10 & 1 \\
 87 & Sr & 3.253e-02 & 3.210e-02 &   ...    &    ...     &   ...     & 3.848e-02 & 2.155e-01 &   ...    &    ...     &   ...     &    92$\pm$4   &   101.60 &  126.00 & 1 \\
 88 & Sr & 7.662e-01 & 6.421e-02 &   ...    &    ...     &   ...     & 9.063e-01 & 4.310e-01 &   ...    &    ...     &   ...     &   6.2$\pm$0.3 &     6.60 &    6.00 & 1 \\
    &    & 84.54\%   & 14.90\%   &          &            & 0.56\%    &           &           &          &            &           &               &          &         &   \\
 89 &  Y & 9.203e-01 &   ...     &   ...    &  8.000e-02 &   ...     & 1.000e00  &   ...     &   ...    &  1.000e00  &   ...     &  19.0$\pm$0.6 &    19.60 &   24.50 & 1 \\
    &    & 92.03\%   &           &          & 8.00\%     &           &           &           &          &            &           &               &          &         &   \\
 90 & Zr & 3.716e-01 &   ...     &   ...    &  1.429e-01 &   ...     & 4.477e-01 &   ...     &   ...    &  8.993e-01 &   ...     &    21$\pm$2   &    20.80 &   18.90 & 1 \\
 91 & Zr & 1.078e-01 &   ...     &   ...    &  4.785e-03 &   ...     & 1.299e-01 &   ...     &   ...    &  3.012e-02 &   ...     &    60$\pm$8   &    64.70 &   90.50 & 1 \\
 92 & Zr & 1.604e-01 &   ...     &   ...    &  1.122e-02 &   ...     & 1.932e-01 &   ...     &   ...    &  7.062e-02 &   ...     &    33$\pm$4   &    34.60 &   47.20 & 1 \\
 94 & Zr & 1.748e-01 &   ...     &   ...    &    ...     &   ...     & 2.106e-01 &   ...     &   ...    &    ...     &   ...     &    26$\pm$1   &    26.60 &   30.10 & 1 \\
 96 & Zr & 1.542e-02 & 1.100e-02 &   ...    &    ...     &   ...     & 1.859e-02 & 1.000e00  &   ...    &    ...     &   ...     &  10.7$\pm$0.5 &    11.10 &   18.10 & 1 \\
    &    &  83.00\%  & 1.10\%    &          & 15.89\%    &           &           &           &          &            &           &               &          &         &   \\
 92 & Mo &   ...     &   ...     &   ...    &    ...     & 1.509e-01 &   ...     &   ...     &   ...    &    ...     & 6.156e-01 &    70$\pm$10  &    ...   &   ...   & 1 \\
 94 & Mo & 5.995e-04 &   ...     &   ...    &    ...     & 9.165e-02 & 1.205e-03 &   ...     &   ...    &    ...     & 3.844e-01 &   102$\pm$20  &   103.70 &  120.30 & 1 \\
 95 & Mo & 8.816e-02 &   ...     &   ...    &  7.093e-02 &   ...     & 1.772e-01 &   ...     &   ...    &  2.710e-01 &   ...     &   292$\pm$12  &   301.40 &  322.00 & 1 \\
 96 & Mo & 1.665e-01 &   ...     &   ...    &    ...     &   ...     & 3.347e-01 &   ...     &   ...    &    ...     &   ...     &   112$\pm$8   &   118.00 &  150.40 & 1 \\
 97 & Mo & 5.604e-02 &   ...     &   ...    &  3.958e-02 &   ...     & 1.126e-01 &   ...     &   ...    &  1.512e-01 &   ...     &   339$\pm$14  &   352.90 &  387.80 & 1 \\
 98 & Mo & 1.826e-01 &   ...     &   ...    &  5.839e-02 &   ...     & 3.669e-01 &   ...     &   ...    &  2.231e-01 &   ...     &    99$\pm$7   &   100.10 &  111.80 & 1 \\
100 & Mo & 3.691e-03 &   ...     &   ...    &  9.237e-02 &   ...     & 7.418e-03 &   ...     &   ...    &  3.548e-01 &   ...     &   108$\pm$14  &   106.90 &  110.90 & 1 \\
    &    & 49.76\%   &           &          &  26.18\%   & 24.06\%   &           &           &          &            &           &               &          &         &   \\
130 & Ba &   ...     &   ...     &   ...    &    ...     & 1.060e-03 &   ...     &   ...     &   ...    &    ...     & 5.124e-01 &   760$\pm$110 &    ...   &   ...   & 1 \\
132 & Ba &   ...     &   ...     &   ...    &    ...     & 1.009e-03 &   ...     &   ...     &   ...    &    ...     & 4.876e-01 &   379$\pm$137 &    ...   &   ...   & 1 \\
134 & Ba & 2.427e-02 &   ...     &   ...    &    ...     &   ...     & 3.003e-02 &   ...     &   ...    &    ...     &   ...     & 176.0$\pm$5.6 &   172.10 &  157.90 & 1 \\
135 & Ba & 1.726e-02 &   ...     &   ...    &  4.877e-02 &   ...     & 2.135e-02 &   ...     &   ...    &  2.570e-01 &   ...     &   455$\pm$15  &   463.50 &  498.40 & 1 \\
136 & Ba & 7.860e-02 &   ...     &   ...    &    ...     &   ...     & 9.726e-02 &   ...     &   ...    &    ...     &   ...     &  61.2$\pm$2.0 &    62.20 &   68.60 & 1 \\
137 & Ba & 7.348e-02 &   ...     &   ...    &  3.875e-02 &   ...     & 9.092e-02 &   ...     &   ...    &  2.042e-01 &   ...     &  76.3$\pm$2.4 &    77.40 &   81.90 & 1 \\
138 & Ba & 6.146e-01 &   ...     &   ...    &  1.022e-01 &   ...     & 7.604e-01 &   ...     &   ...    &  5.387e-01 &   ...     &   4.0$\pm$0.2 &     4.00 &    4.80 & 1 \\
    &    & 80.82\%   &           &          & 18.97\%    & 0.21\%    &           &           &          &            &           &               &          &         &   \\
138 & La &   ...     &   ...     &   ...    &    ...     & 8.893e-04 &   ...     &   ...     &   ...    &    ...     & 1.000e00  &  ...          &    ...   &   ...   & 1 \\
139 & La & 6.205e-01 &   ...     &   ...    &  3.786e-01 &   ...     & 1.000e00  &   ...     &   ...    &  1.000e00  &   ...     &  31.6$\pm$0.8 &    37.50 &   70.10 & 2 \\
    &    & 62.05\%   &           &          & 37.86\%    & 0.09\%    &           &           &          &            &           &               &          &         &   \\
136 & Ce &   ...     &   ...     &   ...    &    ...     & 1.901e-03 &   ...     &   ...     &   ...    &    ...     & 4.320e-01 &   328$\pm$21  &    ...   &   ...   & 1 \\
138 & Ce &   ...     &   ...     &   ...    &    ...     & 2.500e-03 &   ...     &   ...     &   ...    &    ...     & 5.680e-01 &   179$\pm$5   &    ...   &   ...   & 1 \\
140 & Ce & 7.359e-01 &   ...     &   ...    &  1.488e-01 &   ...     & 9.677e-01 &   ...     &   ...    &  6.327e-01 &   ...     &  10.6$\pm$0.5 &    ...   &   ...   & 3 \\
142 & Ce & 2.456e-02 &   ...     &   ...    &  8.636e-02 &   ...     & 3.230e-02 &   ...     &   ...    &  3.673e-01 &   ...     &  28.3$\pm$1.0 &    ...   &   ...   & 3 \\
    &    & 76.05\%   &           &          & 23.52\%    &           & 0.44\%    &           &          &            &           &               &          &         &   \\
141 & Pr & 4.868e-01 &   ...     &   ...    &  5.132e-01 &   ...     & 1.000e00  &   ...     &   ...    &  1.000e00  &   ...     & 111.4$\pm$1.4 &    ...   &   ...   & 3 \\
    &    & 48.68\%   &           &          & 51.32\%    &           &           &           &          &            &           &               &          &         &   \\
142 & Nd & 2.515e-01 &   ...     &   ...    &    ...     & 2.056e-02 & 4.535e-01 &   ...     &   ...    &    ...     & 2.669e-01 &  35.0$\pm$0.7 &    ...   &   ...   & 1 \\
143 & Nd & 3.821e-02 &   ...     &   ...    &  8.271e-02 &   ...     & 6.890e-02 &   ...     &   ...    &  2.246e-01 &   ...     &   245$\pm$3   &    ...   &   ...   & 1 \\
144 & Nd & 1.209e-01 &   ...     &   ...    &  1.171e-01 &   ...     & 2.180e-01 &   ...     &   ...    &  3.178e-01 &   ...     &  81.3$\pm$1.5 &    ...   &   ...   & 1 \\
145 & Nd & 2.285e-02 &   ...     &   ...    &  6.022e-02 &   ...     & 4.121e-02 &   ...     &   ...    &  1.635e-01 &   ...     &   425$\pm$5   &    ...   &   ...   & 1 \\
146 & Nd & 1.102e-01 &   ...     &   ...    &  6.155e-02 &   ...     & 1.986e-01 &   ...     &   ...    &  1.671e-01 &   ...     &  91.2$\pm$1.0 &    ...   &   ...   & 1 \\
148 & Nd & 1.094e-02 &   ...     &   ...    &  4.680e-02 &   ...     & 1.973e-02 &   ...     &   ...    &  1.270e-01 &   ...     &   147$\pm$2   &    ...   &   ...   & 1 \\
150 & Nd &   ...     &   ...     &   ...    &    ...     & 5.647e-02 &   ...     &   ...     &   ...    &    ...     & 7.331e-01 &   159$\pm$10  &    ...   &   ...   & 1 \\
    &    & 55.46\%   &           &          &  36.84\%   & 7.70\%    &           &           &          &            &           &               &          &         &   \\
144 & Sm &   ...     &   ...     &   ...    &    ...     & 3.096e-02 &   ...     &   ...     &   ...    &    ...     & 1.000e00  &    92$\pm$6   &    ...   &   ...   & 1 \\
147 & Sm & 3.193e-02 &   ...     &   ...    &  1.227e-01 &   ...     & 1.082e-01 &   ...     &   ...    &  1.821e-01 &   ...     &   973$\pm$1   &    ...   &   ...   & 1 \\
148 & Sm & 1.091e-01 &   ...     &   ...    &    ...     &   ...     & 3.697e-01 &   ...     &   ...    &    ...     &   ...     &   241$\pm$2   &    ...   &   ...   & 1 \\
149 & Sm & 1.722e-02 &   ...     &   ...    &  1.208e-01 &   ...     & 5.835e-02 &   ...     &   ...    &  1.792e-01 &   ...     &  1820$\pm$17  &    ...   &   ...   & 1 \\
150 & Sm & 7.392e-02 &   ...     &   ...    &    ...     &   ...     & 2.504e-01 &   ...     &   ...    &    ...     &   ...     &   422$\pm$4   &    ...   &   ...   & 1 \\
152 & Sm & 6.115e-02 &   ...     &   ...    &  2.055e-01 &   ...     & 2.072e-01 &   ...     &   ...    &  3.050e-01 &   ...     &   473$\pm$4   &    ...   &   ...   & 1 \\
154 & Sm & 1.815e-03 &   ...     &   ...    &  2.249e-01 &   ...     & 6.149e-03 &   ...     &   ...    &  3.337e-01 &   ...     &   206$\pm$12  &    ...   &   ...   & 1 \\
    &    & 29.51\%   &           &          & 67.39\%    &  3.1\%    &           &           &          &            &           &               &          &         &   \\
151 & Eu & 3.124e-02 &   ...     &   ...    &  4.471e-01 &   ...     & 5.409e-01 &   ...     &   ...    &  4.744e-01 &   ...     &  3775$\pm$150 &    ...   &   ...   & 1 \\
153 & Eu & 2.652e-02 &   ...     &   ...    &  4.954e-01 &   ...     & 4.591e-01 &   ...     &   ...    &  5.256e-01 &   ...     &  2780$\pm$100 &    ...   &   ...   & 1 \\
    &    & 5.78\%    &           &          & 94.25\%    &           &           &           &          &            &           &               &          &         &   \\
152 & Gd & 1.767e-03 &   ...     &   ...    &    ...     & 2.334e-04 & 1.151e-02 &   ...     &   ...    &    ...     & 1.846e-01 &  1049$\pm$17  &    ...   &   ...   & 1 \\
154 & Gd & 2.076e-02 &   ...     &   ...    &    ...     & 1.031e-03 & 1.352e-01 &   ...     &   ...    &    ...     & 8.153e-01 &  1028$\pm$12  &    ...   &   ...   & 1 \\
155 & Gd & 8.730e-03 &   ...     &   ...    &  1.391e-01 &   ...     & 5.684e-02 &   ...     &   ...    &  1.646e-01 &   ...     &  2648$\pm$30  &    ...   &   ...   & 1 \\
156 & Gd & 3.486e-02 &   ...     &   ...    &  1.700e-01 &   ...     & 2.270e-01 &   ...     &   ...    &  2.011e-01 &   ...     &   615$\pm$5   &    ...   &   ...   & 1 \\
157 & Gd & 1.676e-02 &   ...     &   ...    &  1.397e-01 &   ...     & 1.091e-01 &   ...     &   ...    &  1.653e-01 &   ...     &  1369$\pm$15  &    ...   &   ...   & 1 \\
158 & Gd & 6.820e-02 &   ...     &   ...    &  1.804e-01 &   ...     & 4.440e-01 &   ...     &   ...    &  2.133e-01 &   ...     &   324$\pm$3   &    ...   &   ...   & 1 \\
160 & Gd & 2.507e-03 &   ...     &   ...    &  2.161e-01 &   ...     & 1.632e-02 &   ...     &   ...    &  2.556e-01 &   ...     &   154$\pm$28  &    ...   &   ...   & 1 \\
    &    & 15.36\%   &           &          & 84.53\%    & 0.13\%    &           &           &          &            &           &               &          &         &   \\
156 & Dy &   ...     &   ...     &   ...    &    ...     & 5.601e-04 &   ...     &   ...     &   ...    &    ...     & 1.256e-01 &  1567$\pm$145 &    ...   &   ...   & 1 \\
158 & Dy &   ...     &   ...     &   ...    &    ...     & 9.580e-04 &   ...     &   ...     &   ...    &    ...     & 2.149e-01 &  1060$\pm$400 &    ...   &   ...   & 1 \\
160 & Dy & 2.043e-02 &   ...     &   ...    &    ...     & 2.940e-03 & 1.385e-01 &   ...     &   ...    &    ...     & 6.595e-01 &   890$\pm$12  &    ...   &   ...   & 1 \\
161 & Dy & 1.044e-02 &   ...     &   ...    &  1.784e-01 &   ...     & 7.079e-02 &   ...     &   ...    &  2.104e-01 &   ...     &  1964$\pm$19  &    ...   &   ...   & 1 \\
162 & Dy & 4.156e-02 &   ...     &   ...    &  2.144e-01 &   ...     & 2.818e-01 &   ...     &   ...    &  2.528e-01 &   ...     &   446$\pm$4   &    ...   &   ...   & 1 \\
163 & Dy & 8.921e-03 &   ...     &   ...    &  2.400e-01 &   ...     & 6.048e-02 &   ...     &   ...    &  2.830e-01 &   ...     &  1112$\pm$11  &    ...   &   ...   & 1 \\
164 & Dy & 6.615e-02 &   ...     &   ...    &  2.152e-01 &   ...     & 4.485e-01 &   ...     &   ...    &  2.537e-01 &   ...     & 212.0$\pm$3.0 &    ...   &   ...   & 1 \\
    &    & 14.75\%   &           &          &  84.80\%   & 0.45\%    &           &           &          &            &           &               &          &         &   \\
204 & Pb & 1.841e-02 &   ...     & 1.119e-03&    ...     &   ...     & 3.997e-02 &   ...     & 2.074e-03&    ...     &   ...     &  89.0$\pm$5.5 &    ...   &   ...   & 1 \\
206 & Pb & 1.096e-01 &   ...     & 7.992e-02&    ...     &   ...     & 2.380e-01 &   ...     & 1.481e-01&    ...     &   ...     &  15.8$\pm$0.8 &    ...   &   ...   & 1 \\
207 & Pb & 1.311e-01 &   ...     & 7.481e-02&    ...     &   ...     & 2.845e-01 &   ...     & 1.387e-01&    ...     &   ...     &   9.7$\pm$1.3 &    ...   &   ...   & 1 \\
208 & Pb & 2.014e-01 &   ...     & 3.836e-01&    ...     &   ...     & 4.372e-01 &   ...     & 7.111e-01&    ...     &   ...     &  0.36$\pm$0.03&    ...   &   ...   & 1 \\
    &    & 46.05\%   &           & 53.94\%  &            &           &           &           &          &            &           &               &          &         &   \\
\noalign{\smallskip}                         
\hline
\end{tabular} 
$$                                           
}
\end{table*}

\subsection{s-, r- and p-processes in barium stars}
Barium stars are enriched in neutron capture elements, and the excess of heavy elements 
can be deduced from a comparison with normal stars of similar metallicities. If the
excess is due to the s-process main component, one can consider that abundances
due to other processes (r, p, and other s-process components) are similar
to those of normal stars of same metallicity, i.e., 
$\epsilon_r$(X) = $\epsilon_r$(X)$_{nor}$, $\epsilon_p$(X) = $\epsilon_p$(X)$_{nor}$,
$\epsilon_{sw}$(X) = $\epsilon_{sw}$(X)$_{nor}$ and $\epsilon_{st}$(X) = $\epsilon_{st}$(X)$_{nor}$. 
In this way, $\epsilon_o$(X) (line 5 of Table \ref{normais}) was attributed to each
barium star as the fraction corresponding to processes other than the s-process 
main component. The logarithmic mean abundances ($\log\epsilon$(X)) shown in 
Tables 13 and 14 of paper I were used to compute the fractions corresponding to 
each process.

The abundance fraction corresponding to the s-process main component of an element is
\begin{equation}
\label{partes}
\epsilon_s(X)=\epsilon(X)-\epsilon_o(X).
\end{equation}

Table \ref{fracporc} shows abundances relative to the s-process main component calculated 
with equation \ref{partes}. In order to characterise the overabundance of neutron
capture elements in barium stars, $\epsilon_s$(X) values were compared to 
$\epsilon_s$(X)$_{nor}$ from Table \ref{normais}, which is the fraction relative to the
s-process main component in a normal star of same metallicity. Table \ref{fracporc} 
also shows the percentages $\epsilon_s$(X)/$\epsilon_s(X)_{nor}$ $\times$ 100. 
The r-elements, Sm, Eu, Gd and Dy, also show large overabundances,
in some cases, similar to s-elements.

Normal stars are expected to have lower abundances of heavy elements than barium stars
($\epsilon_{nor}$(X) $<$ $\epsilon$(X)), however, for some elements in some stars, 
abundances in barium stars obtained for the present sample were
much lower than those of normal stars. It is the case of
Mo (HD 27271, HD 116869, HD 123396, HD 210709, HD 210910, HD 223938), 
Eu (HR 107), Gd (HD 210709), Dy (HD 89948) and Pb (HD 22589, HD 210910).
It is not clear why these barium stars show such low abundances of these 
elements. For these cases, the equation \ref{partes}
does not apply, and there is no $\epsilon_s$(X) for them (Table \ref{fracporc}).

After computing the abundance relative to the s-process main component for an element
in barium stars using equation \ref{partes}, the abundance of each isotope
and for each process can be determined with
\begin{equation}
\label{partesiso}
\epsilon_s^i(X)=f_s^i\epsilon_s(X); \hskip 1.5cm \sigma_{\epsilon^is(X)}=f_s^i\sigma_{\epsilon s(X)}
\end{equation}
\begin{equation}
\epsilon_{sw}^i(X)=f_{sw}^i\epsilon_{sw}(X); \hskip 1cm \sigma_{\epsilon^isw(X)}=f_{sw}^i\sigma_{\epsilon sw(X)}
\end{equation}
\begin{equation}
\epsilon_{st}^i(X)=f_{st}^i\epsilon_{st}(X); \hskip 1.22cm \sigma_{\epsilon^ist(X)}=f_{st}^i\sigma_{\epsilon st(X)}
\end{equation}
\begin{equation}
\epsilon_r^i(X)=f_r^i\epsilon_r(X); \hskip 1.5cm \sigma_{\epsilon^ir(X)}=f_r^i\sigma_{\epsilon r(X)}
\end{equation}
\begin{equation}
\epsilon_p^i(X)=f_p^i\epsilon_p(X); \hskip 1.34cm \sigma_{\epsilon^ip(X)}=f_p^i\sigma_{\epsilon p(X)}
\end{equation}
where $f_s^i$, $f_{sw}^i$, $f_{st}^i$, $f_r^i$ and $f_p^i$ are the abundance fractions 
of the corresponding isotope ``$i$'', respectively, of the three s-process components, the
r- and p-processes relative to total abundance of the involved process, shown in columns 8 to 12
of Table \ref{sechoque}. The total abundance can be described by the equation
\begin{eqnarray}
\epsilon(X)= & \sum_{i}^nf_s^i\epsilon_s(X)+\sum_{i}^nf_{sw}^i\epsilon_{sw}(X) +\sum_{i}^nf_{st}^i\epsilon_{st}(X)+ \nonumber \\
& \sum_{i}^nf_r^i\epsilon_r(X)+\sum_{i}^nf_p^i\epsilon_p(X).
\end{eqnarray}

Abundances relative to the s-process main component for each nuclide of the sample 
barium stars are shown in column 3 of Table \ref{resproc}, and for normal stars, in 
column 13. The difference between these two columns is shown in column 15.

\begin{table*}
\caption{Abundance fractions corresponding to the s-process main component for
barium stars calculated with equation \ref{partes} (upper table) and percentage of abundance
(lower table) relative to the s-process main component of a barium star compared to normal stars 
of a same metallicity ($\epsilon_s$(X)/$\epsilon_s$(X)$_{nor}$ $\times$ 100).}
\label{fracporc}
{\scriptsize
   $$ 
\setlength\tabcolsep{2pt}
\begin{tabular}{lrrrrrrrrrrrrrr}
\hline\hline
\noalign{\smallskip}
star & $\epsilon_s$(Sr) & $\epsilon_s$(Y) & $\epsilon_s$(Zr) & $\epsilon_s$(Mo) & $\epsilon_s$(Ba) & $\epsilon_s$(La) &
 $\epsilon_s$(Ce) & $\epsilon_s$(Pr) & $\epsilon_s$(Nd) & $\epsilon_s$(Sm) & $\epsilon_s$(Eu) & $\epsilon_s$(Gd) &
 $\epsilon_s$(Dy) & $\epsilon_s$(Pb) \\
\noalign{\smallskip}
\hline
\noalign{\smallskip}
HD 749    &   3781$\pm$2150 & 2260$\pm$994 & 9600$\pm$4450 &  63$\pm$64  & 1771$\pm$784  & 207$\pm$94  & 1364$\pm$600 & 28$\pm$18 & 512$\pm$241 &  70$\pm$35  &    3$\pm$3    & 14$\pm$21 &  106$\pm$74   &  144$\pm$126 \\
HR 107    &   2237$\pm$791  &  294$\pm$28  &  614$\pm$74   & 127$\pm$45  &  512$\pm$60   &  24$\pm$4   &   53$\pm$7   &  5$\pm$1  &  20$\pm$4   &   6$\pm$2   &   ...         & 16$\pm$5  &  ...          &  288$\pm$135 \\
HD 5424   &   1012$\pm$579  &  522$\pm$230 & 2517$\pm$1169 &  26$\pm$23  & 1132$\pm$498  & 131$\pm$58  & 1027$\pm$450 & 30$\pm$18 & 410$\pm$190 &  52$\pm$25  &  1.4$\pm$1.4  &  7$\pm$9  &  197$\pm$129  &  303$\pm$211 \\
HD 8270   &   2790$\pm$980  &  580$\pm$54  & 1475$\pm$173  & 110$\pm$39  &  652$\pm$76   &  48$\pm$7   &  127$\pm$15  &  4$\pm$1  &  52$\pm$8   &   6$\pm$2   & 0.92$\pm$0.61 &  5$\pm$2  &    2$\pm$1    &   89$\pm$47  \\
HD 12392  &   4179$\pm$2362 & 2147$\pm$944 & 6854$\pm$3182 & 169$\pm$118 & 3331$\pm$1465 & 406$\pm$181 & 1792$\pm$786 & 83$\pm$49 & 647$\pm$302 & 226$\pm$106 &    5$\pm$4    & 28$\pm$32 &   60$\pm$44   &  918$\pm$638 \\
HD 13551  &   3366$\pm$1178 &  758$\pm$70  & 1481$\pm$174  & 176$\pm$58  &  704$\pm$82   &  45$\pm$7   &  144$\pm$17  &  5$\pm$1  &  50$\pm$8   &   8$\pm$2   & 0.33$\pm$0.47 & 13$\pm$4  &    3$\pm$1    &   85$\pm$45  \\
HD 22589  &   3748$\pm$1318 &  619$\pm$58  & 2463$\pm$288  &  49$\pm$26  &  543$\pm$64   &  32$\pm$5   &   71$\pm$9   &  3$\pm$1  &  24$\pm$4   &   3$\pm$1   & 0.70$\pm$0.69 &  2$\pm$2  &    3$\pm$1    &   ...        \\
HD 27271  &   3177$\pm$1812 & 1075$\pm$475 & 3459$\pm$1620 &  ...        &  816$\pm$365  &  51$\pm$26  &  182$\pm$82  &  9$\pm$7  &  67$\pm$36  &  16$\pm$10  &  2.4$\pm$2.8  & 10$\pm$17 &   12$\pm$14   &  109$\pm$100 \\
HD 48565  &   2451$\pm$857  &  421$\pm$39  & 1462$\pm$170  &  53$\pm$20  &  620$\pm$72   &  77$\pm$11  &  474$\pm$55  & 13$\pm$2  & 136$\pm$19  &  20$\pm$4   & 0.59$\pm$0.42 & 13$\pm$4  &   16$\pm$3    &  467$\pm$209 \\
HD 76225  &   8244$\pm$2868 & 1259$\pm$117 & 3524$\pm$410  & 142$\pm$50  & 1457$\pm$169  & 105$\pm$15  &  281$\pm$33  & 11$\pm$2  &  74$\pm$11  &  13$\pm$3   & 0.85$\pm$0.69 & 12$\pm$4  &   10$\pm$2    &  285$\pm$135 \\
HD 87080  &   3366$\pm$1178 &  812$\pm$75  & 3196$\pm$371  &  61$\pm$24  & 1476$\pm$171  & 265$\pm$37  &  977$\pm$113 & 27$\pm$4  & 365$\pm$51  &  46$\pm$9   &  3.8$\pm$1.3  & 34$\pm$9  &   75$\pm$11   &  346$\pm$159 \\
HD 89948  &   5506$\pm$1923 &  915$\pm$85  & 2078$\pm$244  & 112$\pm$42  &  651$\pm$76   &  54$\pm$8   &  125$\pm$15  &  7$\pm$1  &  56$\pm$9   &  10$\pm$3   & 0.33$\pm$0.58 &  6$\pm$3  &   ...         &   76$\pm$44  \\
HD 92545  &   3670$\pm$1300 &  563$\pm$53  & 1624$\pm$193  & 128$\pm$52  & 1113$\pm$130  &  47$\pm$7   &  147$\pm$18  &  7$\pm$1  &  46$\pm$8   &   8$\pm$3   &  2.4$\pm$1.2  &  5$\pm$3  &    7$\pm$2    &  302$\pm$148 \\
HD 106191 &   2088$\pm$743  &  723$\pm$67  & 2352$\pm$275  & 114$\pm$43  &  515$\pm$61   &  28$\pm$4   &  102$\pm$12  & 14$\pm$2  &  36$\pm$6   &  11$\pm$3   & 0.55$\pm$0.64 & 11$\pm$4  &   10$\pm$2    &  179$\pm$90  \\
HD 107574 &   3277$\pm$1144 &  442$\pm$41  &  978$\pm$115  & 175$\pm$56  & 1923$\pm$222  &  53$\pm$8   &  145$\pm$17  &  5$\pm$1  &  54$\pm$8   &  10$\pm$2   & 1.45$\pm$0.65 &  5$\pm$2  &    6$\pm$1    &  268$\pm$123 \\
HD 116869 &    735$\pm$440  &  314$\pm$140 &  869$\pm$417  &  ...        &  660$\pm$294  &  55$\pm$26  &  187$\pm$84  &  8$\pm$6  &  92$\pm$45  &  14$\pm$8   &  0.3$\pm$1.3  &  5$\pm$10 &   21$\pm$18   &  279$\pm$202 \\
HD 123396 &    144$\pm$84   &   56$\pm$25  &  394$\pm$184  &  ...        &  170$\pm$75   &  16$\pm$7   &  134$\pm$59  &  4$\pm$3  &  64$\pm$30  &   9$\pm$5   & 0.26$\pm$0.36 &  5$\pm$5  &   15$\pm$10   &   88$\pm$61  \\
HD 123585 &   4987$\pm$1737 & 1248$\pm$115 & 2980$\pm$346  & 262$\pm$83  & 2716$\pm$314  & 199$\pm$28  &  804$\pm$93  & 30$\pm$4  & 236$\pm$33  &  51$\pm$10  &  5.9$\pm$1.7  & 12$\pm$4  &   19$\pm$3    & 1031$\pm$458 \\
HD 147609 & 112924$\pm$4479 & 2291$\pm$211 & 5047$\pm$584  & 144$\pm$48  & 1755$\pm$203  & 203$\pm$28  &  768$\pm$89  & 25$\pm$4  & 204$\pm$29  &  42$\pm$8   &  4.8$\pm$1.5  & 25$\pm$7  &   24$\pm$4    &  173$\pm$83  \\
HD 150862 &   3616$\pm$1283 & 1647$\pm$153 & 4128$\pm$482  & 177$\pm$67  & 1116$\pm$131  &  60$\pm$9   &  134$\pm$16  &  8$\pm$1  &  38$\pm$7   &   8$\pm$3   & 1.15$\pm$0.98 &  9$\pm$4  &    9$\pm$3    &  317$\pm$155 \\
HD 188985 &   4358$\pm$1527 &  900$\pm$83  & 2957$\pm$345  & 112$\pm$42  & 1070$\pm$125  &  95$\pm$14  &  425$\pm$49  & 14$\pm$2  & 149$\pm$21  &  23$\pm$5   & 1.14$\pm$0.76 & 17$\pm$5  &    7$\pm$2    &  374$\pm$174 \\
HD 210709 &    944$\pm$587  &  526$\pm$235 & 2158$\pm$1024 &  ...        &  639$\pm$289  &  50$\pm$26  &  242$\pm$109 &  8$\pm$6  &  97$\pm$50  &  13$\pm$9   &  0.3$\pm$2.1  &  ...      &   11$\pm$14   &  185$\pm$154 \\
HD 210910 &   2807$\pm$1580 &  254$\pm$114 & 357 $\pm$180  &  ...        &  547$\pm$244  &  29$\pm$14  &   93$\pm$42  & 13$\pm$8  &  37$\pm$20  &   8$\pm$5   &  3.1$\pm$2.4  & 16$\pm$18 &    8$\pm$9    &   ...        \\
HD 222349 &   2442$\pm$945  &  439$\pm$185 & 1536$\pm$948  &  52$\pm$22  &  758$\pm$434  &  69$\pm$29  &  297$\pm$108 &  7$\pm$4  & 118$\pm$70  &  16$\pm$12  & 0.20$\pm$0.32 &  9$\pm$12 &   12$\pm$15   &  578$\pm$248 \\
BD+18 5215&   3880$\pm$1353 &  515$\pm$48  & 2072$\pm$241  & 111$\pm$37  & 1151$\pm$133  &  59$\pm$8   &  189$\pm$22  &  9$\pm$1  &  52$\pm$8   &  13$\pm$3   &  0.3$\pm$0.4  & 52$\pm$13 &   13$\pm$2    &   60$\pm$33  \\
HD 223938 &   1654$\pm$853  &  418$\pm$41  & 2024$\pm$179  &  ...        &  982$\pm$88   &  63$\pm$10  &  242$\pm$34  &  6$\pm$1  & 145$\pm$17  &  22$\pm$3   &  1.5$\pm$1.8  &  9$\pm$3  &   18$\pm$2    &  350$\pm$258 \\
\noalign{\smallskip}
\hline\hline                                                                                                                                                                                   
\noalign{\smallskip}
star  & Sr(\%) & Y(\%) & Zr(\%) & Mo(\%) & Ba(\%) & La(\%) & Ce(\%) & Pr(\%) & Nd(\%) & Sm(\%) & Eu(\%) & Gd(\%) & Dy(\%) & Pb(\%) \\  
\noalign{\smallskip}
\hline
HD 749    & 636  & 1652 & 3123 & 185  & 2020 & 1654 & 6526  & 1166 & 3650 & 2987  & 1508 & 795  & 6715  & 404  \\
HR 107    & 758  & 464  & 382  & 728  & 1159 & 440  & 514   & 392  & 274  & 421   & ...  & 1755 & ...   & 1609 \\
HD 5424   & 534  & 1337 & 2362 & 224  & 3955 & 4015 & 15334 & 3274 & 8525 & 5487  & 1657 & 1293 & 32236 & 2619 \\
HD 8270   & 1087 & 1066 & 1045 & 724  & 1691 & 1029 & 1398  & 341  & 813  & 514   & 888  & 609  & 268   & 570  \\
HD 12392  & 809  & 1830 & 2539 & 568  & 4357 & 3820 & 9851  & 3890 & 5253 & 10711 & 2604 & 1844 & 4262  & 2953 \\
HD 13551  & 1374 & 1466 & 1096 & 1206 & 1912 & 1019 & 1664  & 439  & 821  & 671   & 334  & 1755 & 358   & 570  \\
HD 22589  & 1029 & 775  & 1261 & 231  & 1000 & 452  & 553   & 183  & 276  & 176   & 513  & 142  & 317   & ...  \\
HD 27271  & 573  & 849  & 1201 & ...  & 996  & 439  & 934   & 396  & 506  & 732   & 1280 & 595  & 804   & 326  \\
HD 48565  & 1522 & 1290 & 1596 & 545  & 2542 & 2862 & 8336  & 1591 & 3306 & 2377  & 824  & 2621 & 2914  & 4749 \\
HD 76225  & 2485 & 1746 & 1968 & 731  & 2942 & 1666 & 2404  & 756  & 915  & 910   & 673  & 1251 & 981   & 1422 \\
HD 87080  & 1374 & 1570 & 2365 & 419  & 4011 & 5984 & 11296 & 2417 & 5971 & 3973  & 3848 & 4627 & 9942  & 2322 \\
HD 89948  & 1621 & 1236 & 1136 & 560  & 1285 & 828  & 1042  & 497  & 671  & 676   & 259  & 606  & ...   & 369  \\
HD 92545  & 710  & 480  & 601  & 430  & 1456 & 444  & 810   & 343  & 374  & 402   & 1336 & 342  & 483   & 972  \\
HD 106191 & 601  & 953  & 1258 & 560  & 993  & 418  & 831   & 946  & 429  & 746   & 417  & 1093 & 1002  & 854  \\
HD 107574 & 1729 & 1134 & 918  & 1532 & 6716 & 1604 & 2164  & 605  & 1118 & 1083  & 1773 & 943  & 1023  & 2322 \\
HD 116869 & 227  & 447  & 496  & ...  & 1364 & 896  & 1641  & 559  & 1156 & 977   & 241  & 539  & 2257  & 1422 \\
HD 123396 & 337  & 737  & 1474 & ...  & 2565 & 2817 & 8848  & 1743 & 5382 & 3318  & 1004 & 2514 & 8375  & 3328 \\
HD 123585 & 2235 & 2673 & 2404 & 1964 & 8085 & 5023 & 10199 & 2859 & 4219 & 4747  & 6372 & 1755 & 2702  & 495  \\
HD 147609 & 5400 & 4546 & 3815 & 1010 & 4879 & 4725 & 9087  & 2297 & 3414 & 3666  & 4903 & 3533 & 3192  & 1193 \\
HD 150862 & 668  & 1334 & 1464 & 569  & 1395 & 534  & 702   & 356  & 298  & 388   & 623  & 542  & 613   & 972  \\
HD 188985 & 1283 & 1216 & 1616 & 560  & 2112 & 1461 & 3552  & 903  & 1791 & 1547  & 881  & 1691 & 672   & 1820 \\
HD 210709 & 152  & 365  & 672  & ...  & 696  & 381  & 1106  & 299  & 660  & 517   & 137  & ...  & 673   & 81   \\
HD 210910 & 973  & 411  & 227  & ...  & 1267 & 533  & 914   & 986  & 512  & 606   & 2718 & 1808 & 982   & ...  \\
HD 222349 & 1553 & 1380 & 1714 & 545  & 3176 & 2643 & 5333  & 958  & 2931 & 2008  & 282  & 1869 & 2203  & 6009 \\
BD+18 5215& 1954 & 1255 & 1862 & 930  & 3840 & 1717 & 2692  & 980  & 1041 & 1307  & 353  & 8625 & 1978  & 7596 \\
HD 223938 & 547  & 642  & 1233 & ...  & 2173 & 1107 & 2274  & 454  & 1950 & 1629  & 1288 & 979  & 1988  & 495  \\
\noalign{\smallskip}                         
\hline
\end{tabular} 
$$                                           
}
\end{table*}

\begin{table*}[ht!]
\caption{Results relative to s-, r- and p-processes for barium stars. 
$\epsilon_s$(X), $\epsilon_{sw}$(X) and $\epsilon_{st}$(X): abundance fractions of s-process
main, weak and strong components;
$\epsilon_r$(X) and $\epsilon_p$(X): abundance fractions of r- and p-processes;
$\sigma$N and $\sigma$N (Si): cross section times abundance fraction corresponding to 
s-process main component taking into account the overabundance of barium stars in an usual
scale and Si scale; 
$\epsilon_s$(X)$_{nor}$: abundance fractions of s-process main component for normal stars;
diff: $\epsilon_s$(X) - $\epsilon_s$(X)$_{nor}$;
$\sigma$N$_{gs}$: cross section times abundance fraction corresponding to 
s-process main component without the overabundance of barium stars.
Full table is only available in electronic form.}
{\scriptsize
\label{resproc}
 $$
\setlength\tabcolsep{3pt}
\begin{tabular}{rrcccccccccccccccc}
\hline\hline
A & el & $\epsilon_s$(X) & $\sigma_{\epsilon_s(X)}$ & $\epsilon_{sw}$(X) & $\epsilon_{st}$(X) & 
$\epsilon_r$(X) & $\sigma_{\epsilon_r(X)}$ & $\epsilon_p$(X) & $\sigma_{\epsilon_p(X)}$ & 
$\sigma$N & $\sigma_{\sigma N}$ & $\epsilon_s$(X)$_{nor}$ & $\sigma_{\epsilon s(X){nor}}$ & diff & 
$\sigma$N (Si) & $\sigma_{\sigma N(Si)}$ & $\sigma$N$_{gs}$ \\
\hline 
\noalign{\smallskip}
 &&&&&&&&&& HD 749 &&&&&& \\
\hline
 84 & Sr &  ...   &  ...   &   ...      &   ...        & ...   & ...   &  4.2  & 1.8   & ...   & ...   &  ...   &  ...    &   ...   &     ...   &     ...   &    ...    \\
 86 & Sr &  208.7 &  118.7 & 37$\pm$16  &   ...        & ...   & ...   & ...   & ...   & 13357 &  7621 &  32.80 &  13.85  &  175.89 & 0.373E+03 & 0.213E+03 & 0.324E+03 \\
 87 & Sr &  145.5 &   82.7 & 23$\pm$9   &   ...        & ...   & ...   & ...   & ...   & 13387 &  7635 &  22.87 &   9.66  &  122.64 & 0.374E+03 & 0.213E+03 & 0.325E+03 \\
 88 & Sr & 3427.1 & 1948.9 & 45$\pm$19  &   ...        & ...   & ...   & ...   & ...   & 21248 & 12127 & 538.70 & 227.51  & 2888.40 & 0.593E+03 & 0.339E+03 & 0.516E+03 \\
 89 &  Y & 2260.2 &  994.0 &   ...      &   ...        & 11.9  &  5.5  & ...   & ...   & 42944 & 18935 & 136.79 &  63.17  & 2123.40 & 0.120E+04 & 0.529E+03 & 0.111E+04 \\
 90 & Zr & 4297.9 & 1992.2 &   ...      &   ...        & 53.0  & 20.7  & ...   & ...   & 90255 & 42711 & 137.63 &  53.75  & 4160.22 & 0.252E+04 & 0.119E+04 & 0.211E+04 \\
 91 & Zr & 1247.0 &  578.0 &   ...      &   ...        &  1.8  &  0.7  & ...   & ...   & 74821 & 36089 &  39.93 &  15.60  & 1207.09 & 0.209E+04 & 0.101E+04 & 0.175E+04 \\
 92 & Zr & 1854.7 &  859.7 &   ...      &   ...        &  4.2  &  1.6  & ...   & ...   & 61205 & 29325 &  59.39 &  23.20  & 1795.30 & 0.171E+04 & 0.820E+03 & 0.143E+04 \\
 94 & Zr & 2021.7 &  937.2 &   ...      &   ...        & ...   & ...   & ...   & ...   & 52565 & 24450 &  64.74 &  25.29  & 1956.98 & 0.147E+04 & 0.683E+03 & 0.123E+04 \\
 96 & Zr &  178.5 &   82.7 & 4.1$\pm$1.6&   ...        & ...   & ...   & ...   & ...   &  1910 &   890 &   5.72 &   2.23  &  172.75 & 0.533E+02 & 0.249E+02 & 0.445E+02 \\
\noalign{\smallskip}                         
\hline
\end{tabular} 
$$                                           
}
\end{table*}

\begin{figure*}[ht!]
\centerline{\includegraphics[totalheight=15.0cm]{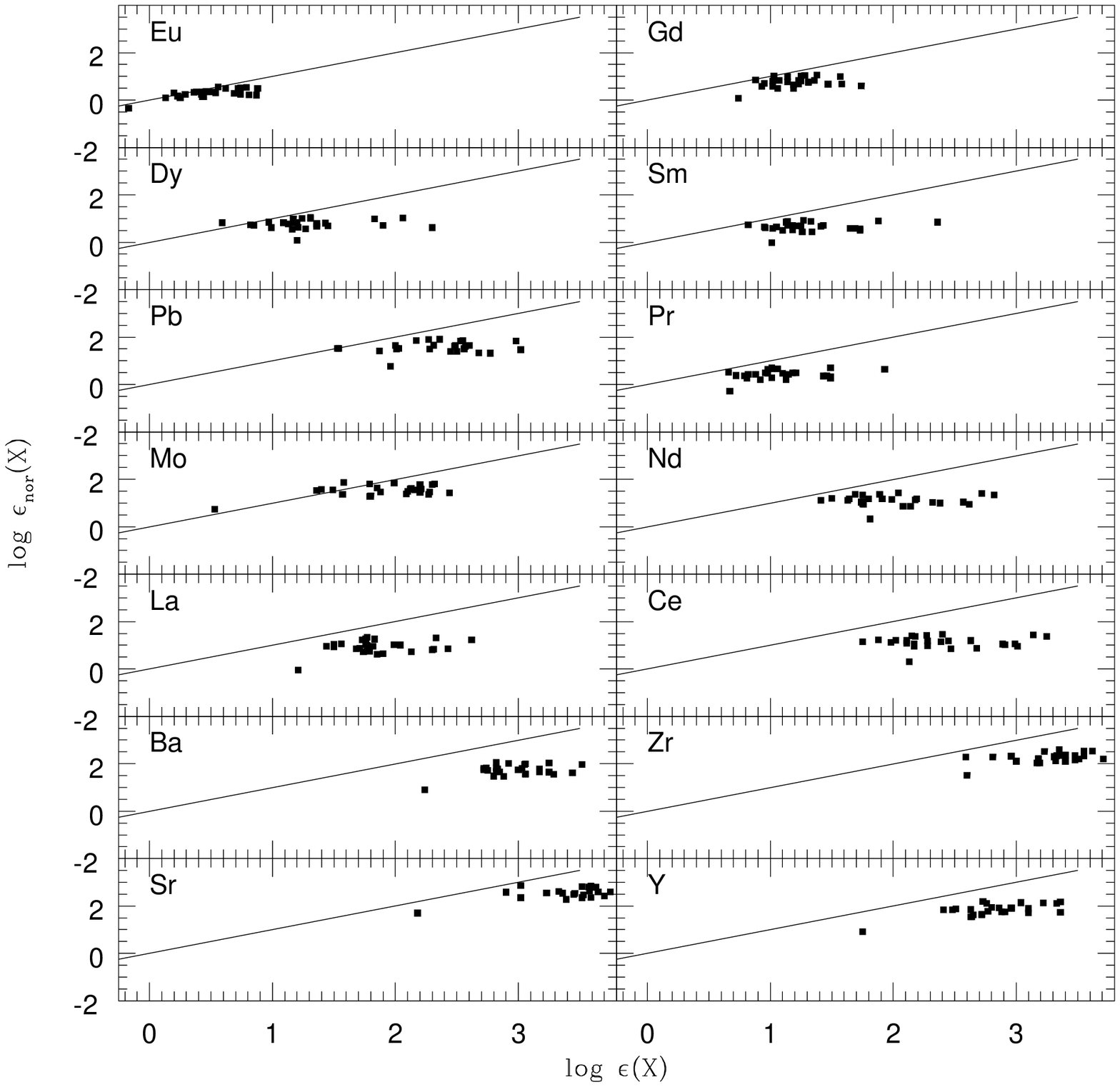}}
\caption{\label{norBaeleps} Total abundances of
heavy elements for the barium ($\log\epsilon$(X)) and normal ($\log\epsilon_{nor}$(X)) stars 
with the same metallicities. Solid lines
indicate $\log\epsilon_{nor}$(X) = $\log\epsilon$(X).}
\end{figure*}

\begin{figure*}[ht!]
\centerline{\includegraphics[totalheight=15.0cm]{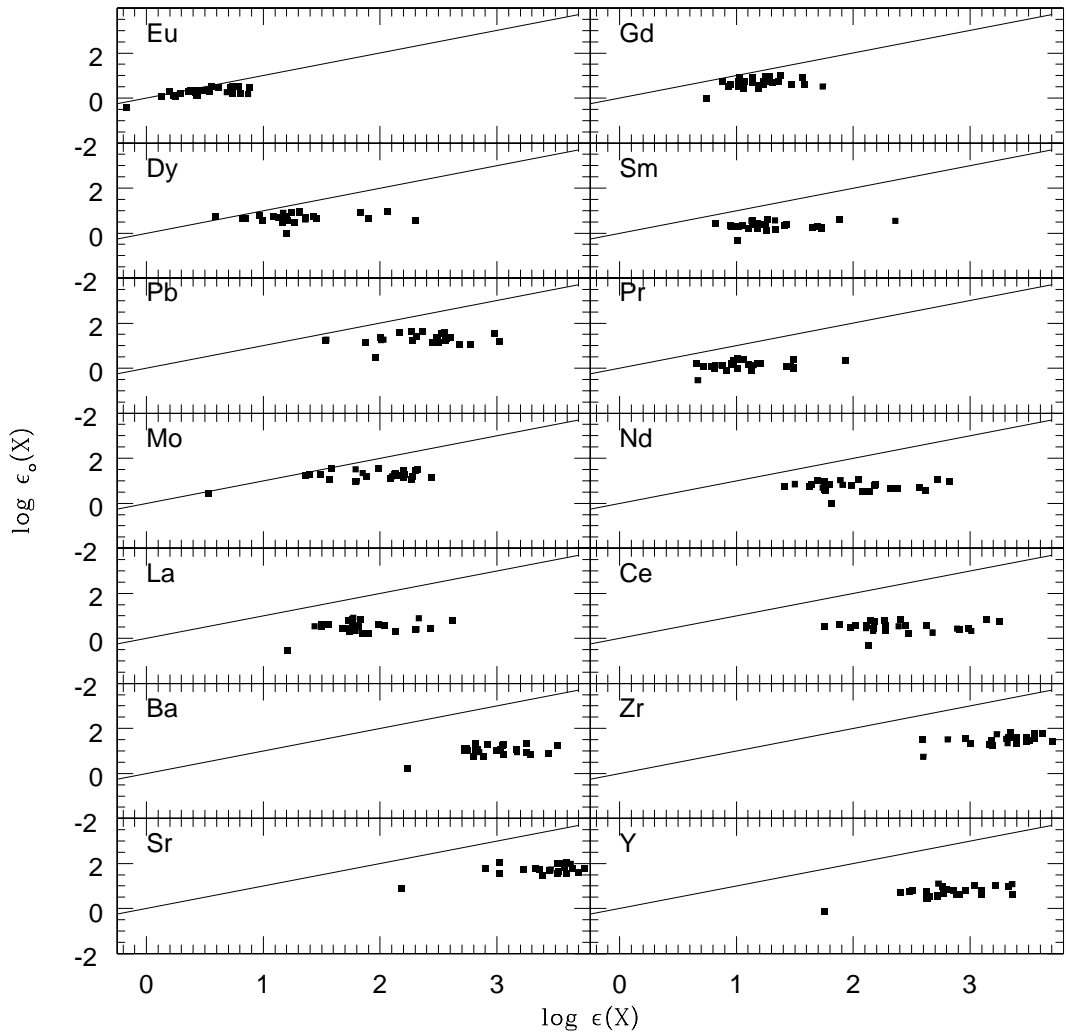}}
\caption{\label{norBaelrp} Abundance fraction of heavy elements 
due to all processes except the s-process main component ($\log\epsilon_o$(X)) vs. total 
abundance of barium stars ($\log\epsilon$(X)). Solid lines
indicate $\log\epsilon_o$(X) = $\log\epsilon$(X)}
\end{figure*}

\begin{figure*}[ht!]
\centerline{\includegraphics[totalheight=15.0cm]{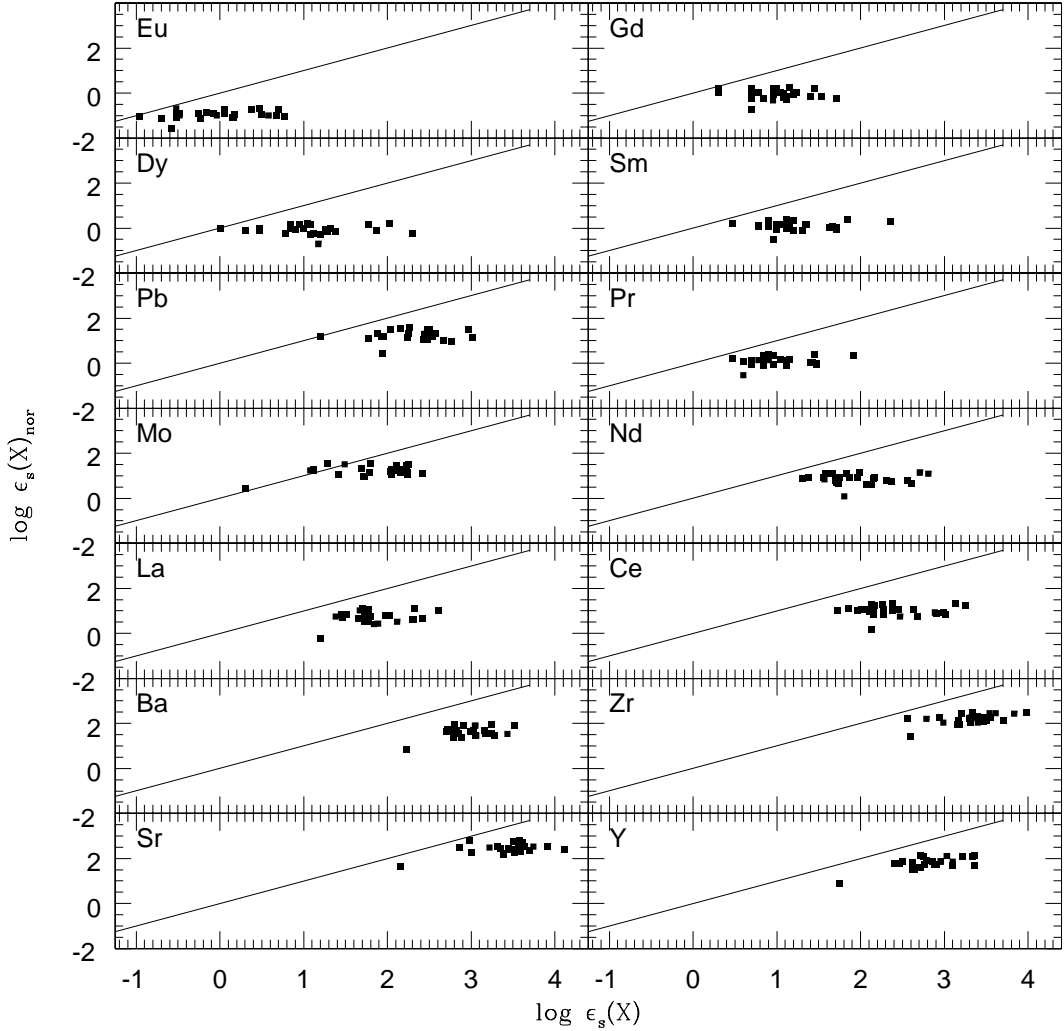}}
\caption{\label{norBaels} Abundances corresponding to the s-process
main component for heavy elements of barium stars ($\log\epsilon_s$(X)) vs. normal stars
($\log\epsilon_s$(X)$_{nor}$) with the same metallicities.
Solid lines indicate $\log\epsilon_s$(X)$_{nor}$ = $\log\epsilon_s$(X).}
\end{figure*}

\begin{figure*}[ht!]
\centerline{\includegraphics[totalheight=15.0cm]{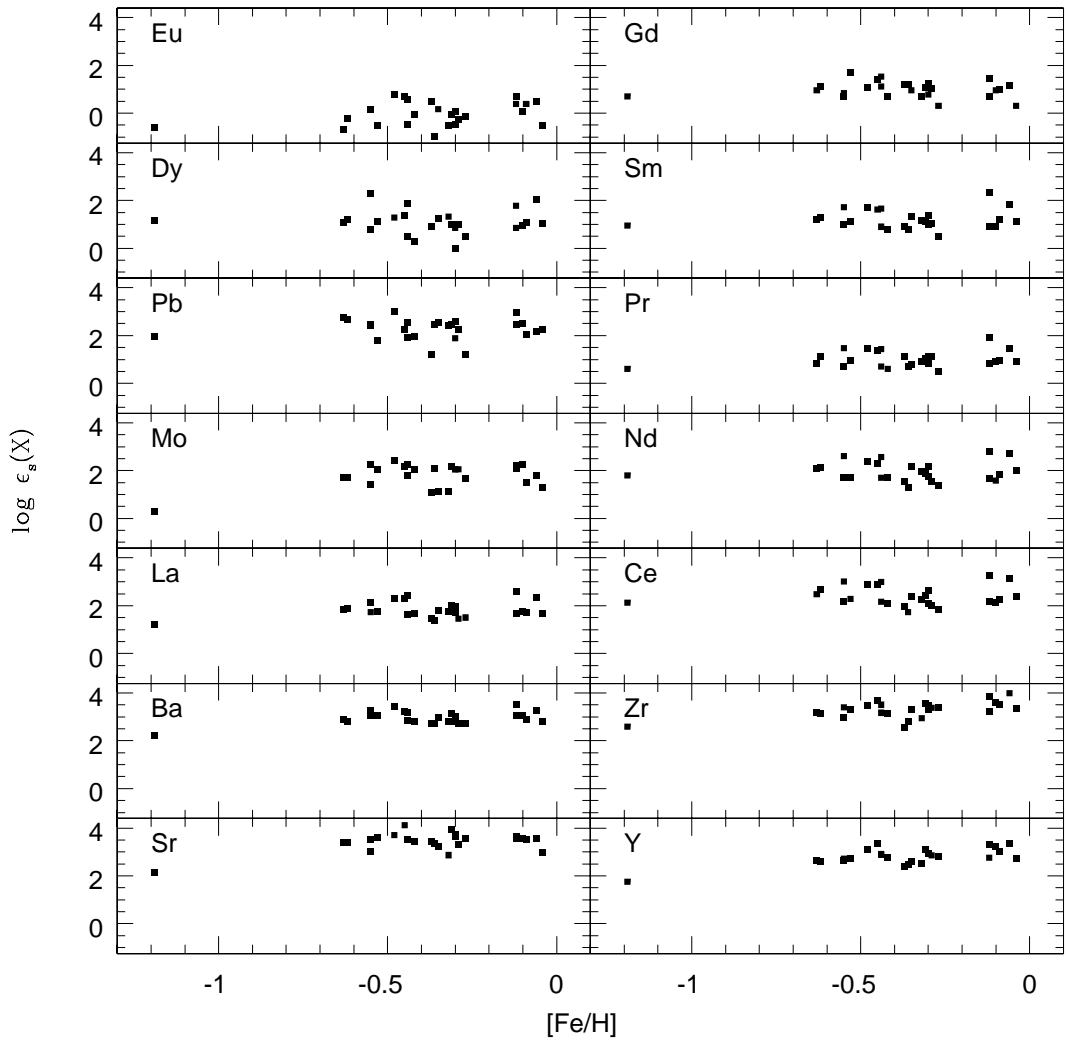}}
\caption{\label{basmet} $\log\epsilon_s$(X) vs. [Fe/H], where $\log\epsilon_s$(X)
are the abundances corresponding to the s-process main component of heavy elements in barium stars.}
\end{figure*}

\begin{figure*}[h!]
\centerline{\includegraphics[totalheight=15.0cm]{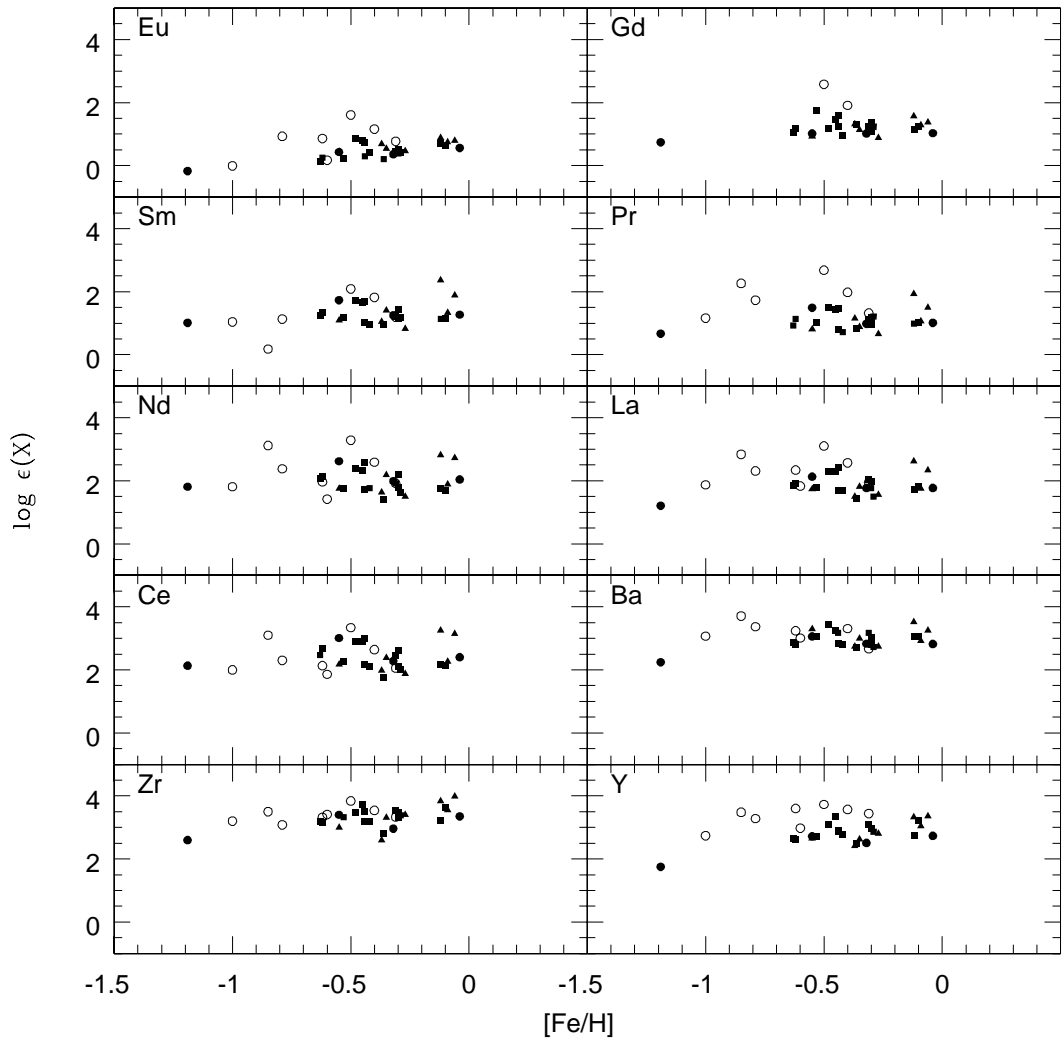}}
\caption{\label{lepsbaagb} $\log\epsilon$(X) vs. [Fe/H] for present sample 
barium stars and post-AGBs from \citet{reyniers04} and \citet{winckel00}.
Filled symbols indicate barium stars: squares: $\log g \ge 3.7$;
triangles: $2.4 < \log g < 3.7$; circles: $\log g \le 2.4$. 
Open circles are post-AGBs.}
\end{figure*}

\begin{figure*}[h!]
\centerline{\includegraphics[totalheight=19.0cm]{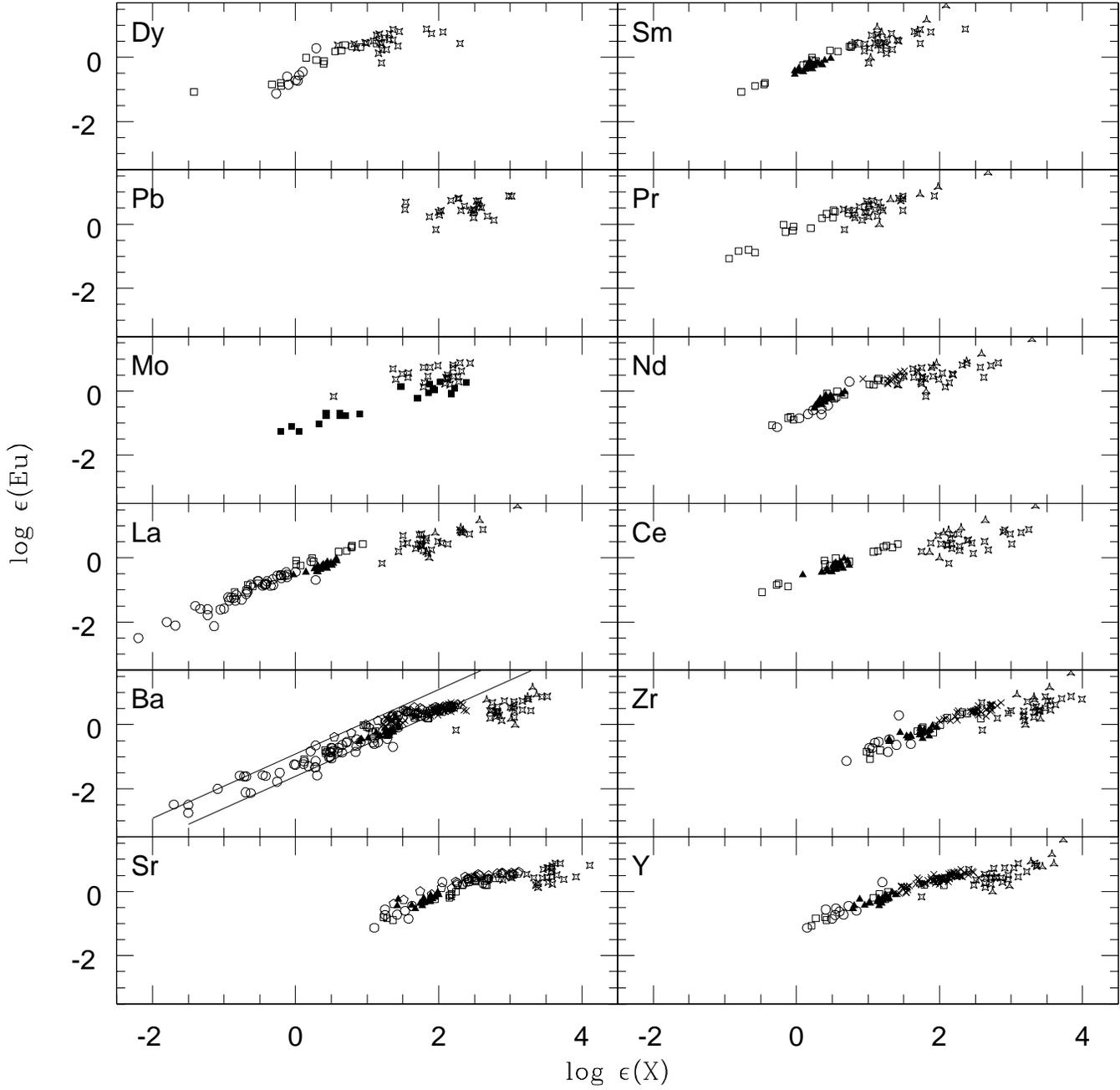}}
\caption{\label{lepseulepsba}  $\log\epsilon$(Eu) vs. $\log\epsilon$(X) for normal, 
post-AGB and barium stars. Normal stars from the literature are represented by symbols
as in Figure \ref{litpatm1}. Starred squares are the present sample barium stars. Starred 
triangles are post-AGB from \citet{reyniers04} and \citet{winckel00}.
Solid lines in the Ba panel represent r-only production of Ba (upper line) and solar 
[Ba/Eu] (lower line).}
\end{figure*}

\begin{figure*}[ht!]
\centerline{\includegraphics[totalheight=10.0cm]{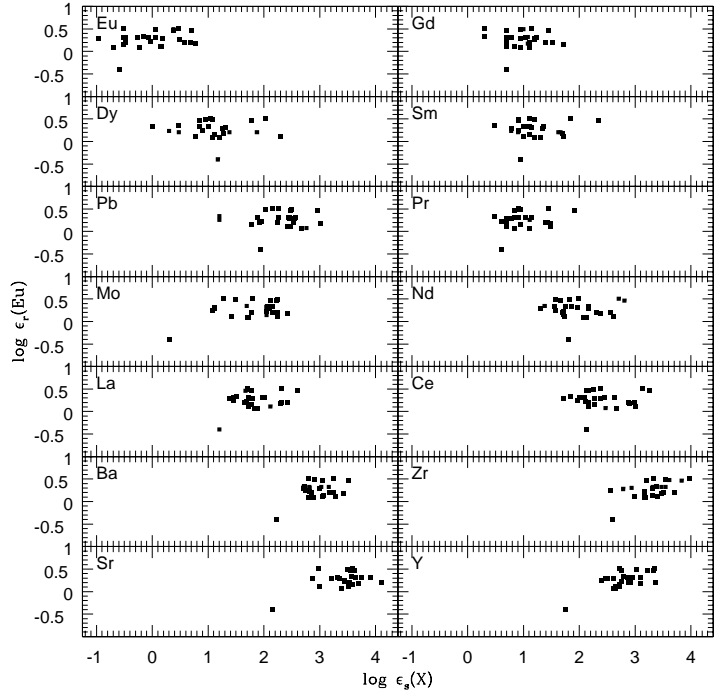}}
\caption{\label{norBaelrEusEba} $\log\epsilon_r$(Eu) vs. $\log\epsilon_s$(X),
where $\log\epsilon_r$(Eu) is the abundance fraction due to r-process for Eu  
and $\log\epsilon_s$(X) is that due to s-process main component for other heavy elements 
in barium stars.}
\end{figure*}

\begin{figure*}[ht!]
\centerline{\includegraphics[totalheight=10.0cm]{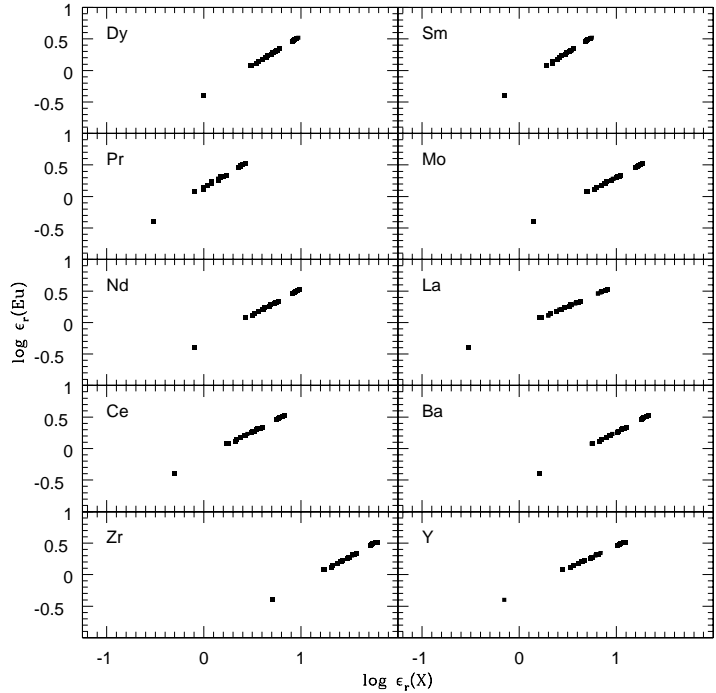}}
\caption{\label{norBaelr} $\log\epsilon_r$(Eu) vs. $\log\epsilon_r$(X),
where $\log\epsilon_r$(Eu) and $\log\epsilon_r$(X) are the abundance fractions due to r-process 
for Eu and other heavy elements in barium stars, respectively.}
\end{figure*}

Some elements of Table \ref{normais} are formed in larger amount through the r-process. They are
Eu (94.2\%), Gd (84.54\%), Dy (84.8\%) and Sm (67.4\%), according to \citet{arlandini99}. 
For these elements, barium stars are expected to have abundances values closer to normal 
stars than for s-elements. It can be verified that
this supposition is valid for Eu by comparing data from Tables 13 and 14 of 
paper I to those from Table \ref{normais}. Figure \ref{norBaeleps} shows the behaviour
of $\log\epsilon_{nor}$(X) of normal stars calculated by least-squares fitting with
$\log\epsilon$(X) of
barium stars, and Figure \ref{norBaelrp} shows the behaviour of $\log\epsilon$(X) of barium
stars with $\log\epsilon_o$(X). It is important to point out that Figures 
\ref{norBaeleps} to \ref{norBaelr}, the elements were arranged in increasing order of
contribution of the s-process main component, following \citet{arlandini99}, 
Eu, Gd, Dy, Sm, Pb, Pr, Mo, Nd, La, Ce, Ba, Zr, Sr and Y. Behaviours tend to be approximately
constant, however, Eu data are very close to a straight line with tangent = 1, 
differently from the other elements. The larger distance of the data from tangent = 1, the larger 
is the abundance of barium stars as compared to normal stars. If the fraction corresponding to 
the s-process main component
is withdrawn as in Figure \ref{norBaelrp}, the behaviours of Eu, Gd and Dy are almost
unaltered relative to Figure \ref{norBaeleps}, however the change in ordinates 
due to the missing abundance is remarkable for the other elements.

Figure \ref{norBaels} shows the behaviour of $\log\epsilon_s$(X) with $\log\epsilon_s$(X)$_{nor}$, 
which are the abundance fractions of the s-process main component of barium stars
(equation \ref{partes}) and normal stars of same metallicity, respectively.
Differently from Figure \ref{norBaelrp}, the changes in ordinates due to the missing
fraction r in Eu, Gd and Dy abundances in Figure \ref{norBaels} are remarkable, in comparison
with Figure \ref{norBaeleps}. Figure \ref{norBaels} shows similar differences from tangent = 1 
as Figures \ref{norBaeleps} and \ref{norBaelrp}, making evident that the fraction s of the 
abundance
of barium stars leads the behaviour. The difference between the maximum and minimum values of
the abundance fraction corresponding to the s-process main component of normal stars is
$\Delta\log\epsilon_s$(X)$_{nor}$ $\approx$ 1 in the range of metallicities of the sample stars.
For barium stars this difference is larger, $\Delta\log\epsilon_s$(X) $\approx$ 2.

According to Figure \ref{basmet}, the abundance relative to the s-process main component of 
heavy elements is essentially independent of [Fe/H] for the present sample barium stars.

\begin{figure*}[ht!]
\centerline{\includegraphics[totalheight=15.0cm]{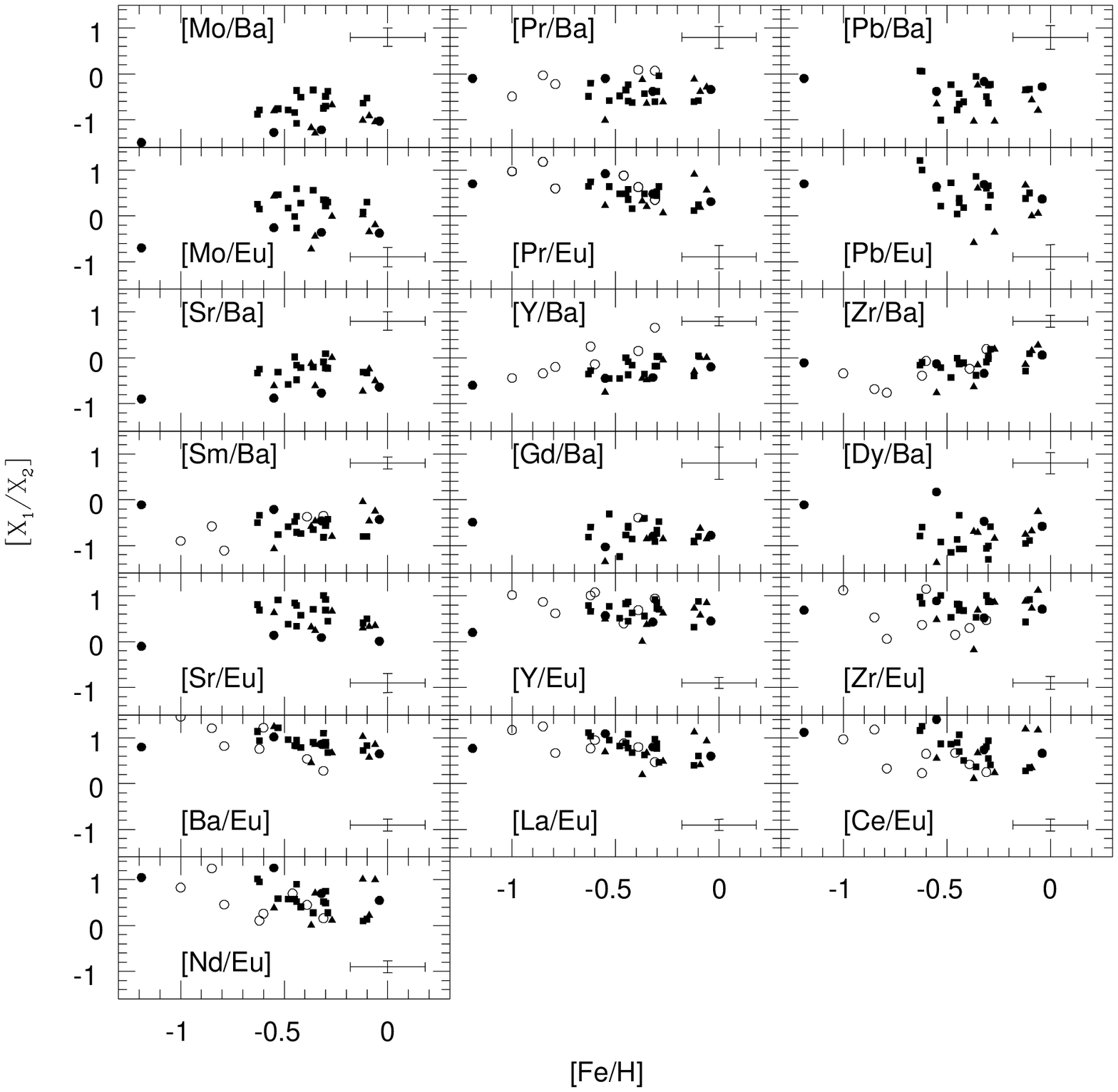}}
\caption{\label{relacsr} Abundance ratios involving s- and r-processes, light and heavy
s-elements. Uncertainties indicated are the largest values of Table \ref{relacsrt} for each
panel. Symbols are the same as in Figure \ref{lepsbaagb}.}
\end{figure*}

\subsection{Abundance ratios involving s- and r-processes}
The r-process is related to final stages of evolution of massive stars
(M $>$ 8 M$\sb \odot$) whereas the s-process main component occurs in AGB stars of low
(1-3 M$\sb \odot$) or intermediate (4-8 M$\sb \odot$) masses. The timescale for
stars to reach the SN II (t $<$ 10$\sp 8$ yr) stage is lower than that for stars to reach
the AGB phase (t $>$ 10$\sp 8$ yr), therefore the first heavy elements ejected in the
interstellar medium of the Galaxy, observed in very metal-poor stars, are expected
to be mainly due to the r-process \citep{truran02}. The products of nucleosynthesis of AGB 
stars to the interstellar medium, particularly the s-process main component appeared
latter.

In order to investigate when the s-process contribution starts, usually 
ratios involving s- and r-processes by using the best representatives of each one,
Eu for r- and Ba for s-processes are studied. Figure 6 from \citet{burr00} shows the
behaviour of Ba relative to Eu for normal stars of metallicities -3 $<$ [Fe/H] $<$+0.5.
From the moment that s-process starts to produce Ba, an increase
on its abundance relative to Eu can be seen, since the s-process is 
responsible for $\sim$ 81 \% of its production, according to \citet{arlandini99}.

Figure \ref{lepsbaagb} shows $\log\epsilon$(X) vs. [Fe/H] for barium stars from the 
present sample and post-AGB stars from \citet{reyniers04} and \citet{winckel00}.
Abundances of post-AGBs are larger or similar to those of barium stars, 
except for Sm in one star. This is expected considering that barium stars were 
enriched by an AGB companion. 

Figure \ref{lepseulepsba} shows $\log\epsilon$(Eu) vs. $\log\epsilon$(X), where X are 
heavy elements other than Eu, for the present sample barium stars and stars from the literature.
From Nd to Y, for which the s-process contribution is larger, 
barium stars and post-AGBs are clearly overabundant, located on the high abundance end.

In the Ba panel of Figure \ref{lepseulepsba}, the lowest values of 
$\log\epsilon$(Eu) corresponding to normal stars, are close to the upper line where 
[Ba/Eu] = -0.70, representing Ba production by r-process only \citep{mashonk03}.
As $\log\epsilon$(Eu) increases, data become closer to the lower line, where data are 
compatible with [Ba/Eu] values for the solar ratio. Both barium and post-AGB stars values
are very different from the solar ratio line.

In Figure \ref{norBaelrEusEba}, the 
r-process fraction of Eu abundance [$\log (0.942\epsilon$(Eu))] correlates with
the s-process main component abundance fraction of other elements in barium stars. In 
Figure \ref{norBaelr}, the r-process fraction of Eu is plotted against fractions
of r-process of other elements, for data given in Table \ref{normais}. 
In the latter, the correlation is linear, with no
dispersion, indicating that the s-process main component 
contribution causes a scatter in earlier figures. This linear correlation
in Figure \ref{norBaelr} is expected, given that only the r part of the 
abundances for all involved elements were used. Sr was not included 
in this figure because it does not have an r-process contribution 
(see Table \ref{sechoque}).

\begin{table*}
\caption{[X/Eu] (upper table) and [X/Ba] (lower table) for the barium stars.}
{\footnotesize
\label{relacsrt}
   $$ 
\setlength\tabcolsep{2.5pt}
\begin{tabular}{lrrrrrrrrrr}
\noalign{\smallskip}
\hline\hline
\noalign{\smallskip}
star   &  [Sr/Eu] &[Y/Eu]& [Zr/Eu] & [Mo/Eu] & [Ba/Eu] &[La/Eu]& [Ce/Eu] &[Nd/Eu]& [Pr/Eu] & [Pb/Eu] \\
\noalign{\smallskip}
\hline
\noalign{\smallskip}
HD 749     &  0.35$\pm$0.21 & 0.85$\pm$0.12 &  1.12$\pm$0.14 & -0.20$\pm$0.21 & 0.85$\pm$0.13 & 0.93$\pm$0.12 & 1.17$\pm$0.13 & 1.00$\pm$0.13 & 0.56$\pm$0.25 &  0.05$\pm$0.27 \\
HR 107     &  0.71$\pm$0.18 & 0.56$\pm$0.10 &  0.53$\pm$0.10 &  0.56$\pm$0.16 & 0.91$\pm$0.11 & 0.63$\pm$0.10 & 0.37$\pm$0.10 & 0.28$\pm$0.10 & 0.48$\pm$0.18 &  0.86$\pm$0.22 \\
HD 5424    &  0.14$\pm$0.21 & 0.57$\pm$0.12 &  0.89$\pm$0.14 & -0.26$\pm$0.21 & 1.02$\pm$0.13 & 1.09$\pm$0.12 & 1.40$\pm$0.13 & 1.26$\pm$0.13 & 0.92$\pm$0.25 &  0.64$\pm$0.27 \\
HD 8270    &  0.58$\pm$0.18 & 0.63$\pm$0.10 &  0.68$\pm$0.10 &  0.28$\pm$0.16 & 0.79$\pm$0.11 & 0.68$\pm$0.10 & 0.51$\pm$0.10 & 0.41$\pm$0.10 & 0.16$\pm$0.18 &  0.18$\pm$0.22 \\
HD 12392   &  0.30$\pm$0.21 & 0.73$\pm$0.12 &  0.88$\pm$0.14 &  0.02$\pm$0.21 & 1.03$\pm$0.13 & 1.13$\pm$0.12 & 1.19$\pm$0.13 & 1.01$\pm$0.13 & 0.91$\pm$0.25 &  0.67$\pm$0.27 \\
HD 13551   &  0.79$\pm$0.18 & 0.87$\pm$0.10 &  0.81$\pm$0.10 &  0.59$\pm$0.16 & 0.95$\pm$0.11 & 0.78$\pm$0.10 & 0.70$\pm$0.10 & 0.52$\pm$0.10 & 0.36$\pm$0.18 &  0.29$\pm$0.22 \\
HD 22589   &  0.67$\pm$0.18 & 0.62$\pm$0.10 &  0.86$\pm$0.10 & -0.01$\pm$0.16 & 0.67$\pm$0.11 & 0.49$\pm$0.10 & 0.24$\pm$0.10 & 0.11$\pm$0.10 & 0.06$\pm$0.18 & -0.36$\pm$0.22 \\
HD 27271   &  0.33$\pm$0.21 & 0.58$\pm$0.12 &  0.73$\pm$0.14 & -0.35$\pm$0.21 & 0.57$\pm$0.13 & 0.41$\pm$0.12 & 0.35$\pm$0.13 & 0.22$\pm$0.13 & 0.18$\pm$0.25 &  0.00$\pm$0.27 \\
HD 48565   &  0.69$\pm$0.18 & 0.66$\pm$0.10 &  0.84$\pm$0.10 &  0.15$\pm$0.16 & 0.94$\pm$0.11 & 1.04$\pm$0.10 & 1.25$\pm$0.10 & 0.96$\pm$0.10 & 0.74$\pm$0.18 &  1.00$\pm$0.22 \\
HD 76225   &  1.01$\pm$0.18 & 0.92$\pm$0.10 &  1.01$\pm$0.10 &  0.35$\pm$0.16 & 1.10$\pm$0.11 & 0.97$\pm$0.10 & 0.81$\pm$0.10 & 0.52$\pm$0.10 & 0.50$\pm$0.18 &  0.60$\pm$0.22 \\
HD 87080   &  0.34$\pm$0.18 & 0.45$\pm$0.10 &  0.69$\pm$0.10 & -0.26$\pm$0.16 & 0.82$\pm$0.11 & 1.08$\pm$0.10 & 1.07$\pm$0.10 & 0.90$\pm$0.10 & 0.58$\pm$0.18 &  0.39$\pm$0.22 \\
HD 89948   &  0.92$\pm$0.18 & 0.86$\pm$0.10 &  0.87$\pm$0.10 &  0.34$\pm$0.16 & 0.83$\pm$0.11 & 0.77$\pm$0.10 & 0.55$\pm$0.10 & 0.49$\pm$0.10 & 0.44$\pm$0.18 &  0.19$\pm$0.22 \\
HD 92545   &  0.41$\pm$0.18 & 0.32$\pm$0.10 &  0.43$\pm$0.10 &  0.08$\pm$0.16 & 0.72$\pm$0.11 & 0.40$\pm$0.10 & 0.28$\pm$0.10 & 0.10$\pm$0.10 & 0.12$\pm$0.18 &  0.38$\pm$0.22 \\
HD 106191  &  0.45$\pm$0.18 & 0.71$\pm$0.10 &  0.87$\pm$0.10 &  0.30$\pm$0.16 & 0.68$\pm$0.11 & 0.46$\pm$0.10 & 0.41$\pm$0.10 & 0.28$\pm$0.10 & 0.64$\pm$0.18 &  0.45$\pm$0.22 \\
HD 107574  &  0.63$\pm$0.18 & 0.49$\pm$0.10 &  0.48$\pm$0.10 &  0.43$\pm$0.16 & 1.24$\pm$0.11 & 0.69$\pm$0.10 & 0.55$\pm$0.10 & 0.39$\pm$0.10 & 0.23$\pm$0.18 &  0.58$\pm$0.22 \\
HD 116869  &  0.09$\pm$0.21 & 0.43$\pm$0.12 &  0.52$\pm$0.14 & -0.36$\pm$0.21 & 0.86$\pm$0.13 & 0.80$\pm$0.12 & 0.74$\pm$0.13 & 0.70$\pm$0.13 & 0.48$\pm$0.25 &  0.69$\pm$0.27 \\
HD 123396  & -0.10$\pm$0.21 & 0.20$\pm$0.12 &  0.69$\pm$0.14 & -0.70$\pm$0.21 & 0.80$\pm$0.13 & 0.77$\pm$0.12 & 1.12$\pm$0.13 & 1.05$\pm$0.13 & 0.70$\pm$0.25 &  0.70$\pm$0.27 \\
HD 123585  &  0.38$\pm$0.18 & 0.51$\pm$0.10 &  0.53$\pm$0.10 &  0.17$\pm$0.16 & 0.96$\pm$0.11 & 0.82$\pm$0.10 & 0.86$\pm$0.10 & 0.58$\pm$0.10 & 0.48$\pm$0.18 &  0.72$\pm$0.22 \\
HD 147609  &  0.85$\pm$0.18 & 0.83$\pm$0.10 &  0.82$\pm$0.10 & -0.01$\pm$0.16 & 0.83$\pm$0.11 & 0.89$\pm$0.10 & 0.90$\pm$0.10 & 0.58$\pm$0.10 & 0.48$\pm$0.18 &  0.04$\pm$0.22 \\
HD 150862  &  0.50$\pm$0.18 & 0.88$\pm$0.10 &  0.92$\pm$0.10 &  0.30$\pm$0.16 & 0.83$\pm$0.11 & 0.60$\pm$0.10 & 0.35$\pm$0.10 & 0.14$\pm$0.10 & 0.25$\pm$0.18 &  0.50$\pm$0.22 \\
HD 188985  &  0.69$\pm$0.18 & 0.73$\pm$0.10 &  0.89$\pm$0.10 &  0.21$\pm$0.16 & 0.91$\pm$0.11 & 0.87$\pm$0.10 & 0.94$\pm$0.10 & 0.75$\pm$0.10 & 0.53$\pm$0.18 &  0.66$\pm$0.22 \\
HD 210709  &  0.01$\pm$0.21 & 0.45$\pm$0.12 &  0.71$\pm$0.14 & -0.38$\pm$0.21 & 0.65$\pm$0.13 & 0.60$\pm$0.12 & 0.66$\pm$0.13 & 0.55$\pm$0.13 & 0.31$\pm$0.25 &  0.37$\pm$0.27 \\
HD 210910  &  0.32$\pm$0.21 & 0.00$\pm$0.12 & -0.18$\pm$0.14 & -0.73$\pm$0.21 & 0.45$\pm$0.13 & 0.20$\pm$0.12 & 0.11$\pm$0.13 & 0.01$\pm$0.13 & 0.32$\pm$0.25 & -0.58$\pm$0.27 \\
HD 222349  &  0.81$\pm$0.18 & 0.79$\pm$0.10 &  0.98$\pm$0.10 &  0.26$\pm$0.16 & 1.14$\pm$0.11 & 1.11$\pm$0.10 & 1.16$\pm$0.10 & 1.02$\pm$0.10 & 0.65$\pm$0.18 &  1.21$\pm$0.22 \\
BD+18 5215 &  0.91$\pm$0.18 & 0.77$\pm$0.10 &  1.01$\pm$0.10 &  0.46$\pm$0.16 & 1.22$\pm$0.11 & 0.95$\pm$0.10 & 0.87$\pm$0.10 & 0.59$\pm$0.10 & 0.64$\pm$0.18 &  0.21$\pm$0.22 \\
HD 223938  &  0.24$\pm$0.21 & 0.37$\pm$0.12 &  0.69$\pm$0.14 & -0.45$\pm$0.21 & 0.85$\pm$0.13 & 0.67$\pm$0.12 & 0.67$\pm$0.13 & 0.71$\pm$0.13 & 0.20$\pm$0.25 &  0.60$\pm$0.27 \\
\noalign{\smallskip}
\hline\hline
\noalign{\smallskip}
star & [Sr/Ba] &[Y/Ba] &[Zr/Ba] & [Mo/Ba] & [Pb/Ba] &[Pr/Ba] &[Sm/Ba] & [Gd/Ba] & [Dy/Ba] & \\
\noalign{\smallskip}
\hline
\noalign{\smallskip}
HD 749     & -0.50$\pm$0.20 &  0.00$\pm$0.10 &  0.27$\pm$0.13 & -1.05$\pm$0.20 & -0.80$\pm$0.26 & -0.29$\pm$0.24 & -0.25$\pm$0.13 & -0.86$\pm$0.35 & -0.26$\pm$0.23 &\\
HR 107     & -0.20$\pm$0.17 & -0.35$\pm$0.07 & -0.38$\pm$0.07 & -0.35$\pm$0.14 & -0.05$\pm$0.21 & -0.43$\pm$0.17 & -0.65$\pm$0.11 & -0.40$\pm$0.12 &   ...          &\\
HD 5424    & -0.88$\pm$0.20 & -0.45$\pm$0.10 & -0.13$\pm$0.13 & -1.28$\pm$0.20 & -0.38$\pm$0.26 & -0.10$\pm$0.24 & -0.21$\pm$0.13 & -1.03$\pm$0.35 &  0.17$\pm$0.23 &\\
HD 8270    & -0.21$\pm$0.17 & -0.16$\pm$0.07 & -0.11$\pm$0.07 & -0.51$\pm$0.14 & -0.61$\pm$0.21 & -0.63$\pm$0.17 & -0.74$\pm$0.11 & -0.86$\pm$0.12 & -1.07$\pm$0.09 &\\
HD 12392   & -0.73$\pm$0.20 & -0.30$\pm$0.10 & -0.15$\pm$0.13 & -1.01$\pm$0.20 & -0.36$\pm$0.26 & -0.12$\pm$0.24 & -0.04$\pm$0.13 & -0.94$\pm$0.35 & -0.76$\pm$0.23 &\\
HD 13551   & -0.16$\pm$0.17 & -0.08$\pm$0.07 & -0.14$\pm$0.07 & -0.36$\pm$0.14 & -0.66$\pm$0.21 & -0.59$\pm$0.17 & -0.71$\pm$0.11 & -0.61$\pm$0.12 & -1.07$\pm$0.09 &\\
HD 22589   &  0.00$\pm$0.17 & -0.05$\pm$0.07 &  0.19$\pm$0.07 & -0.68$\pm$0.14 & -1.03$\pm$0.21 & -0.61$\pm$0.17 & -0.80$\pm$0.11 & -0.85$\pm$0.12 & -0.84$\pm$0.09 &\\
HD 27271   & -0.24$\pm$0.20 &  0.01$\pm$0.10 &  0.16$\pm$0.13 & -0.92$\pm$0.20 & -0.57$\pm$0.26 & -0.39$\pm$0.24 & -0.47$\pm$0.13 & -0.63$\pm$0.35 & -0.68$\pm$0.23 &\\
HD 48565   & -0.25$\pm$0.17 & -0.28$\pm$0.07 & -0.10$\pm$0.07 & -0.79$\pm$0.14 &  0.06$\pm$0.21 & -0.20$\pm$0.17 & -0.34$\pm$0.11 & -0.60$\pm$0.12 & -0.60$\pm$0.09 &\\
HD 76225   & -0.09$\pm$0.17 & -0.18$\pm$0.07 & -0.09$\pm$0.07 & -0.75$\pm$0.14 & -0.50$\pm$0.21 & -0.60$\pm$0.17 & -0.82$\pm$0.11 & -0.91$\pm$0.12 & -1.06$\pm$0.09 &\\
HD 87080   & -0.48$\pm$0.17 & -0.37$\pm$0.07 & -0.13$\pm$0.07 & -1.08$\pm$0.14 & -0.43$\pm$0.21 & -0.24$\pm$0.17 & -0.36$\pm$0.11 & -0.58$\pm$0.12 & -0.34$\pm$0.09 &\\
HD 89948   &  0.09$\pm$0.17 &  0.03$\pm$0.07 &  0.04$\pm$0.07 & -0.49$\pm$0.14 & -0.64$\pm$0.21 & -0.39$\pm$0.17 & -0.56$\pm$0.11 & -0.74$\pm$0.12 & -1.30$\pm$0.09 &\\
HD 92545   & -0.31$\pm$0.17 & -0.40$\pm$0.07 & -0.29$\pm$0.07 & -0.64$\pm$0.14 & -0.34$\pm$0.21 & -0.60$\pm$0.17 & -0.80$\pm$0.11 & -0.90$\pm$0.12 & -0.95$\pm$0.09 &\\
HD 106191  & -0.23$\pm$0.17 &  0.03$\pm$0.07 &  0.19$\pm$0.07 & -0.38$\pm$0.14 & -0.23$\pm$0.21 & -0.04$\pm$0.17 & -0.42$\pm$0.11 & -0.48$\pm$0.12 & -0.59$\pm$0.09 &\\
HD 107574  & -0.61$\pm$0.17 & -0.75$\pm$0.07 & -0.76$\pm$0.07 & -0.81$\pm$0.14 & -0.66$\pm$0.21 & -1.01$\pm$0.17 & -1.07$\pm$0.11 & -1.35$\pm$0.12 & -1.37$\pm$0.09 &\\
HD 116869  & -0.77$\pm$0.20 & -0.43$\pm$0.10 & -0.34$\pm$0.13 & -1.22$\pm$0.20 & -0.17$\pm$0.26 & -0.38$\pm$0.24 & -0.46$\pm$0.13 & -0.80$\pm$0.35 & -0.47$\pm$0.23 &\\
HD 123396  & -0.90$\pm$0.20 & -0.60$\pm$0.10 & -0.11$\pm$0.13 & -1.50$\pm$0.20 & -0.10$\pm$0.26 & -0.10$\pm$0.24 & -0.11$\pm$0.13 & -0.49$\pm$0.35 & -0.11$\pm$0.23 &\\
HD 123585  & -0.58$\pm$0.17 & -0.45$\pm$0.07 & -0.43$\pm$0.07 & -0.79$\pm$0.14 & -0.24$\pm$0.21 & -0.48$\pm$0.17 & -0.59$\pm$0.11 & -1.24$\pm$0.12 & -1.15$\pm$0.09 &\\
HD 147609  &  0.02$\pm$0.17 &  0.00$\pm$0.07 & -0.01$\pm$0.07 & -0.84$\pm$0.14 & -0.79$\pm$0.21 & -0.35$\pm$0.17 & -0.48$\pm$0.11 & -0.77$\pm$0.12 & -0.87$\pm$0.09 &\\
HD 150862  & -0.33$\pm$0.17 &  0.05$\pm$0.07 &  0.09$\pm$0.07 & -0.53$\pm$0.14 & -0.33$\pm$0.21 & -0.58$\pm$0.17 & -0.80$\pm$0.11 & -0.80$\pm$0.12 & -0.89$\pm$0.09 &\\
HD 188985  & -0.22$\pm$0.17 & -0.18$\pm$0.07 & -0.02$\pm$0.07 & -0.70$\pm$0.14 & -0.25$\pm$0.21 & -0.38$\pm$0.17 & -0.48$\pm$0.11 & -0.66$\pm$0.12 & -1.01$\pm$0.09 &\\
HD 210709  & -0.64$\pm$0.20 & -0.20$\pm$0.10 &  0.06$\pm$0.13 & -1.03$\pm$0.20 & -0.28$\pm$0.26 & -0.34$\pm$0.24 & -0.43$\pm$0.13 & -0.78$\pm$0.35 & -0.58$\pm$0.23 &\\
HD 210910  & -0.13$\pm$0.20 & -0.45$\pm$0.10 & -0.63$\pm$0.13 & -1.18$\pm$0.20 & -1.03$\pm$0.26 & -0.13$\pm$0.24 & -0.58$\pm$0.13 & -0.43$\pm$0.35 & -0.69$\pm$0.23 &\\
HD 222349  & -0.33$\pm$0.17 & -0.35$\pm$0.07 & -0.16$\pm$0.07 & -0.88$\pm$0.14 &  0.07$\pm$0.21 & -0.49$\pm$0.17 & -0.50$\pm$0.11 & -0.81$\pm$0.12 & -0.79$\pm$0.09 &\\
BD+18 5215 & -0.31$\pm$0.17 & -0.45$\pm$0.07 & -0.21$\pm$0.07 & -0.76$\pm$0.14 & -1.01$\pm$0.21 & -0.58$\pm$0.17 & -0.76$\pm$0.11 & -0.31$\pm$0.12 & -0.92$\pm$0.09 &\\
HD 223938  & -0.61$\pm$0.20 & -0.48$\pm$0.10 & -0.16$\pm$0.13 & -1.30$\pm$0.20 & -0.25$\pm$0.26 & -0.65$\pm$0.24 & -0.47$\pm$0.13 & -0.85$\pm$0.35 & -0.71$\pm$0.23 &\\
\hline
\end{tabular}
   $$ 
}
\end{table*}

[X/Eu] vs. [Fe/H] for barium and post-AGB stars are shown
in Table \ref{relacsrt} and Figure \ref{relacsr}. The star HD 210910 shows lower
values of [X/Eu] for Y, Zr, Ba, La, Ce, Nd, Pr and Pb, reaching negative values
for [Zr/Eu] = -0.18. The value of [Sr/Eu] is low for the star HD 123396, 
[Sr/Eu] = -0.1. This result is expected from 
Figure 6 of paper I, which shows that the stars HD 123396 and HD 210910 
present the lowest values of [SrII/Fe] and [ZrII/Fe], respectively, and their 
[Eu/Fe] = 0.50 and 0.54, thus very close. For other stars, [X/Eu] is
in the range of 0 $\leq$ [X/Eu] $\leq$ 1.3. The post-AGBs are also
mainly in this range, the lowest value being [Zr/Eu] = -0.01. For all elements,
[X/Eu] vs. [Fe/H] is approximately constant in the range of metallicities of 
the sample stars.

For Mo, that has a contribution of 49.76\% of the s-process main component, 
26.18\% of the r-process and 24.06\% of the p-process, [Mo/Eu] are mainly in the range 
from -0.4 to 0.6. Relative to Ba, the values are lower, in the range 
-1.5 $\leq$ [Mo/Ba] $\leq$ -0.35.
Regarding the ratios involving Pr, that has 49\% of contribution from the s-process main
component and 51\% from the r-process, data are mainly in the ranges
-1.01 $\leq$ [Pr/Ba] $\leq$ -0.04 and 0.12 $\leq$ [Pr/Eu] $\leq$ 0.92.
There is no r-process contribution in the Pb abundance, according to
\citet{arlandini99}. 46\% of contribution comes from
s-process main component and 54\% has been attributed in previous work to the strong
s-process component. Data are in the ranges -1.03 $\leq$ [Pb/Ba] $\leq$ 0.07
and -0.58 $\leq$ [Pb/Eu] $\leq$ 0.86. Abundances of Pb obtained for barium stars are
usually larger than those of Mo, which are larger than those of Pr, and the same
is true for solar abundances. Only for 5 stars the Mo abundance is
larger than those of Pb and for one star Pr abundance is larger than that of Mo.
However, [X/Fe] values are such that the values of [Mo/Ba,Eu] are usually lower than
[Pr,Pb/Ba,Eu]. As a consequence, the range for Mo involves lower values in 
Figure \ref{relacsr}.

Elements formed mainly by the s-process main component can be divided into two groups:
light s-elements around magic neutron number 50 and heavy s-elements 
around magic neutron number 82. In this work, Sr, Y, Zr were included in
light s-elements and Ba, La, Ce, Nd in the heavy s-elements groups. For these
elements, s-process main component contribution is larger than 50\% according 
to \citet{arlandini99}. It is
worth to emphasize that Sm is rather an r-element because its s-process contribution
is less than 30\% while r-process contributes with 67.4\% of its production. The 
presence of light s-elements in very metal-poor stars cannot be entirely 
explained by an r-process contribution. 
For instance, there is no production of Sr by r-process, as shown in 
Table \ref{sechoque}, and it is observed in the most metal-poor stars. Another
nucleosynthetic process related to massive stars is needed in order to explain
such observed abundances. Figure 9 from \citet{burr00} shows an increasing trend of
[Sr/Ba] toward lower metallicities [Fe/H] $<$ -1. It has been suggested 
that beyond the observational uncertainties, another
nucleosynthesis source would explain the increasing [Sr/Ba] at lower
metallicities, and this source could be the s-process weak component.
According to Table \ref{sechoque}, 14.9\% of solar abundance of Sr come from
the s-process weak component and 0.56\% from the p-process. At metallicities as
low as [Fe/H] $<$ -3, the Sr production is expected to be low, given that
the s-process weak component is secondary and in such environment there is
a lack of pre-existing seed nuclei.

Abundance ratios of
light s-elements relative to Ba are shown in Table \ref{relacsrt} and 
Figure \ref{relacsr}. In the range of metallicities studied, the ratios 
are approximately constant in the ranges -1 $\leq$ [Sr/Ba] $\leq$ 0.1,
-0.75 $\leq$ [Y/Ba] $\leq$ 0.05 and -0.80 $\leq$ [Zr/Ba] $\leq$ 0.30.
AGB stars are included in the same range for [Zr/Ba], while for
[Y/Ba] two of them show higher values, 0.25 and 0.66. Figure 9 from
\citet{burr00} shows a dispersion around [Fe/H] $\approx$ -1
with -0.55 $\leq$ [Sr/Ba] $\leq$ 0.1, similarly to the present sample
barium stars.

%

\section{Neutron Exposures}

\subsection{Theoretical predictions of Malaney (1987)}

s-element nucleosynthesis depends on
the neutron exposure to which the seed nuclei were submitted inside the AGB star.
Considering the scenario of material transfer for the barium star
formation, it is reasonable to expect that the abundance patterns of barium
stars show signatures of the neutron exposure during the occurrence of s-process
in the AGB companion.

Models trying to reproduce the resulting abundances of s-process, were presented 
by \citet{cd80}, who calculated theoretical abundances by using exact solutions 
from \citet{clayton74} for models of exponential distribution, and the
approximate solution from \citet{clayton61} for a model of single exposure.

\citet{malaney87a} presented theoretical predictions of
abundances from the s-process starting on iron seeds, considering single neutron
exposure for several values of $\tau_o$ and neutron density
n$\sb n$ = 10$\sp 8$ cm$\sp {-3}$. \citet{malaney87b} also provides
theoretical predictions, but considering an exponential neutron distribution
for several values of $\tau_o$ and two different values of neutron density,
10$^8$cm$\sp {-3}$ and 10$^{12}$cm$\sp {-3}$. 

Abundances resulting from theoretical predictions are usually normalized for
$\log\epsilon_c$(Sr) = 20. \citet{tl83} provide an expression to transform
observational data to this scale

\begin{equation}
\label{abaju}
\epsilon_c(X)=C(y-1)\epsilon_\odot(X)
\end{equation}
where y is the relative abundance between barium and normal stars of same metallicity
$$\log y = \log\epsilon(X) - \log\epsilon_{nor}(X)$$
where $\log\epsilon$(X) of barium stars are those from Tables 13 and 14
of paper I and $\log\epsilon_{nor}$(X) are those from Table \ref{normais}.
The value of C for each star is known by setting $\log\epsilon_c$(Sr) = 20. 

In order to check the fit, \citet{cd80} used an expression largely
used in the literature \citep[e.g.][]{winckel00,claudio03}.
\begin{equation}
\label{quaju}
S^2={1\over N}\sum_{i=1}^N{(O_i-P_i)^2\over \sigma_i^2}
\end{equation}
where O$\sb i$ are abundances $\log\epsilon_c$(X) for each element ``$i$'',
$P\sb i$ are theoretical predictions, $N$ is the number of elements and
$\sigma$ is the uncertainty on $\log\epsilon_c$(X) calculated by

\begin{equation}
\label{errlepsc}
\sigma_{\log\epsilon c}=\sqrt{\sigma_{\log C}^2+{\sigma^2_{\log y}\over (1-10^{-\log y})^2} + \sigma^2_{\log\epsilon\odot(X)}}
\end{equation}
with
\begin{equation}
\sigma_{\log y}=\sqrt{\sigma^2_{\log\epsilon(X)}+\sigma^2_{\log\epsilon nor(X)}}
\end{equation}
and the uncertainty on C
\begin{equation}
\sigma_{\log C}=\sqrt{{\sigma^2_{\log y}\over (1-10^{-\log y})^2} + \sigma^2_{\log\epsilon\odot(Sr)}}
\end{equation}
where $y$ represents Sr, considering $\sigma_{\log\epsilon c(Sr)}$ = 0.

\begin{figure}[ht!]
\centerline{\includegraphics[totalheight=9.5cm]{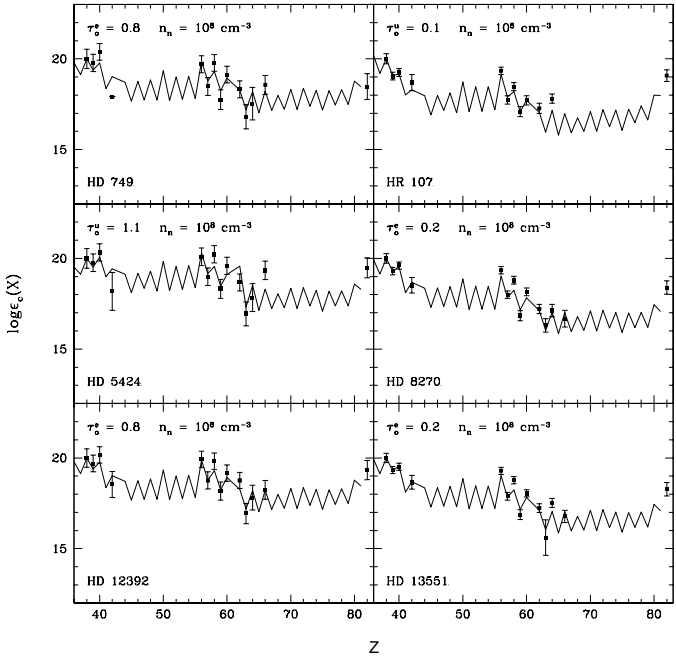}}
\caption{\label{expo1} Fittings of observed data to
theoretical predictions by \citet{malaney87a,malaney87b}.}
\end{figure}

\begin{figure}[ht!]
\centerline{\includegraphics[totalheight=9.5cm]{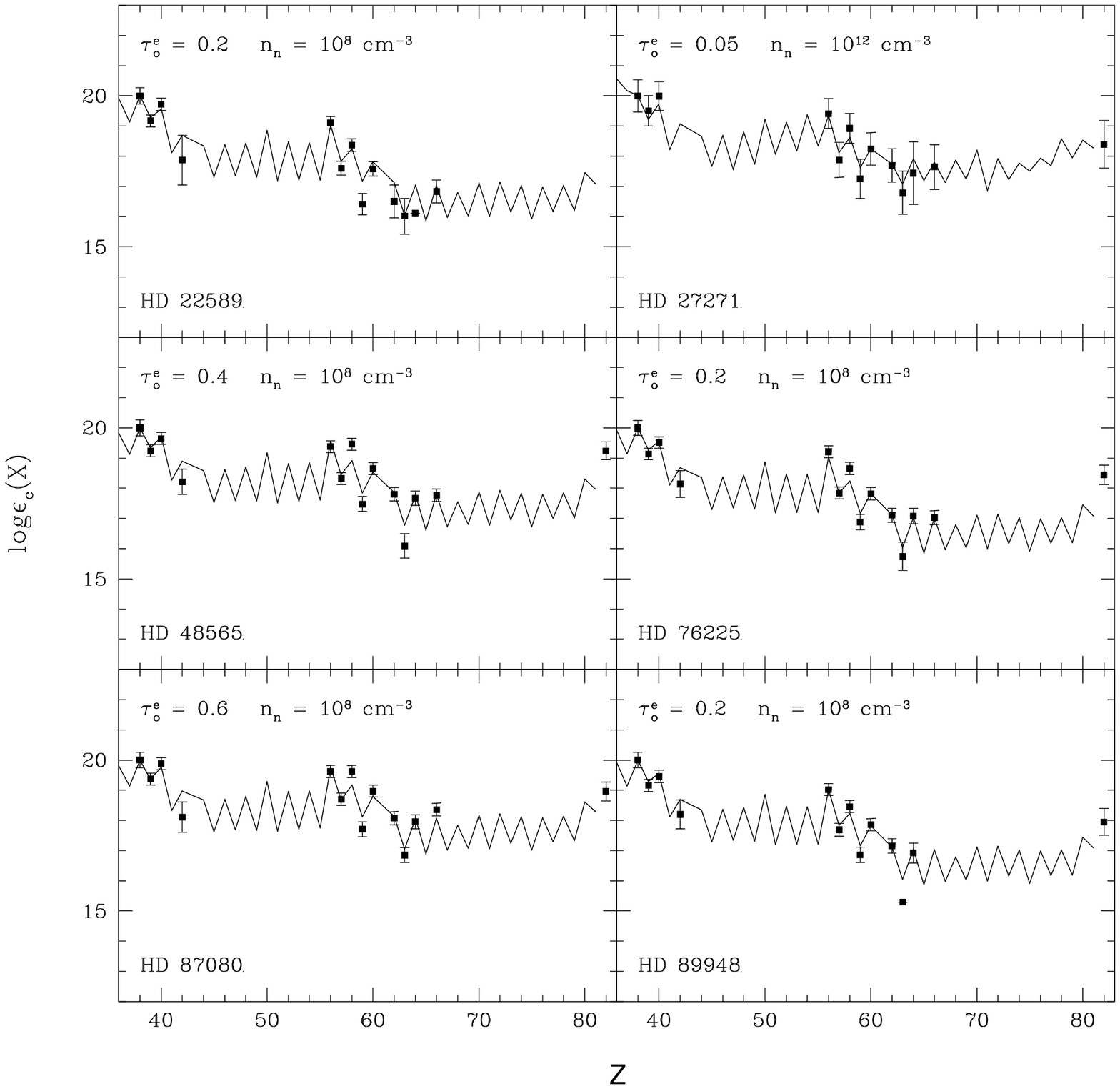}}
\caption{\label{expo2} Same as Figure \ref{expo1} for other 6 sample stars.}
\end{figure}

\begin{figure}[ht!]
\centerline{\includegraphics[totalheight=9.5cm]{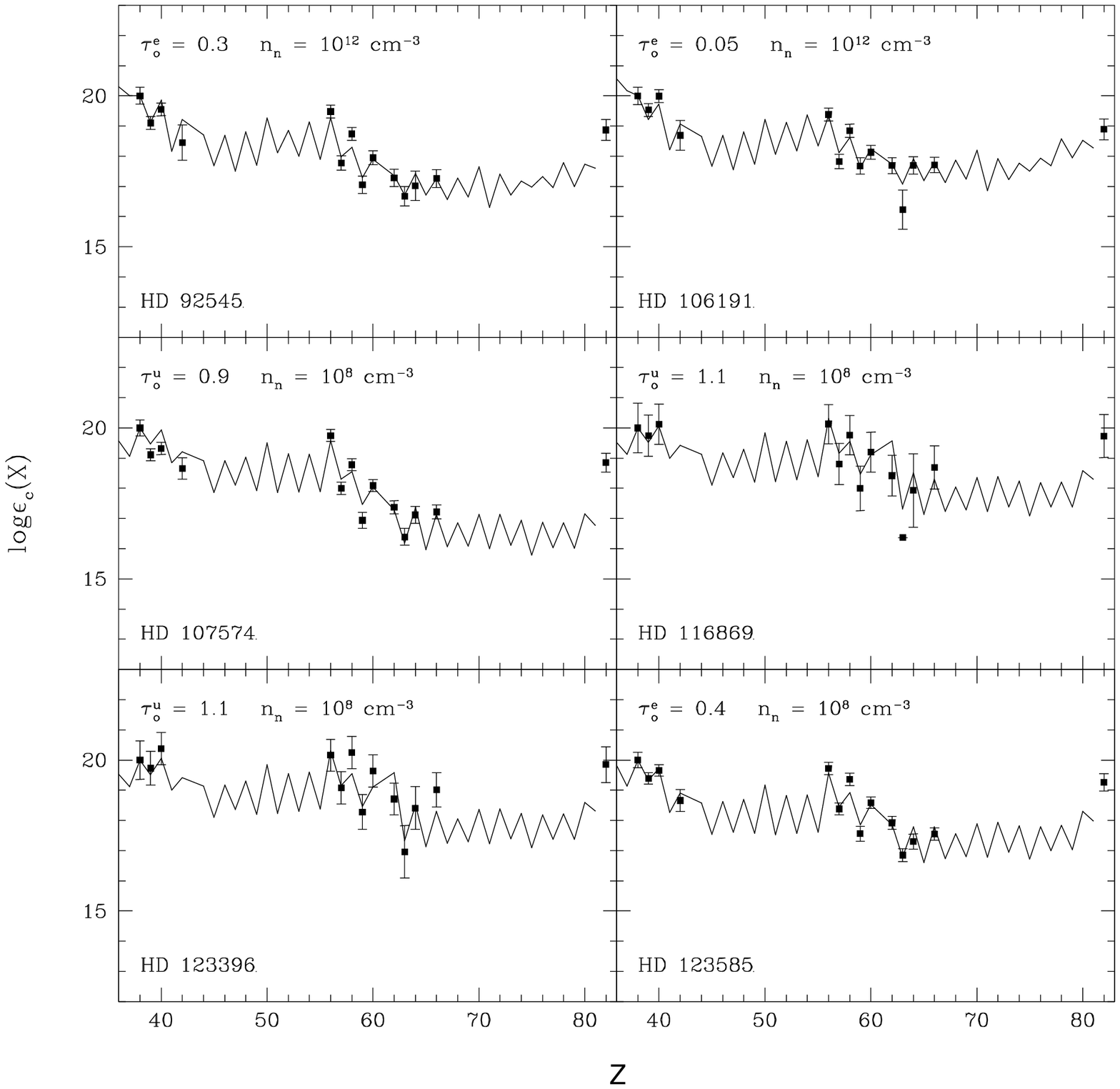}}
\caption{\label{expo3} Same as Figure \ref{expo1} for other 6 sample stars.}
\end{figure}

\begin{figure}[ht!]
\centerline{\includegraphics[totalheight=9.5cm]{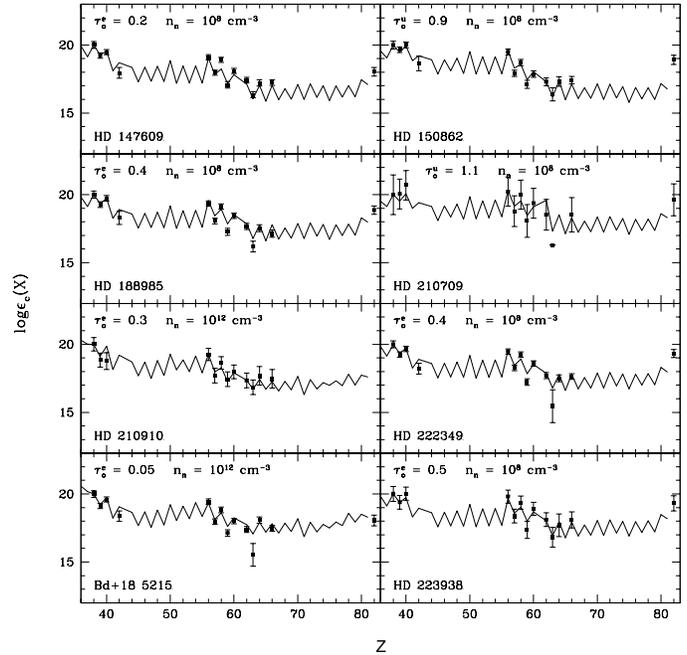}}
\caption{\label{expo4} Same as Figure \ref{expo1} for other 8 sample stars.}
\end{figure}

The best fit between theoretical predictions and observational data correspond
to the smallest value of S$\sp 2$. S$\sp 2$ was calculated for all P values from
\citet{malaney87a} and \citet{malaney87b} tables. Table \ref{tau} shows the best 
fittings of S$\sp 2$ and $\tau_o$ for each Malaney's table.
Figures \ref{expo1} to \ref{expo4} show the best fittings obtained for each present 
sample barium star, showing the values of $\tau_o$ and
neutron density n$\sb n$, in columns 8 and 9 of Table \ref{tau}.

For some elements of some barium stars the differences between abundances 
of barium and normal stars were too small, although 
$\epsilon_{nor}$(X) $<$ $\epsilon$(X). It is the case of
Mo (HD 749), Eu (HD 89948, HD 116869, HD 210709) and Gd (HD 22589).
In these cases, the uncertainty on $\log\epsilon_c$(X) is too large, decreasing 
the quality of the fit. They were withdrawn before using equation \ref{quaju}, 
and are easily seen in Figures \ref{expo1} to \ref{expo4} due their discrepant 
values with respect to theoretical predictions, and there are no error bars for them.
Except for such cases, there is a good agreement between observed and 
theoretical data shown in Figures \ref{expo1} to \ref{expo4}. For most stars,
the best fit was that for which theoretical predictions consider an exponential
neutron exposure distribution, with neutron density 10$\sp {12}$ cm$\sp {-3}$ 
for 5 stars and 10$\sp 8$ cm$\sp {-3}$ for 14 stars. For 7 stars, the best fit was
found for a single neutron exposure, and
neutron density 10$\sp 8$ cm$\sp {-3}$. This result is curious since the
exponential neutron exposure distribution is more well accepted in the literature.
However, \citet{busso99} suggested that s-process is a result of a number of 
single exposures instead of an 
exponential distribution. \citet{claudio03} also obtained a single exposure
and neutron density 10$\sp 8$ cm$\sp {-3}$ as the best fit for 2 barium stars. 
One of them, HD 87080, for which they found $\tau$ = 1.0, is in common with the
present sample, for which we found a best fit with an exponential distribution of
$\tau$ = 0.6 and neutron density 10$\sp 8$ cm$\sp {-3}$. The reason of this
difference is unclear. The metallicities adopted were very close (see Tables 9 and 11
from paper I), however, values of $\log\epsilon$(X) in this work are usually
higher than those from \citet{claudio03}. Least-squares fittings were used here
to compute abundances of normal stars (equation \ref{abaju}), while 
\citet{claudio03} used the sum of solar abundance with
star metallicity as a reference.

\begin{table*}
\caption{Results on neutron exposures. Columns 2 and 3 correspond to results of the
fittings to theoretical predictions of \citet{malaney87a} and columns 4 to 7, those of \citet{malaney87b}. 
S$\sp 2$(u), S$\sp 2$(e8), S$\sp 2$(e12) are the fit quality, respectively, for single exposure,
exponential exposure under neutron density 10$\sp 8$ cm$\sp {-3}$ and exponential exposure under
neutron density 10$\sp {12}$ cm$\sp {-3}$. $\tau_o^a$, $\tau_o^b$ and $\tau_o^c$ are the mean
distributions of neutron exposure corresponding to S$\sp 2$(u), S$\sp 2$(e8) and S$\sp 2$(e12),
respectively. $\tau_o$(f) and $n\sb n$ are results corresponding to smaller S$\sp 2$. Column 10
shows if the best fit corresponds to exponential (exp) or single (sing) exposure. Columns 11 to 13
correspond to results of fit of $\sigma$N curves, being $\chi_{red}$ the quality of the fit.
Numbers in parenthesis are errors in last decimals, and were estimated directly from the fittings.
The uncertainties on $\tau_o^a$, $\tau_o^b$ and $\tau_o^c$ were estimated in 0.1 mb$\sp {-1}$. 
For all cases, $\tau_o$ is given in mb$\sp {-1}$. Number in parenthesis are errors in last decimals.}
\label{tau}
\begin{center}
\begin{tabular}{lcccccclllcllc}
\hline
\noalign{\smallskip}
star & S$^2$(u) & $\tau_o^a$ & S$^2$(e8) & $\tau_o^b$ & S$^2$(e12) & $\tau_o^c$ & $\tau_o$(f) & $n_n$ &
exp & $\tau_o$($\sigma$N) & G(\%) & $\chi_{red}$ \\
\noalign{\smallskip}
\hline
\noalign{\smallskip}
HD 749	    & 0.69 & 1.00 & 0.52 & 0.80 & 0.77 & 0.80  & 0.8  & 10$^8$  & exp  & 0.406(5)  & 0.568(19)  & 0.73 \\
HR 107      & 0.98 & 0.10 & 1.47 & 0.20 & 1.30 & 0.30  & 0.1  & 10$^8$  & sing & 0.43(6)   & 0.083(20)  & 0.56 \\
HD 5424     & 0.85 & 1.10 & 1.15 & 0.80 & 1.71 & 0.80  & 1.1  & 10$^8$  & sing & 1.05(1)   & 0.086(1)   & 0.94 \\
HD 8270     & 1.33 & 0.90 & 1.21 & 0.20 & 1.46 & 0.30  & 0.2  & 10$^8$  & exp  & 0.27(1)   & 0.35(3)    & 0.41 \\
HD 12392    & 0.46 & 1.10 & 0.38 & 0.80 & 0.79 & 0.80  & 0.8  & 10$^8$  & exp  & 0.40(2)   & 0.73(3)    & 0.59 \\
HD 13551    & 1.57 & 0.90 & 1.25 & 0.20 & 1.44 & 0.30  & 0.2  & 10$^8$  & exp  & 0.22(1)   & 1.06(56)   & 0.50 \\
HD 22589    & 1.10 & 0.80 & 0.89 & 0.20 & 1.27 & 0.20  & 0.2  & 10$^8$  & exp  & 0.187(23) & 1.25(54)   & 0.52 \\
HD 27271    & 0.26 & 0.90 & 0.24 & 0.40 & 0.16 & 0.05  & 0.05 & 10$^{12}$ & exp  & 0.272(14) & 0.568(19)  & 0.35 \\
HD 48565    & 3.14 & 1.00 & 1.34 & 0.40 & 2.66 & 0.05  & 0.4  & 10$^8$  & exp  & 0.659(10) & 0.0631(23) & 0.32 \\
HD 76225    & 0.80 & 0.10 & 0.71 & 0.20 & 1.83 & 0.30  & 0.2  & 10$^8$  & exp  & 0.25(1)   & 0.83(12)   & 0.38 \\
HD 87080    & 2.22 & 1.00 & 1.17 & 0.60 & 2.49 & 0.80  & 0.6  & 10$^8$  & exp  & 0.71(1)   & 0.145(5)   & 0.48 \\
HD 89948    & 0.67 & 0.10 & 0.45 & 0.20 & 1.65 & 0.30  & 0.2  & 10$^8$  & exp  & 0.24(2)   & 0.77(36)   & 0.30 \\
HD 92545    & 1.42 & 0.90 & 1.22 & 0.20 & 0.85 & 0.30  & 0.3  & 10$^{12}$ & exp  & 0.291(3)  & 0.261(14)  & 0.40 \\
HD 106191   & 1.36 & 0.90 & 1.16 & 0.40 & 0.72 & 0.05  & 0.05 & 10$^{12}$ & exp  & 0.383(1)  & 0.130(1)   & 0.38 \\
HD 107574   & 1.91 & 0.90 & 2.28 & 0.20 & 2.05 & 0.30  & 0.9  & 10$^8$  & sing & 0.298(2)  & 0.245(5)   & 0.42 \\
HD 116869   & 0.14 & 1.10 & 0.18 & 0.80 & 0.33 & 0.80  & 1.1  & 10$^8$  & sing & 0.55(2)   & 0.0646(22) & 0.55 \\
HD 123396   & 0.46 & 1.10 & 0.84 & 0.80 & 1.24 & 0.80  & 1.1  & 10$^8$  & sing & 0.54(2)   & 0.0315(22) & 0.54 \\
HD 123585   & 2.17 & 1.00 & 0.98 & 0.40 & 2.14 & 0.50  & 0.4  & 10$^8$  & exp  & 0.46(1)   & 0.275(15)  & 0.46 \\
HD 147609   & 2.33 & 0.10 & 1.90 & 0.20 & 2.42 & 0.30  & 0.2  & 10$^8$  & exp  & 0.288(7)  & 1.07(11)   & 0.41 \\
HD 150862   & 0.66 & 0.90 & 1.57 & 0.20 & 1.11 & 0.30  & 0.9  & 10$^8$  & sing & 0.271(3)  & 0.489(26)  & 0.47 \\
HD 188985   & 1.83 & 0.90 & 1.77 & 0.40 & 1.94 & 0.40  & 0.4  & 10$^8$  & exp  & 0.401(4)  & 0.225(5)   & 0.39 \\
HD 210709   & 0.11 & 1.10 & 0.19 & 0.80 & 0.26 & 0.80  & 1.1  & 10$^8$  & sing & 0.314(3)  & 0.303(9)   & 0.38 \\
HD 210910   & 0.55 & 0.10 & 0.63 & 0.20 & 0.38 & 0.30  & 0.3  & 10$^{12}$ & exp  & 0.44(4)   & 0.075(13)  & 0.52 \\
HD 222349   & 3.02 & 0.90 & 1.28 & 0.40 & 2.00 & 0.05  & 0.4  & 10$^8$  & exp  & 0.379(8)  & 0.170(12)  & 0.46 \\
BD+18 5215  & 2.41 & 0.90 & 2.74 & 0.40 & 1.50 & 0.05  & 0.05 & 10$^{12}$ & exp  & 0.274(1)  & 0.503(2)   & 0.63 \\
HD 223938   & 0.26 & 1.00 & 0.20 & 0.50 & 0.29 & 0.70  & 0.5  & 10$^8$  & exp  & 0.384(5)  & 0.229(9)   & 0.34 \\

\hline
\end{tabular}
\end{center}
\end{table*}

\subsection{$\sigma$N Curve}
Another way to evaluate the neutron exposure nature is through $\sigma$N curves.
For such, the cross section values ($\sigma_c$ (30keV), column 13 of 
Table \ref{sechoque}) and abundances related to the s-process main component of 
each nuclide (from equation \ref{partesiso}) are needed.
Results of $\sigma$N curve for each barium star are shown in column 11 of Table \ref{resproc}.

The uncertainty on $\sigma$N is given by the expression

\begin{equation}
\sigma_{\sigma N}=\sigma_c\epsilon_s(X)\sqrt{\biggl({\sigma_{\epsilon s}\over \epsilon_s}\biggr)^2+\biggl({\sigma_{\sigma c}\over \sigma_c}\biggr)^2}.
\end{equation}

In the literature, $\sigma$N is found in Si scale, where $\epsilon$(Si) = 10$\sp 6$.
\citet{andgre89} provide the following relation between the Si scale and the scale where
$\log\epsilon$(X)=$\log(n_x/n_H)$ + 12:
\begin{equation}
\epsilon(X)_{Si} = 10^{\log\epsilon(X) - x}
\end{equation}
where $x$ = 1.554 $\pm$ 0.020, and the uncertainty is 
\begin{equation}
\sigma_{\epsilon Si}=\epsilon(X)_{Si}\sqrt{(\ln{10}\sigma_{\log\epsilon Si})^2+\sigma_x^2}.
\end{equation}

To compute abundances relative to the s-process main component in the Si scale it is 
necessary to transform abundances of normal stars to this scale. Then the distribution
relative to each process in the Si scale is done in the same way to usual scales.
Column 16 of Table \ref{resproc} shows the results of $\sigma$N = $\sigma_c\epsilon_s$(X)$_{Si}$ 
for nuclides of barium stars.

For comparison purposes, the distribution of abundance in several nucleosynthetic 
processes were also done. The total abundance of an element of a
barium star was distributed in solar proportions, with fractions corresponding to
s-, r- and p-processes, shown in columns 3 to 7 in Table \ref{sechoque}.
Results for this $\sigma$N curve are shown in column 18 of Table \ref{resproc}, and
they are represented by open symbols in Figures \ref{abiso1} to \ref{abiso4}.
These results are lower than others because the overabundance derived from the s-process
is neglected.

\begin{figure*}[ht!]
\centerline{\includegraphics[totalheight=6.0cm]{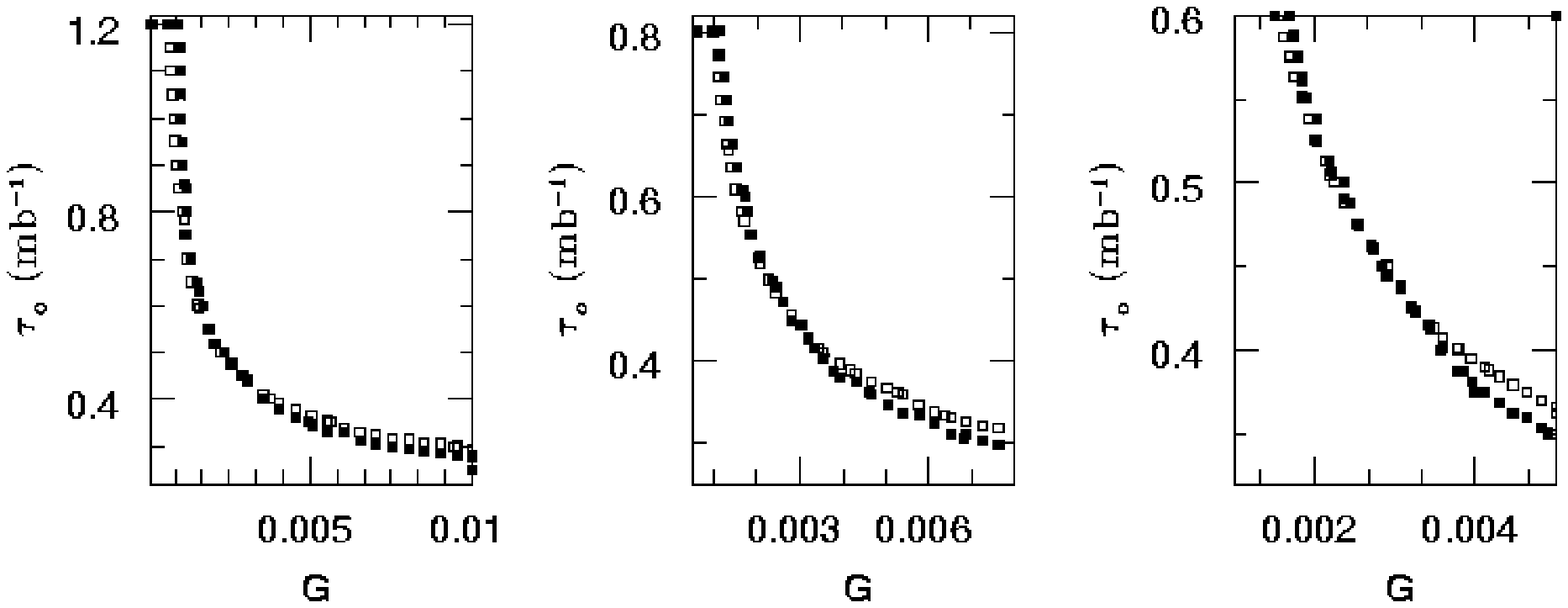}}
\caption{\label{ajusign} Example of fit of parameters G and $\tau_o$ for
the barium star HD 123585.
Open symbols correspond to the use of ${\partial\chi\over \partial G}=0$ 
and filled symbols, ${\partial\chi\over \partial\tau_o}=0$. From left to right the
plot is zoomed in order to better visualize the crossing of the two functions.}
\end{figure*}

A theoretical $\sigma$N curve is calculated with equation \ref{cursin}, where
two parameters ($\tau_o$ and G) must be determined by fitting observed data.
A robust statistics was used, which, in this work, consists in finding values of 
G and $\tau_o$ that simultaneously minimize the sum of absolute deviations 
represented by
$$\chi = \sum\mid{f(x_k)-y_k}\mid,$$
where $f(x_k)$ is the theoretical value calculated with equation \ref{cursin} and
$y_k$ is the value calculated directly for $\sigma$N by using observed data.
Data very far from the curve were neglected. Figure \ref{ajusign} shows the 
functions that minimize $\chi$ relative to each parameter. The crossing point 
of these functions provides the values of G and $\tau_o$ that minimize
simultaneously both functions. More details on this
procedure are found in \citet{celo00}. Table \ref{tau} shows results of the
fittings and their $\chi_{red}=\chi/(n-2)$, where $n$ is the number of data,
representing the goodness of fits.

\begin{figure*}[ht!]
\centerline{\includegraphics[totalheight=8cm]{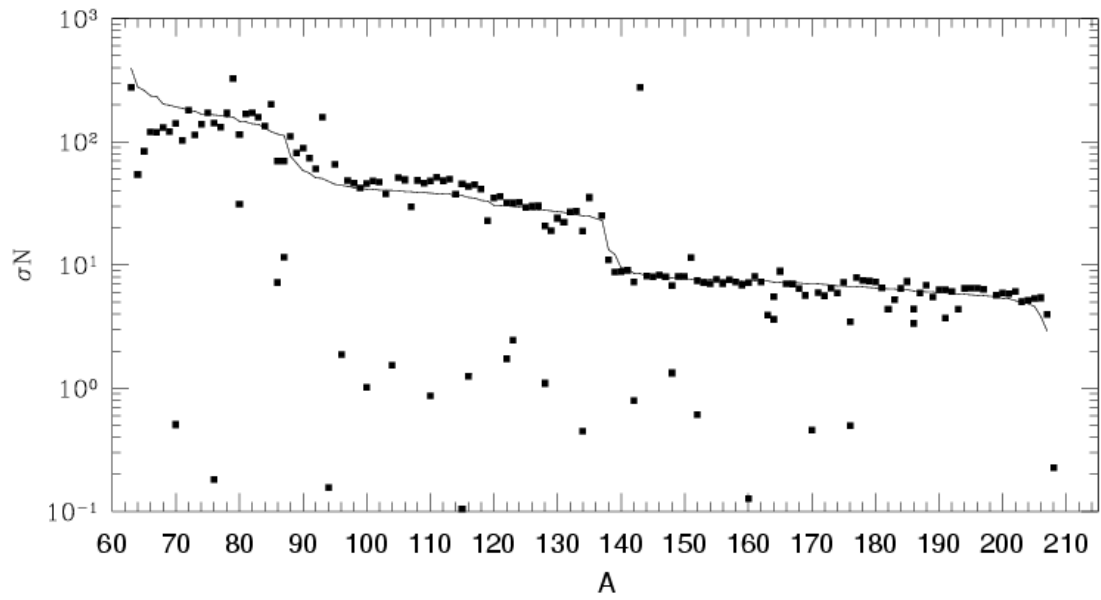}}
\caption{\label{sigNsolteo} Solar $\sigma$N curve. Abundances related to
s-process main component were taken from \citet{arlandini99}.}
\end{figure*}

\begin{figure*}[ht!]
\centerline{\includegraphics[totalheight=19.5cm]{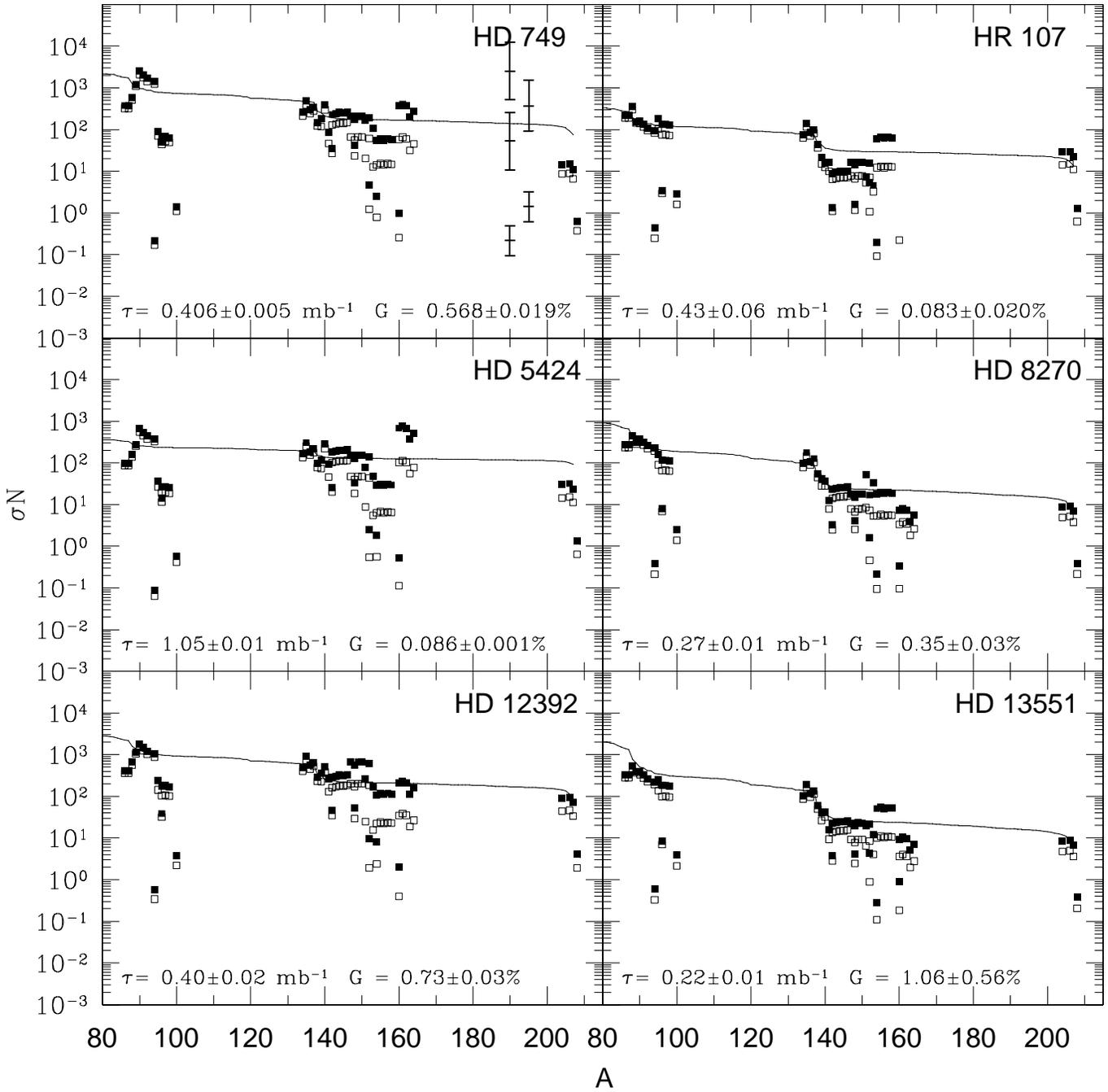}}
\caption{\label{abiso1} $\sigma$N curves for sample barium stars. Open squares represent
solar distribution of abundances (column 18 of Table \ref{resproc}) and filled
squares represent the distribution taking into account the overabundance of
barium stars (column 16 of Table \ref{resproc}). Error bars in the HD 749 panel
at A=190 and A=195, indicate the typical errors in each region of the logarithmic scale.}
\end{figure*}

\begin{figure*}[ht!]
\centerline{\includegraphics[totalheight=19.5cm]{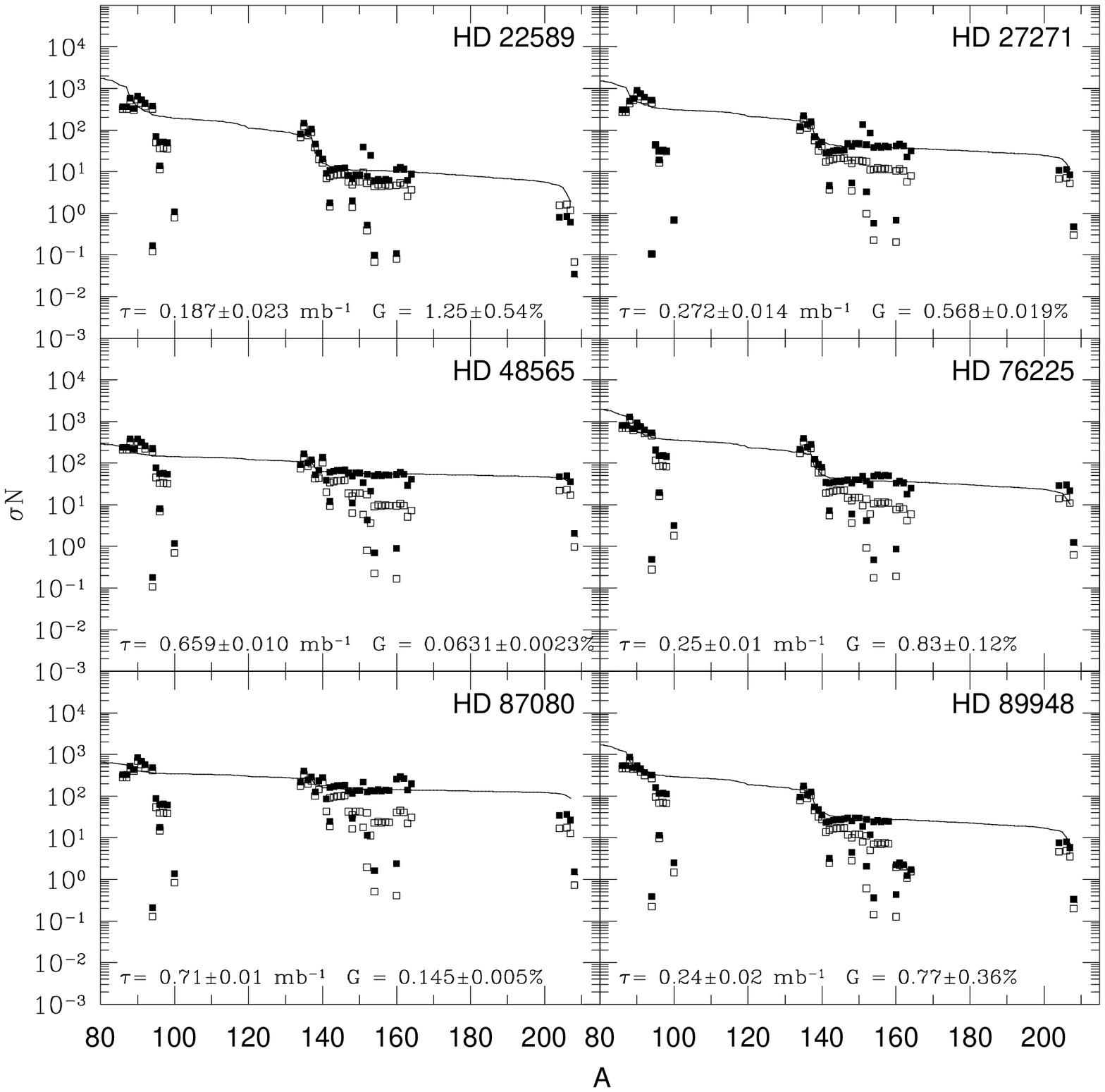}}
\caption{\label{abiso2} Same as Figure \ref{abiso1} for other 6 sample stars.}
\end{figure*}

\begin{figure*}[ht!]
\centerline{\includegraphics[totalheight=19.5cm]{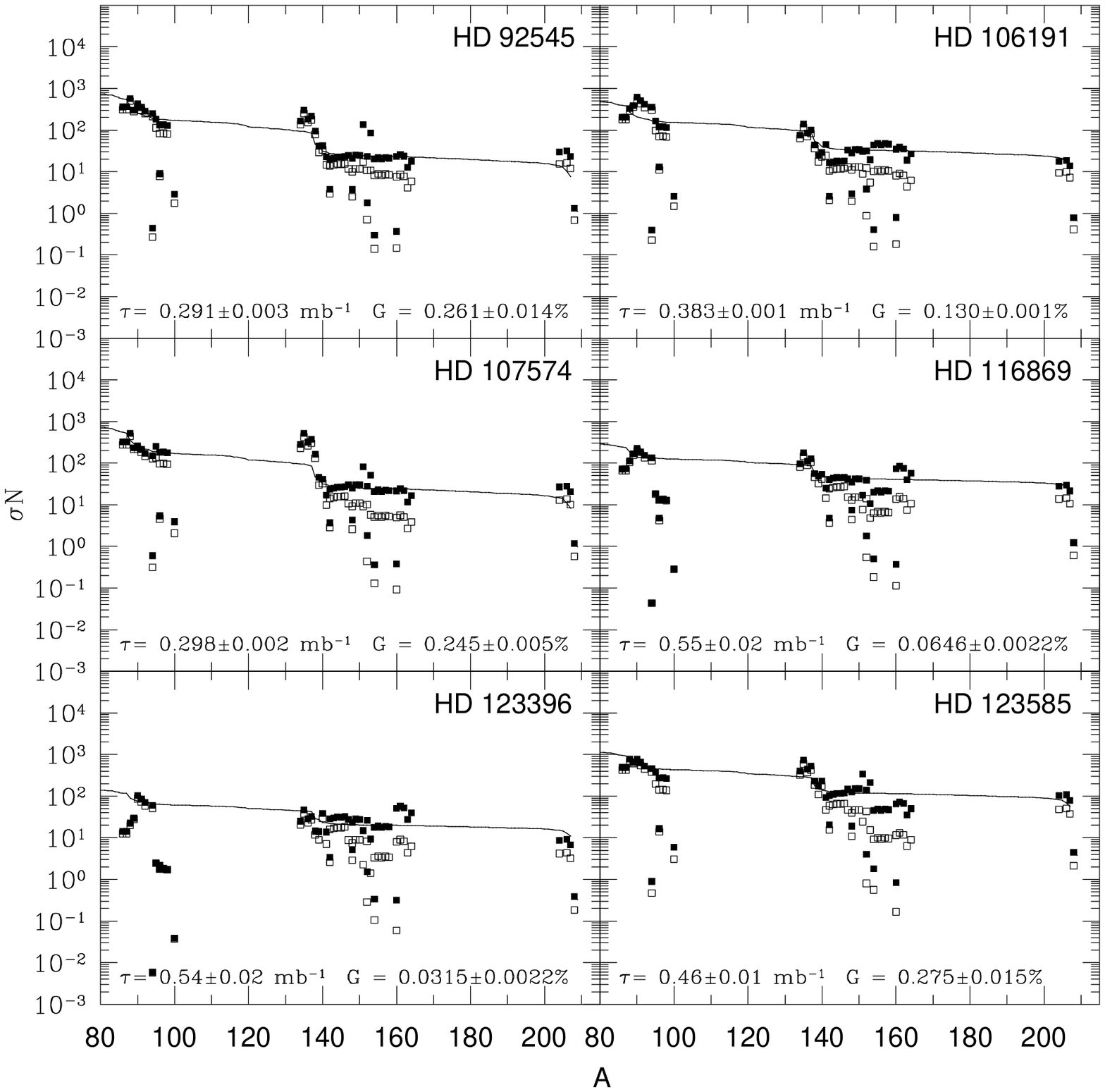}}
\caption{\label{abiso3} Same as Figure \ref{abiso1} for other 6 sample stars.}
\end{figure*}

\begin{figure*}[ht!]
\centerline{\includegraphics[totalheight=19.5cm]{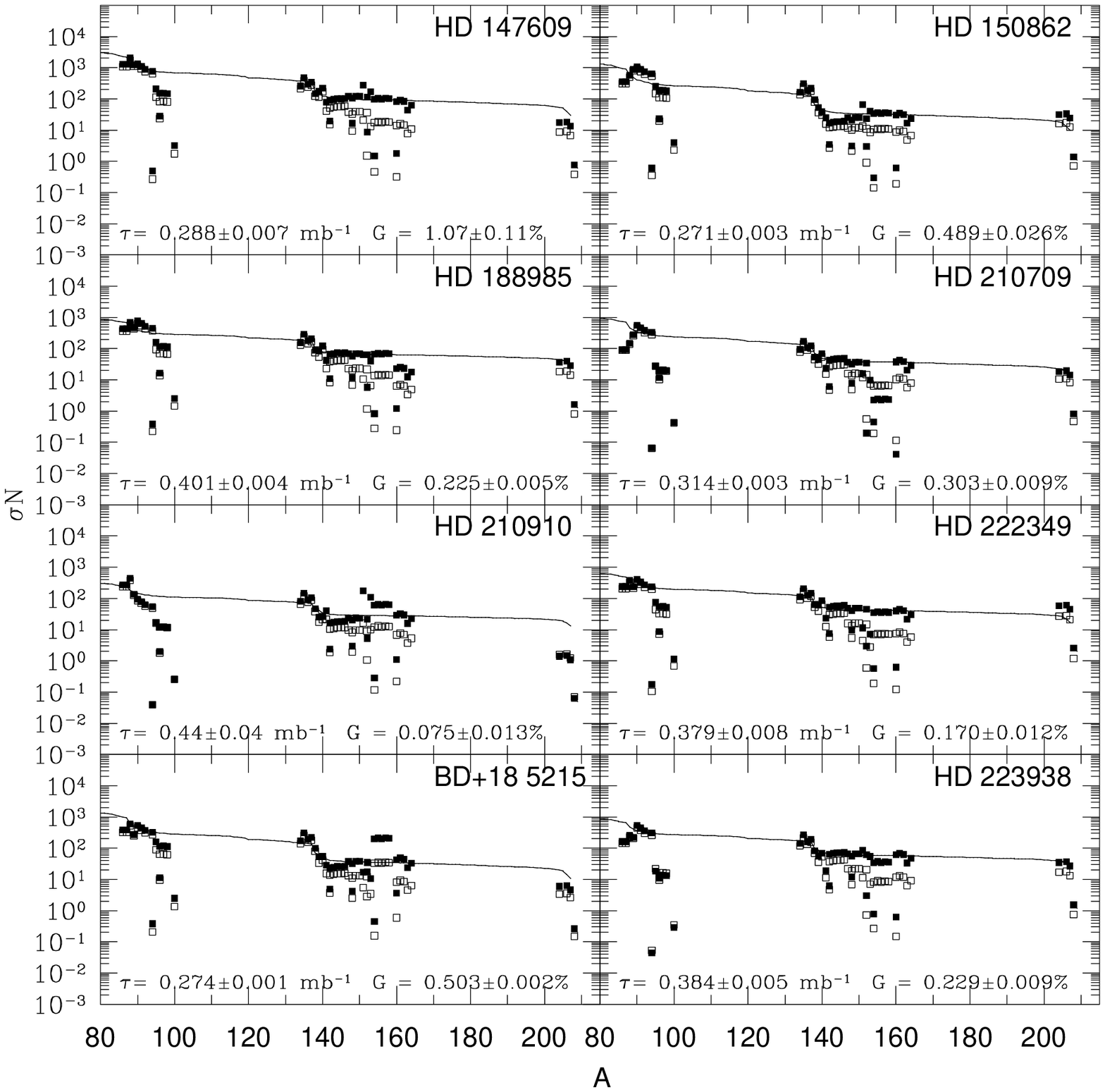}}
\caption{\label{abiso4} Same as Figure \ref{abiso1} for other 8 sample stars.}
\end{figure*}

The method was tested for the Sun, by using data from \citet{arlandini99}. The 
resulting line from the fitting is shown in Figure \ref{sigNsolteo}, in which
$\tau_o$ = 0.345 $\pm$  0.005 mb$\sp{-1}$ and G = 0.043 $\pm$  0.002\%. 
Figure 19b from \citet{kapp89} shows their solar $\sigma$N curve,
resulting $\tau_o$ = 0.30 $\pm$  0.01 mb$\sp{-1}$ and G = 0.043 $\pm$  0.002\%. 
Figures \ref{abiso1} to \ref{abiso4} show that the uncertainties, 
represented by error bars, are very large
for barium stars, and for this reason they were neglected in computing the quality 
of the fit. Branching and too discrepant data were also neglected.

For the present sample, theoretical $\sigma$N curves fit very well the observed data,
as shown in Figures \ref{abiso1} to \ref{abiso4}. This confirms that the solar isotopic
composition is adequate for barium stars, and that the transfer of enriched material 
keeps approximately the
solar proportions of each nuclide. In order to build the theoretical curve, one
considers that at the beginning of the neutron capture chain, abundances of all
elements heavier than iron are null. This supposition gets more distant from the
truth as the metallicity increases. Furthermore, the fit brings some
difficulties: a) there is a lack of elements along the curve due to the
difficulty in finding lines in the spectrum, preventing a
higher quality of the fit between theoretical and observed curves; b) barium stars
data show a large dispersion; c) branching values were not considered; d) numerical 
solution of equation \ref{cursin} is difficult, given that two parameters to be
fitted appear in non linear form in the equation. 
The good agreement between theoretical
and observed curves is interesting, taking into account all assumptions
and difficulties found in their building.

\subsection{s-process indices}

\begin{figure*}[ht!]
\centerline{\includegraphics[totalheight=10.0cm]{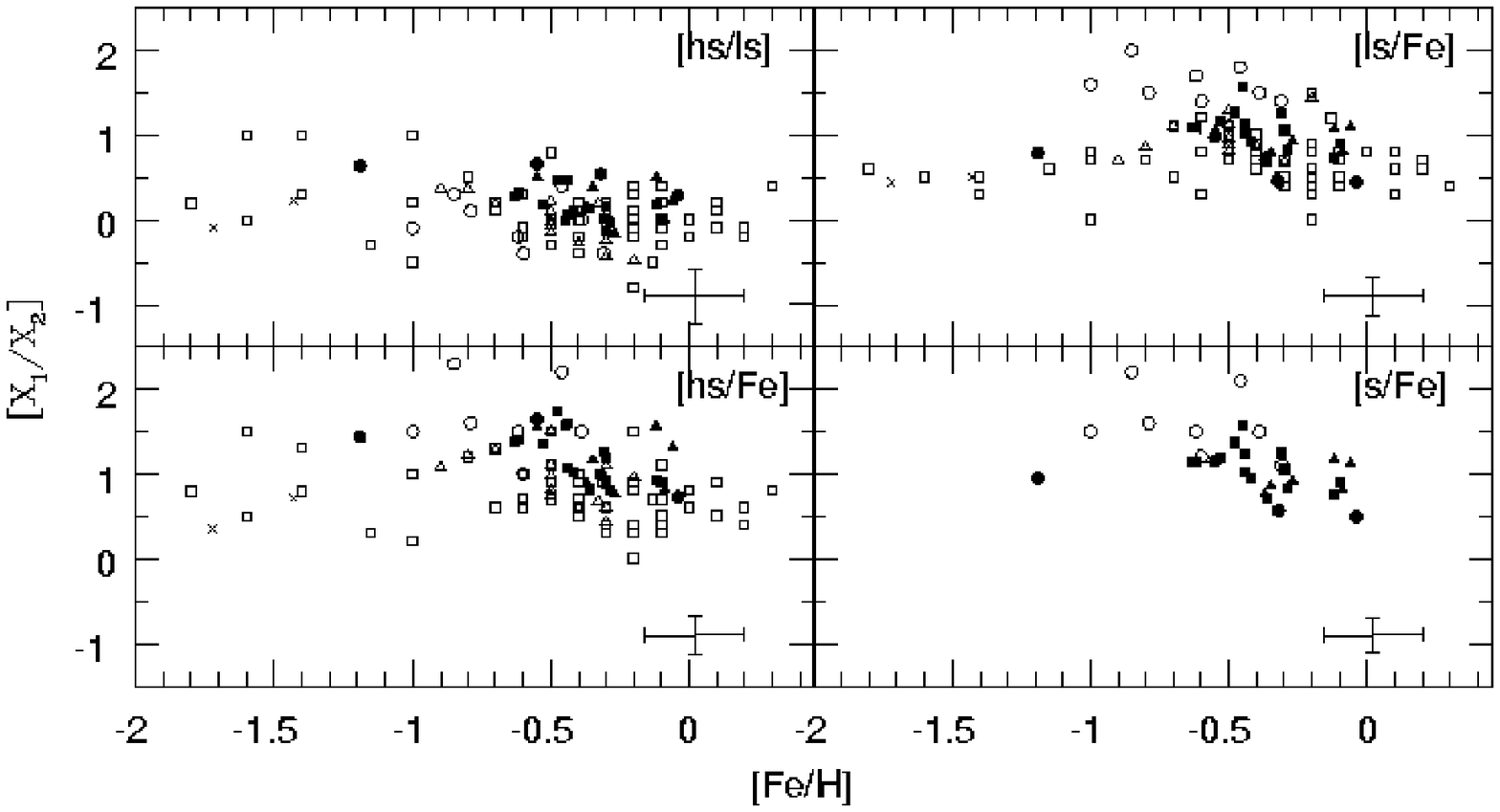}}
\caption{\label{shslsfe} [hs/ls], [ls/Fe], [hs/Fe] and [s/Fe] vs. [Fe/H].
Filled symbols indicate barium stars, as in Figure \ref{lepsbaagb}; 
open circles are post-AGBs; crosses are data from \citet{junqueira01}; 
open squares are data from \citet{lb91};
open triangles are data from  \citet{north94}. The uncertainties indicated 
are the higher values shown in Table \ref{inds}.}
\end{figure*}

One way to relate light and heavy s-elements is through hs and ls indices,
defined as the mean of abundances of light and heavy elements, respectively.
According to section 4.3, Sr, Y and Zr were included into ls, and
Ba, La, Ce and Nd were included into hs.
For the s index, all these heavy and light s-elements were considered.
[hs/ls] = [hs/Fe] - [ls/Fe] was also computed. In the case 
of missing abundances, such element was excluded from the index.
Figure \ref{shslsfe} and Table \ref{inds} show 
[hs/ls], [ls/Fe], [hs/Fe] and [s/Fe] vs. [Fe/H].
Uncertainties on [s/Fe], [ls/Fe] and [hs/Fe] have to take into account
the contribution of the uncertainties on abundances of each element included
in the index. 

A behaviour of [hs/ls] as a function of $\tau_o$ can be inferred from the slope 
of the $\sigma$N curve. The larger $\tau_o$, the smaller is the slope, and, 
at the same time, a smaller slope means
larger abundance of heavy s-elements located on the right end of the curve.
The conclusion is that the larger $\tau_o$, the larger also is [hs/ls], as 
shown in Figure 17 of \citet{wallers97}.
It is reasonable considering that the chain of formation of s-elements cannot 
go far if the neutron flux is low. 

\begin{figure}[ht!]
\centerline{\includegraphics[totalheight=5.0cm]{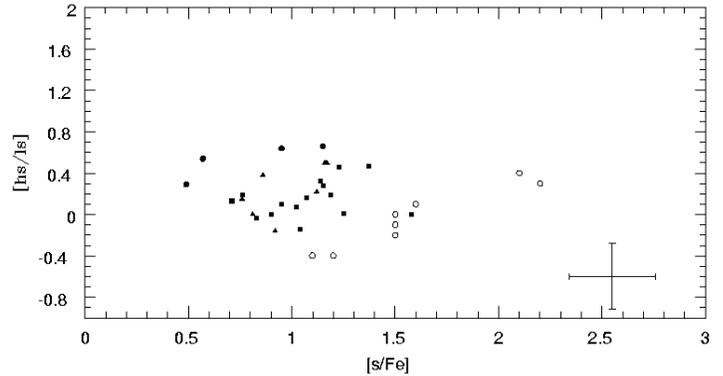}}
\caption{\label{hslss} [hs/ls] vs. [s/Fe]. 
Symbols are the same as in Figure \ref{lepsbaagb}. Error bars in bottom right 
showing the largest uncertainties.}
\end{figure}

\begin{figure}[ht!]
\centerline{\includegraphics[totalheight=9.0cm]{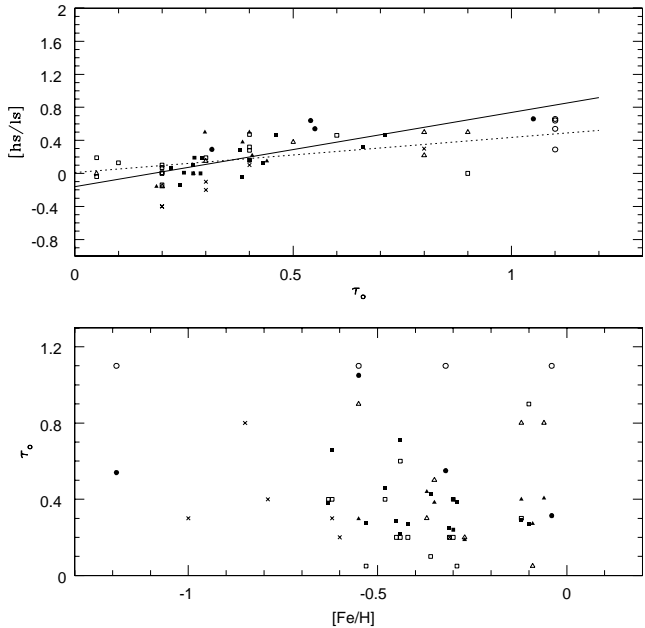}}
\caption{\label{tauhsls} [hs/ls] vs. $\tau_o$ (upper panel), and $\tau_o$ vs. [Fe/H]
(lower panel). 
Crosses are post-AGBs from \citet{reyniers04} and \citet{winckel00}.
Circles, triangles and squares represent the sample barium stars according to
log g, as in Figure \ref{lepsbaagb}. Open symbols correspond to $\tau_o$ values
derived from theoretical predictions by \citet{malaney87a,malaney87b} and
filled symbols are those from $\sigma$N curves. Solid line is the least-squares fitting
with $\tau_o$ from $\sigma$N curves and dotted, with $\tau_o$ from theoretical 
predictions.}
\end{figure}

During the third dredge-up, an amount of protons is introduced in the intershell, a 
region composed of helium located between helium and hydrogen burning shells. These
protons are captured by $^{12}$C and form $^{13}$C through the reaction
$^{12}$C(p,$\gamma$)$^{13}$N($\beta^+\nu$)$^{13}$C or $^{14}$N through 
$^{13}$C(p,$\gamma$)$^{14}$N, creating the $^{13}$C pocket.
If the thermal pulse is independent of metallicity, protons are
introduced in the intershell in the same amounts for higher or lower metallicities. 
It means that the neutron source $\sp {13}$C($\alpha$,n)$\sp {16}$O,
likely the main neutron source in AGB stars, is independent of metallicity. 
However the neutron number by seed iron nucleus
will be larger at low metallicities \citep{clayton88}. If $\tau_o$ is 
proportional to [hs/ls] and inversely proportional to [Fe/H], [hs/ls]
will be also inversely proportional to [Fe/H], so it is expected that
the lower the metallicity, the larger the ratio [hs/ls]. Figure \ref{tauhsls}
confirms a correlation between  $\tau_o$ and [hs/ls] for barium 
and post-AGB stars. However, the anticorrelation between [hs/ls] and [Fe/H] 
is weak for barium and post AGBs stars, as shown in Figure \ref{shslsfe}.
In the same way, the anticorrelation between $\tau_o$ and [Fe/H] is not
confirmed in Figure \ref{tauhsls}. 
The yield of all s-elements to decrease with decreasing metallicities is
compatible with the secondary characteristic of s-process, which requires 
pre-existing seed nuclei.
At intermediate metallicities ($\approx$ -0.8) the Ba peak is dominant among
s-process products in AGB models \citep{busso99}. For higher metallicities the 
Zr peak dominates.
If giant as well as dwarf barium stars have the same physical origin for the
accretion of enriched material from a more evolved companion, it is reasonable to
expect that if the neutron exposure is higher for dwarf more metal deficient stars,
the same occurs for giants.

In Figure \ref{tauhsls}, $\tau_o$ derived from the $\sigma$N curve are mainly
within 1.8 $<$ $\tau_o$ $<$ 0.6, with less spread as compared to those derived
from theoretical predictions by \citet{malaney87a,malaney87b}, however
both show the same trend. For giant stars $\tau_o$ $\approx$ 1 was found from
theoretical predictions, but not from the $\sigma$N curves.

Solid and dashed lines in Figure \ref{tauhsls} are least-squares fittings, with parameters:
\begin{equation}
\label{hltaumal}
[{\rm hs/ls}] = (0.423 \pm 0.095)\tau_o + (0.011 \pm 0.043)
\end{equation}
\begin{equation}
\label{hltausign}
[{\rm hs/ls}] = (0.899 \pm 0.173)\tau_o + (-0.160 \pm 0.068)
\end{equation}
where equation \ref{hltaumal} uses $\tau_o$ values from theoretical
predictions by \citet{malaney87a,malaney87b}, with $\chi^2_{red}$ = 1.710 
and equation \ref{hltausign} uses $\tau_o$ values derived from $\sigma$N curves,
with $\chi^2_{red}$ = 1.464. Regarding $\chi^2_{red}$, in both 
cases the linear and increasing correlation toward higher values of $\tau_o$ have
good quality.

\begin{table}
\caption{s-process indices: s, ls and hs for present sample barium stars.}
{\scriptsize
\label{inds}
   $$ 
\begin{tabular}{lrrrr}
\hline
\noalign{\smallskip}
star & [s/Fe] & [ls/Fe] & [hs/Fe]& [hs/ls] \\
\noalign{\smallskip}
\hline
\noalign{\smallskip}
HD 749     & 1.12$\pm$0.21 & 1.09$\pm$0.23 &  1.31$\pm$0.22 &  0.22$\pm$0.32 \\
HR 107     & 0.71$\pm$0.08 & 0.69$\pm$0.09 &  0.82$\pm$0.07 &  0.13$\pm$0.11 \\
HD 5424    & 1.15$\pm$0.21 & 0.98$\pm$0.23 &  1.64$\pm$0.22 &  0.66$\pm$0.32 \\
HD 8270    & 0.95$\pm$0.08 & 0.93$\pm$0.09 &  1.03$\pm$0.07 &  0.10$\pm$0.11 \\
HD 12392   & 1.17$\pm$0.21 & 1.07$\pm$0.23 &  1.57$\pm$0.22 &  0.50$\pm$0.32 \\
HD 13551   & 1.02$\pm$0.08 & 1.02$\pm$0.09 &  1.08$\pm$0.07 &  0.07$\pm$0.11 \\
HD 22589   & 0.92$\pm$0.08 & 0.93$\pm$0.09 &  0.77$\pm$0.07 & -0.16$\pm$0.11 \\
HD 27271   & 0.81$\pm$0.21 & 0.81$\pm$0.23 &  0.81$\pm$0.22 &  0.00$\pm$0.32 \\
HD 48565   & 1.14$\pm$0.08 & 1.08$\pm$0.09 &  1.40$\pm$0.07 &  0.32$\pm$0.11 \\
HD 76225   & 1.25$\pm$0.08 & 1.25$\pm$0.09 &  1.26$\pm$0.07 &  0.01$\pm$0.11 \\
HD 87080   & 1.23$\pm$0.08 & 1.13$\pm$0.09 &  1.59$\pm$0.07 &  0.46$\pm$0.11 \\
HD 89948   & 1.04$\pm$0.08 & 1.06$\pm$0.09 &  0.92$\pm$0.07 & -0.14$\pm$0.11 \\
HD 92545   & 0.76$\pm$0.08 & 0.73$\pm$0.09 &  0.92$\pm$0.07 &  0.19$\pm$0.11 \\
HD 106191  & 0.83$\pm$0.08 & 0.83$\pm$0.09 &  0.80$\pm$0.07 & -0.04$\pm$0.11 \\
HD 107574  & 1.16$\pm$0.08 & 1.05$\pm$0.09 &  1.55$\pm$0.07 &  0.50$\pm$0.11 \\
HD 116869  & 0.57$\pm$0.21 & 0.45$\pm$0.23 &  0.99$\pm$0.22 &  0.54$\pm$0.32 \\
HD 123396  & 0.95$\pm$0.21 & 0.79$\pm$0.23 &  1.44$\pm$0.22 &  0.64$\pm$0.32 \\
HD 123585  & 1.37$\pm$0.08 & 1.27$\pm$0.09 &  1.74$\pm$0.07 &  0.47$\pm$0.11 \\
HD 147609  & 1.58$\pm$0.08 & 1.58$\pm$0.09 &  1.58$\pm$0.07 &  0.00$\pm$0.11 \\
HD 150862  & 0.90$\pm$0.08 & 0.90$\pm$0.09 &  0.91$\pm$0.07 &  0.00$\pm$0.11 \\
HD 188985  & 1.07$\pm$0.08 & 1.05$\pm$0.09 &  1.20$\pm$0.07 &  0.16$\pm$0.11 \\
HD 210709  & 0.49$\pm$0.21 & 0.44$\pm$0.23 &  0.73$\pm$0.22 &  0.29$\pm$0.32 \\
HD 210910  & 0.76$\pm$0.21 & 0.74$\pm$0.23 &  0.89$\pm$0.22 &  0.15$\pm$0.32 \\
HD 222349  & 1.15$\pm$0.08 & 1.10$\pm$0.09 &  1.38$\pm$0.07 &  0.28$\pm$0.11 \\
BD+18 5215 & 1.19$\pm$0.08 & 1.17$\pm$0.09 &  1.35$\pm$0.07 &  0.19$\pm$0.11 \\
HD 223938  & 0.86$\pm$0.21 & 0.79$\pm$0.23 &  1.17$\pm$0.22 &  0.38$\pm$0.32 \\
\hline
\noalign{\smallskip}
\end{tabular}
   $$ 
}
\end{table}

\begin{table}

\caption{s-process indices s, ls and hs for barium stars collected in the literature.
References: N94 - \citet{north94}; LB91 - \citet{lb91}.}
{\footnotesize\label{indlit}
   $$ 
\begin{tabular}{lrrrrr}
\hline
\noalign{\smallskip}
star & [Fe/H] & [ls/Fe] & [hs/Fe]& [hs/ls] & ref \\
\noalign{\smallskip}
\hline
\noalign{\smallskip}
48565  & -0.90 & 0.70 & 1.07 &  0.37 & N94  \\
76225  & -0.50 & 1.13 & 1.11 & -0.02 & N94  \\
89948  & -0.13 & 1.20 & 0.70 & -0.50 & LB91 \\
92545  & -0.33 & 0.49 & 0.68 &  0.19 & N94  \\
106191 & -0.40 & 0.86 & 0.59 & -0.27 & N94  \\
107574 & -0.80 & 0.85 & 1.22 &  0.37 & N94  \\
123585 & -0.50 & 1.28 & 1.50 &  0.22 & N94  \\
123585 & -0.50 & 1.10 & 1.10 &  0.00 & LB91 \\
147609 & -0.50 & 0.88 & 0.98 &  0.10 & N94  \\
150862 & -0.30 & 1.05 & 0.61 & -0.44 & N94  \\
150862 & -0.20 & 0.60 & 0.40 & -0.20 & LB91 \\
188985 & -0.30 & 1.02 & 1.10 &  0.08 & N94  \\
\hline               
\noalign{\smallskip} 
\end{tabular}        
   $$                
}
\end{table}

Figure 10 from \citet{winckel00} shows [hs/ls] vs. [Fe/H] including
data from several previous work.
Despite the dispersion, [hs/ls] increases toward higher [Fe/H].
The dependence of neutron exposure on [Fe/H] could indicate that other
important parameters affect dredge-up events and nucleosynthetic processes
along the red giant branch evolution.

For the present sample barium stars, [s,ls,hs/Fe] and [hs/ls] slightly decrease
toward higher metallicities, in the range 
0.45 $\leq$ [s/Fe] $\leq$ 1.6, 0.4 $\leq$ [ls/Fe] $\leq$ 1.6,
0.7 $\leq$ [hs/Fe] $\leq$ 1.75 and -0.2 $\leq$ [hs/ls] $\leq$ 0.7.
According to \citet{wheeler89}, [s/Fe] = 0 for less evolved stars
in this same range of metallicities, therefore, the results found for
the present sample show the overabundance of s-elements in barium stars.

One has to be careful in comparing the indices of the present work to those of 
post-AGBs in Figures \ref{relacsr}, \ref{hslss} and \ref{shslsfe}, given that
\citet{reyniers04} and \citet{winckel00} included Sm and not Ce in s and hs,
and only Y and Zr in ls. Figure \ref{hslss} shows that [hs/ls] vs. [s/Fe] 
has a large dispersion for barium stars whereas for post-AGBs there is a linear 
correlation. Figure 8 from \citeauthor{reyniers04} shows the same correlation but
using different combinations of the elements included in the indices. One of
them uses Ce, however, the linear correlation is still present comparing with
using Ba, La, Nd and Sm in hs. Other configurations show larger
dispersion, hence the presence of Ce in hs and s is not the origin of the
dispersion for barium stars.

According to Table \ref{relacsrt} and Figure \ref{relacsr}, there is no difference 
as a function of log g, confirming that the overabundances 
characteristic of a barium star do not depend on luminosity classes.

%

\section{Conclusions}
In this work, s-, r- and p-processes in barium stars were 
studied relative to normal stars. 

Abundances of elements with lower contribution from the s-process main
component are closer to the normal stars, whereas elements with higher
s-contribution are more overabundant. However, r-elements such as 
Sm, Eu Gd and Dy are also enriched in barium stars sometimes at similar 
magnitudes as s-elements, and this occurs because the s-process main component 
chain includes the r-elements with A $<$ 209.

The ratios involving light s-elements (Sr, Y and Zr) and Ba, a heavy s-element,
are approximately constant in the range of metallicities of the present sample,
with a dispersion similar to normal stars and post-AGBs.

Considering that the abundance fraction due to all processes except the 
s-process main component was present in the proto-barium star, it was possible 
to isolate the fraction
corresponding to the s-process main component for barium stars.
The solar isotopic distribution of this abundance fraction was used to
build observed $\sigma$N curves. Their fittings to theoretical curves indicate
that it is possible to use solar isotopic mix to estimate the
contribution of the s-process main component for each isotope of an element.

The derivation of $\tau_o$ was obtained by fitting observed data to theoretical 
predictions and $\sigma$N curves. Theoretical predictions for 
abundances starting with Sr fit very well the observed data for the
present sample, providing an estimation for neutron exposure occurred in
AGB suplying s-process. The [hs/ls] vs. $\tau_o$ considering these theoretical
predictions linearly increases, whereas for
$\tau_o$ vs. [Fe/H] no conclusion was reached.
The $\tau_o$ values from $\sigma$N fittings also provide linear and increasing
$\tau_o$ vs. [hs/ls], but in this case, data are 
in the range 0.2 $\leq$ $\tau_o$ $\leq$ 0.8, showing 
lower dispersion than those for theoretical predictions.

Abundances obtained for barium stars are close to those for AGBs stars, but
they are usually lower. This is reasonable considering that only part of the surface
material of AGBs is transferred to the companion, that becomes a barium star.

%

\begin{acknowledgements}
We acknowledge partial financial support from the Brazilian Agencies CNPq
and FAPESP.
DMA acknowledges a FAPESP PhD fellowship n$^{\circ}$ 00/10405-8
and a FAPERJ post-doctoral fellowship n$^{\circ}$ 152.680/2004.
We are grateful to Marcelo Porto Allen for making available his
robust statistics code, and to the referee, Nils Ryde, for useful comments.
\end{acknowledgements}

%

\end{document}